# The physics of wind-blown sand and dust

**Jasper F. Kok[1], Eric J. R. Parteli[2,3], Timothy I. Michaels[4], and Diana Bou Karam[5]**

[1]Department of Earth and Atmospheric Sciences, Cornell University, Ithaca, NY, USA
[2]Departamento de Física, Universidade Federal do Ceará, Fortaleza, Ceará, Brazil
[3]Institute for Multiscale Simulation, Universität Erlangen-Nürnberg, Erlangen, Germany
[4]Southwest Research Institute, Boulder, CO USA
[5]LATMOS, IPSL, Université Pierre et Marie Curie, CNRS, Paris, France

Email: jasperkok@cornell.edu

**Abstract.** The transport of sand and dust by wind is a potent erosional force, creates sand dunes and ripples, and loads the atmosphere with suspended dust aerosols. This article presents an extensive review of the physics of wind-blown sand and dust on Earth and Mars. Specifically, we review the physics of aeolian saltation, the formation and development of sand dunes and ripples, the physics of dust aerosol emission, the weather phenomena that trigger dust storms, and the lifting of dust by dust devils and other small-scale vortices. We also discuss the physics of wind-blown sand and dune formation on Venus and Titan.

# Table of Contents

# 1. Introduction

The wind-driven emission, transport, and deposition of sand and dust by wind are termed *aeolian processes*, after the Greek god Aeolus, the keeper of the winds. Aeolian processes occur wherever there is a supply of granular material and atmospheric winds of sufficient strength to move them. On Earth, this occurs mainly in deserts, on beaches, and in other sparsely vegetated areas, such as dry lake beds. The blowing of sand and dust in these regions helps shape the surface through the formation of sand dunes and ripples, the erosion of rocks, and the creation and transport of soil particles. Moreover, airborne dust particles can be transported thousands of kilometers from their source region, thereby affecting weather and climate, ecosystem productivity, the hydrological cycle, and various other components of the Earth system.

But aeolian processes are not confined to Earth, and also occur on Mars, Venus, and the Saturnian moon Titan (Greeley and Iversen 1985). On Mars, dust storms occasionally obscure the sun over entire regions of the planet for days at a time, while their smaller cousins, dust devils, punctuate the mostly clear daytime skies elsewhere (Balme and Greeley 2006). The surface of Mars also hosts extensive fields of barchan, transverse, longitudinal, and star-like dunes, as well as other exotic dune shapes that have not been documented on Earth (Bourke *et al.* 2010). On Venus, transverse dunes have been identified by the Magellan orbiter (Weitz *et al.* 1994), while the Cassini orbiter has documented extensive longitudinal sand dunes on Titan (Lorenz *et al.* 2006).

The terms *dust* and *sand* usually refer to solid inorganic particles that are derived from the weathering of rocks. In the geological sciences, sand is defined as mineral (i.e., rock-derived) particles with diameters between 62.5 and 2,000 μm, whereas dust is defined as particles with diameters smaller than 62.5 μm (note that the boundary of 62.5 μm differs somewhat between particle size classification schemes, see Shao 2008, p. 119). In the atmospheric sciences, dust is usually defined as the material that can be readily suspended by wind (e.g., Shao 2008), whereas sand is rarely suspended and can thus form sand dunes and ripples, which are collectively termed *bedforms*.

## *1.1 Modes of wind-blown particle transport*

The transport of particles by wind can occur in several modes, which depend predominantly on particle size and wind speed (Figure 1.1). As wind speed increases, sand particles of ~100 μm diameter are the first to be moved by fluid drag (see Section 2.1.1). After lifting, these particles hop along the surface in a process known as *saltation* (Bagnold 1941, Shao 2008), from the Latin *salto*, which means to leap or spring. The impact of these *saltators* on the soil surface can mobilize particles of a wide range of sizes. Indeed, dust particles are not normally directly lifted by wind because their interparticle cohesive forces are large compared to aerodynamic forces (see Section 2.1.1). Instead, these small particles are predominantly ejected from the soil by the impacts of saltating particles (Gillette *et al.* 1974, Shao *et al.* 1993a). Following ejection, dust particles are susceptible to turbulent fluctuations and thus usually enter short-term (~ 20 - 70 μm diameter) or long-term (< ~20 μm diameter) suspension (Figure 1.1). Long-term suspended dust can remain in the atmosphere up to several weeks and can thus be transported thousands of kilometers from source regions (Gillette and Walker 1977, Zender *et al.* 2003a, Miller *et al.* 2006). As outlined in the next section, these dust aerosols affect the Earth and Mars systems through a wide variety of interactions.

The impacts of saltating particles can also mobilize larger particles. However, the acceleration of particles with diameters in excess of ~500 μm is strongly limited by their large inertia, and these particles generally do not saltate (Shao, 2008). Instead, they usually settle back to the soil after a short hop of generally less than a centimeter, in a mode of transport known as *reptation* (Ungar and Haff 1987). Alternatively, larger particles can roll or slide along the surface, driven by impacts of saltating particles and wind drag forces in a mode of transport known as *creep* (Bagnold 1937). Creep and reptation can account for a substantial fraction of the total wind-blown sand flux (Bagnold 1937, Namikas 2003).

The transport of soil particles by wind can thus be crudely separated into several physical regimes: long-term suspension (< ~20 μm diameter), short-term suspension (~20 – 70 μm), saltation (~70 – 500 μm), and reptation and creep (> ~500 μm) (Figure 1.1). Note that these four transport modes are not discrete: each mode morphs continuously into the next with changing wind speed, particle size, and soil size distribution. The divisions based on particle size between these regimes are thus merely approximate.

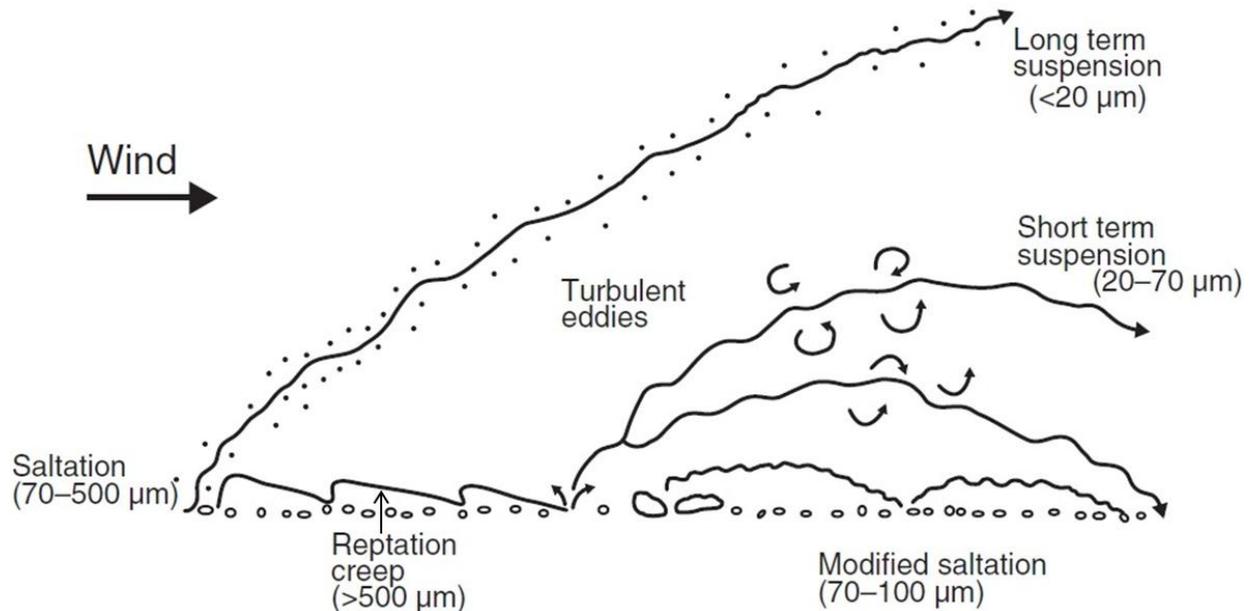

Figure 1.1. Schematic of the different modes of aeolian transport. Reprinted from Nickling and McKenna Neuman (2009), with kind permission from Springer Science+Business Media B.V.

*1.2  Importance of wind-blown sand and dust to the Earth and planetary sciences*
Wind-blown sand has shaped a substantial portion of the Earth's surface by creating sand dunes and ripples in both coastal and arid regions (Bagnold 1941, Pye and Tsoar 1990), and by weathering rocks (Greeley and Iversen 1985), which contributes to the creation of soils over long time periods (Pye 1987). Since aeolian processes arise from the interaction of wind with the surface, the study of aeolian bedforms (such as dunes) and aeolian sediments (such as *loess* soils or aeolian marine sediments) can provide information on the past state of both the atmosphere and the surface (Greeley and Iversen 1985, Pye and Tsoar 1990, Rea 1994). For instance, important constraints on both the ancient and contemporary history of Mars are provided by the inference of formative winds and climate from the morphology and observed time evolution of aeolian surface features (Greeley *et al.* 1992a). Finally, as discussed above, wind-blown sand is also the main source of mineral dust aerosols.

We briefly review the wide range of impacts of wind-blown sand and dust on Earth, Mars, Venus, and Titan in the next sections.

*1.2.1  Impacts of mineral dust aerosols on the Earth system*
Mineral dust aerosols can be entrained by numerous atmospheric phenomena (see Section 5.1.2), the most spectacular of which are probably synoptic scale (~ 1000 km) dust storms (Figure 1.2a). The interaction of dust aerosols with other components of the Earth system produces a wide range of often complex impacts on for instance ecosystems, weather and climate, the hydrological cycle, agriculture, and human health. The importance of many of these impacts has only been recognized over the past few decades, resulting in a strongly increasing interest in the study of aeolian processes (Stout *et al.* 2009). But despite the increasing number of studies addressing mineral dust, many of the impacts on the Earth system remain highly uncertain (Goudie and Middleton 2006, Shao *et al.* 2011a, Mahowald 2011).

The impacts of mineral dust on ecosystems arise predominantly from the delivery of nutrients by dust deposition. Many ocean biota, such as phytoplankton, are iron limited (Martin *et al.* 1991), such that the supply of bioavailable iron by the deposition of dust is hypothesized to be an important control on ocean productivity (Martin *et al.* 1991, Jickells *et al.* 2005). Similarly, the long-term productivity of many land ecosystems is limited by the availability of phosphorus (Chadwick *et al.*, 1999, Okin *et al.* 2004), such that the deposition of dust-borne phosphorus is often critical for ecosystem productivity (i.e., primary biomass production) on long time scales (thousands of years). For example, the productivity of the Amazon rainforest is probably limited by dust-borne phosphorus deposition (Swap *et al.* 1992). The deposition of dust to both land and ocean ecosystems thus stimulates productivity, thereby also affecting biogeochemical cycles of carbon and nitrogen (Mahowald *et al.* 2011). Consequently, global changes in dust deposition to ecosystems are hypothesized to have contributed to changes in $CO_2$ concentrations between glacial and interglacial periods (Martin 1990, Broecker and Henderson 1998) as well as over the past century (Mahowald *et al.* 2010). Moreover, dust-induced changes in $CO_2$ concentrations may also play a role in future climate changes (Mahowald 2011).

Dust aerosols also affect the hydrological cycle in several ways. First, dust redistributes energy by scattering and absorbing both solar and terrestrial radiation. This causes a net heating of the atmosphere, which generally enhances precipitation, but cools the surface, which generally suppresses both evaporation and precipitation (Miller *et al.* 2004, Yoshioka *et al.* 2007, Zhao *et al.* 2011). Second, dust aerosols serve as nuclei for the condensation of water in both the liquid (as cloud condensation nuclei) and solid (as ice nuclei) phases (DeMott *et al.* 2003, Twohy *et al.* 2009). The resulting interactions between dust and clouds are highly complex and poorly understood, and can also either enhance or suppress precipitation (Rosenfeld *et al.* 2001, Ramanathan *et al.* 2001, Toon 2003). Furthermore, the deposition of dust on glaciers and snowpacks decreases the albedo (reflectivity) of these features, which produces a positive (warming) climate forcing and an earlier spring snowmelt (Flanner *et al.* 2009, Painter *et al.* 2010).

As already alluded to in previous paragraphs, dust aerosols affect weather and climate through a wide range of interactions. These include scattering and absorbing radiation, lowering snowpack albedo, altering atmospheric $CO_2$ concentrations by modulating ecosystem productivity, and serving as cloud nuclei and thereby likely increasing cloud lifetime and reflectivity (Twomey 1974, Andreae and Rosenfeld 2008). Since mineral dust therefore affects Earth's radiation balance, changes in the atmospheric dust loading can produce a substantial radiative forcing (Tegen *et al.* 1996, Sokolik and Toon 1996, Mahowald *et al.* 2006, 2010). Conversely, the global dust cycle is also highly sensitive to changes in climate, as evidenced by the several times larger global dust deposition rate during glacial maxima than during interglacials (Rea 1994, Kohfeld and Harrison, 2001), and by an increase in global dust deposition over at least the past 50 years (Prospero and Lamb 2003, Mahowald *et al.* 2010). The radiative forcing resulting from such large changes in the global dust cycle is thought to have played an important role in amplifying past climate changes (Broecker and Hendersen 1998, Jansen *et al.* 2007, Abbot and Halevy 2010). Because of the large uncertainties in both the net radiative forcing of dust aerosols and the response of the global dust cycle to future climate changes, it remains unclear whether this *dust climate feedback* will oppose or enhance future climate changes (Tegen *et al.* 2004, Mahowald *et al.* 2006, Mahowald 2007).

In addition to these diverse impacts of dust on ecosystem productivity, weather and climate, and the hydrological cycle, dust aerosols produce a number of other important impacts. For instance, dust emission reduces soil fertility through the removal of small soil particles rich in nutrients and organic matter, which contributes to desertification and reduces agricultural productivity (Shao 2008, Ravi *et al.* 2011). Furthermore, heterogeneous chemistry occurring on atmospheric mineral dust affects the composition of the troposphere (Sullivan et al., 2007; Cwiertny et al., 2008), inhalation of dust aerosols is a hazard for human health (Prospero *et al.* 1999, O'Hara *et al.* 2000), and dust might inhibit hurricane formation (Evan *et al.* 2006, Sun *et al.* 2008) and force coupled ocean-atmosphere variability in the tropical Atlantic (Evan *et al.* 2011).

*1.2.2 Impacts of aeolian processes on Mars, Venus, and Titan*

Aeolian processes on Mars, Venus, and Titan differ substantially from those on Earth, primarily due to large differences in gravity and air density. On Mars, the lower air density makes it more difficult to move surface material, but dust storms are nonetheless widespread and evidence of active sand transport has been accumulating in recent years (e.g., Cantor *et al.* 2001, Bridges *et al.* 2012a). In contrast, the air density on Venus and Titan is much higher than on Earth. This greatly limits the terminal velocity of sand particles, such that saltation on these planets is probably similar to saltation in water (see Section 2.4.2). These large differences between the physical environment of the different planetary bodies produces correspondingly large differences in the properties of dunes, such as their typical length scales (Figure 1.3).

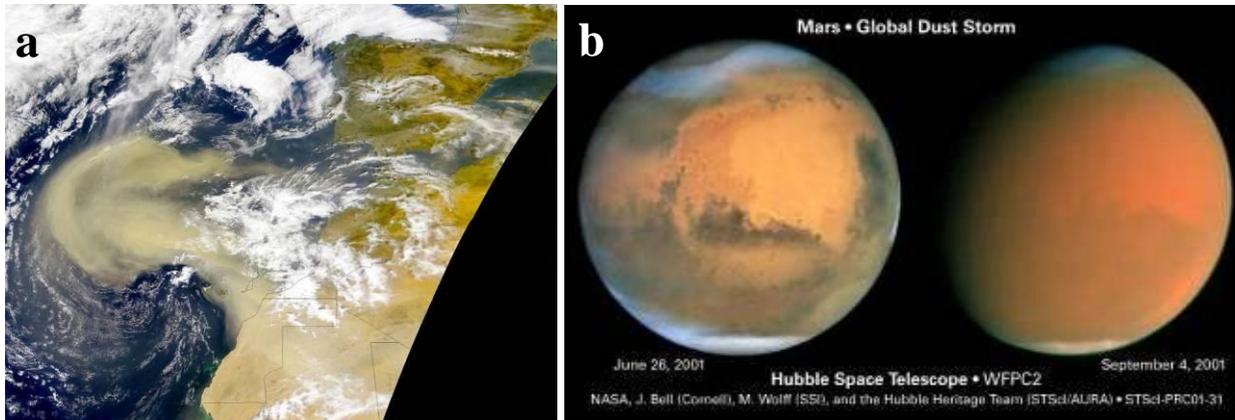

Figure 1.2. (**a**) Satellite image of a massive dust storm blowing off the northwest Sahara desert on February 26, 2000. Image credit: Visible Earth, NASA (http://visibleearth.nasa.gov/). (**b**) Hubble space telescope images of roughly the same hemisphere of Mars, illustrating the effects of the 2001 planet-encircling dust storm. The left image was acquired on 26 June 2001 and shows the dust storm beginning at the lower right of the disk. The right image was acquired on 4 September 2001, and shows dust shrouding most of the planet. Image credit: NASA/James Bell (Cornell University)/Michael Wolff (Space Science Institute)/The Hubble Heritage Team (STScI/AURA).

The cycle of dust aerosols on Mars plays several important roles in the planet's surface-atmosphere system. Significant changes in local and regional surface albedos occur due to dust cover variations caused by local, regional, and planet-encircling dust storms (Figure 1.2b; Thomas and Veverka 1979, Geissler 2005), and even by concentrated areas of dust devil tracks (e.g., Malin and Edgett 2001, Michaels 2006). Such albedo variations affect the surface temperature and potentially local- and regional-scale atmospheric circulations, as well as the overall surface energy balance of the planet (Gierasch and Goody 1968, Kahn *et al.* 1992, Fenton *et al.* 2007). The ever-present dust within the atmosphere strongly modulates the planet's climate by significantly altering the absorption and emission of infrared and visible radiation by the atmosphere (e.g., Gierasch and Goody 1968, Kahn *et al.* 1992). Dust particles also likely serve as condensation nuclei for water- and $CO_2$-ice cloud particles on Mars (Colaprete *et al.* 1999, Colaprete and Toon 2002), which further modulates the planet's climate.

Aeolian processes are also the dominant natural agents shaping the surface of today's Mars (Greeley and Iversen 1985). A diversity of surface features attest to the relevance of aeolian transport for the morphodynamics of the martian landscape, including ephemeral dark streaks associated with craters, modified crater rims, and linear grooves and streamlined ridges or *yardangs*, which have also been observed in Earth's deserts (Fenton and Richardson 2001). Moreover, migration of dunes and ripples has been detected in recent years from satellite images at many different locations on Mars (e.g., Bridges *et al.* 2012a), thus making it evident that saltation occurs during present climatic conditions. Nonetheless, the rate of surficial modification of martian bedforms is thought to be orders of magnitude smaller than for their terrestrial counterparts (e.g., Golombek et al. 2006a), although a recent study has surprisingly found rates of surficial modification within an active martian dune field that are comparable to those on Earth (Bridges *et al.* 2012b). Questions regarding the age, long-term dynamics, and stability of martian dunes have thus motivated a body of modeling work in recent years, both through computer simulations and experiments of underwater bedforms mimicking martian dune shapes (Section 3.3.2).

Because the shape of dunes may serve as a proxy for wind directionality (Section 3.1.5), the morphology of dunes on Mars, Venus and Titan can tell us much about the wind systems on those planetary bodies. For instance, the dominant occurrence of the transverse dune shape in the sand seas of Venus indicates that Venusian sand-moving winds are strongly unidirectional, while the large variety of exotic dune shapes on Mars indicates that a wider spectrum of wind regimes exists on this planet (Section 3.3.2).

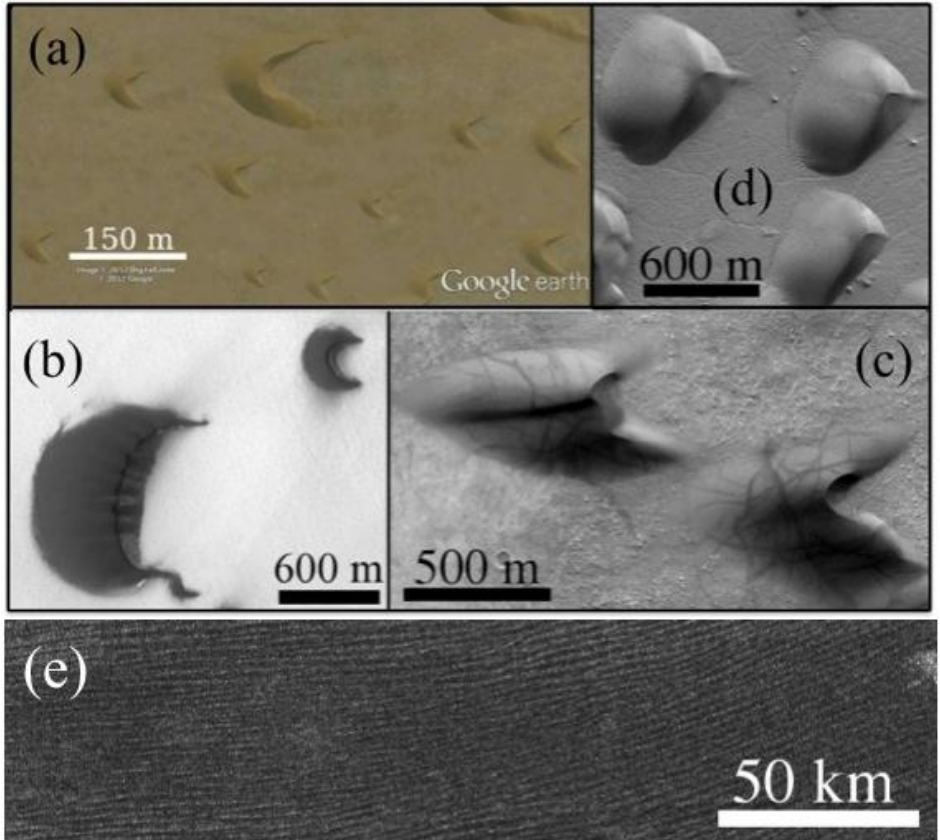

Figure 1.3 Sand dunes in Morocco (a), on Mars (b–d) and on Titan (e). The barchan dune (a–c), which has a crescent shape, is the simplest and most studied type of dune. It forms when the wind is roughly unidirectional and the ground is not completely covered with sand (see Section 3.1.5). Barchans are widespread in Earth deserts and are also a common dune form on Mars, where they can assume a surprisingly large diversity of shapes (Bourke and Goudie 2010). Mars hosts also many other exotic dune forms — such as the wedge dunes in Wirtz crater (d) — which could have been formed by complex wind regimes (see Section 3.3.2). Dunes on Venus and on Titan occur predominantly as sets of long parallel sand ridges, though the wind regimes leading to the formation of dunes on these two celestial bodies are well distinct (see Sections 3.3.3 and 3.3.4). Images credit: Google Earth (a), NASA/JPL/University of Arizona (b, c), NASA/JPL/MSS (d), and NASA/JPL/Caltech (e).

*1.3 Scope and organization of this review*
The first comprehensive review of the physics of wind-blown sand was given by Ralph Bagnold in his seminal 1941 book "The physics of wind-blown sand and desert dunes," which is still a standard reference in the field. The recently updated book "The physics and modelling of wind erosion" by Yaping Shao (2008) provides a more up-to-date review of many aspects of dust emission and wind erosion. Other reviews focusing respectively on aeolian geomorphology (the study of landforms created by aeolian processes), the mechanics of sand transport, and on scaling laws in saltation were given by Nickling and McKenna Neuman (2009), Zheng (2009), and Durán *et al.* (2011a). Recent reviews of the impact of the global dust cycle on various aspects of the Earth system were given by Mahowald *et al.* (2009), Ravi *et al.* (2011), and Shao *et al.* (2011a).

The present article strives to complement these previous reviews by providing a comprehensive review of the physics of wind-blown sand and dust on Earth, Mars, and other planetary bodies, with particular emphasis on the substantial progress made over the past decade. Since saltation drives most wind-blown transport of sand and dust, we start our review with a detailed analysis of the physics of saltation. In Chapter 3, we build on this understanding to review the formation and dynamics of sand dunes and ripples. We then focus on the emission of mineral dust aerosols on Earth and Mars through both saltation impacts and other physical processes in Chapter 4, after which Chapter 5 discusses the meteorological phenomena that produce dust emission. We round out the review in Chapter 6 with a discussion of important remaining questions in the physics of wind-blown sand and dust. This review is structured in such a way that each chapter can be read in isolation.

## 2. The physics of wind-blown sand (saltation)

As discussed in the previous chapter, saltation plays a central role in aeolian processes since it usually initiates the other forms of transport, including the emission of dust aerosols that subsequently travel in suspension (Chapters 4 and 5). Since there is no clear division between particles traveling in saltation and particles traveling in lower energy creep or reptation trajectories, the discussion of saltation in this chapter is implicitly inclusive to the transport of particles in creep and reptation (Figure 1.1).

Saltation is initiated when the wind stress $\tau$ is sufficient to lift surface particles into the fluid stream, which for loose sand occurs around $\tau \sim 0.05$ N/m$^2$ (Greeley and Iversen 1985). Following initiation, the lifted particles are accelerated by wind into ballistic trajectories and the resulting impacts on the soil bed can eject, or *splash*, new saltating particles into the fluid stream. This process produces an exponential increase in the particle concentration (Durán *et al.* 2011a), which leads to increasing drag on the wind, thereby retarding the wind speed in the saltation layer (Bagnold 1936). It is this slowing of the wind that acts as a negative feedback by reducing particle speeds, and thus the splashing of new particles into saltation, which ultimately limits the number of saltating particles (Owen 1964) and thereby partially determines the characteristics of steady state saltation (Section 2.3).

The physics of aeolian saltation can thus be roughly divided into four main physical processes (Anderson and Haff 1991, Kok and Renno 2009a): (i) the initiation of saltation by the aerodynamic lifting of surface particles, (ii) the subsequent trajectories of saltating particles, (iii) the splashing of surface particles into saltation by impacting saltators, and (iv) the modification of the wind profile by the drag of saltating particles. After developing an understanding of these four processes in the following sections, we discuss the transition to and characteristics of steady state saltation produced by the interaction of these four processes in Sections 2.2 and 2.3, respectively. Finally, we investigate the characteristics of saltation on Mars, Venus, and Titan in Section 2.4.

### 2.1 The four main physical processes of aeolian saltation
### 2.1.1 Initiation of saltation: the fluid threshold

Saltation is initiated by the lifting of a small number of particles by wind stress (Greeley and Iversen 1985). The value of the wind stress at which this occurs is termed the *fluid* or *static threshold* (Bagnold 1941). This threshold depends not only on the properties of the fluid, but also on the gravitational and interparticle cohesion forces that oppose the fluid lifting. A schematic of the resulting force balance on a surface particle subjected to wind stress is presented in Figure 2.1. The fluid threshold is distinct from the *dynamic* or *impact threshold*, which is the lowest wind stress at which saltation can be sustained after it has been initiated. For most conditions on Earth and Mars, the impact threshold is smaller than the fluid threshold because the transfer of momentum to the soil bed through particle impacts is more efficient than through fluid drag (see Section 2.4).

An expression for the fluid threshold can be derived from the force balance of a stationary surface particle (Figure 2.1). The surface particle will be entrained by the flow when it pivots around the point of contact with its supporting neighbor (*P* in Figure 2.1). This occurs when the moment of the aerodynamic drag ($F_d$) and lift ($F_l$) forces barely exceeds the moment of the interparticle ($F_{ip}$) and gravitational ($F_g$) forces (Greeley and Iversen 1985, Shao and Lu 2000). At the instant of lifting, we thus have that

$$r_d F_d \approx r_g (F_g - F_l) + r_{ip} F_{ip}, \qquad (2.1)$$

where $r_d$, $r_g$, and $r_{ip}$ are the moment arms in Figure 2.1, which are proportional to the particle diameter $D_p$. The effective gravitational force in a fluid, which includes the buoyancy force, equals

$$F_g = \frac{\pi}{6}(\rho_p - \rho_a)g D_p^3, \qquad (2.2)$$

where $g$ is the constant of gravitational acceleration and $D_p$ is the diameter of a sphere with the same volume as the irregularly-shaped sand particle. The particle density $\rho_p$ depends on the composition of the sand, but equals approximately 2650 kg/m$^3$ for quartz sand on Earth (see Section 2.4). Furthermore, the drag force exerted by the fluid on a surface particle protruding into the flow is given by (Greeley and Iversen 1985, Shao 2008)

$$F_d = K_d \rho_a D_p^2 u_*^2, \tag{2.3}$$

where $\rho_a$ is the air density, $K_d$ is a dimensionless coefficient of the order of ~10 (see Table 3.1 in Greeley and Iversen 1985), and the shear velocity $u_*$ is a scaling parameter proportional to the velocity gradient in boundary layer flow and is defined as (Stull 1988, White 2006)

$$\tau = \rho_a u_*^2. \tag{2.4}$$

The fluid shear stress $\tau$ is equivalent to the flux of horizontal momentum transported downward through the fluid by viscous and turbulent mixing (see Section 2.1.4 for further discussion). A straightforward expression for the fluid threshold shear velocity $u_{*\mathrm{ft}}$ at which saltation is initiated can now be obtained by combining Eqs. (2.1)-(2.3), which yields (Bagnold, 1941)

$$u_{*\mathrm{ft}} = A_\mathrm{ft} \sqrt{\frac{\rho_p - \rho_a}{\rho_a} g D_p}, \tag{2.5}$$

where the constant $A_\mathrm{ft}$ is a function of interparticle forces, the lift force, and the Reynolds number of the flow (Greeley and Iversen 1985). By neglecting these dependencies and fitting Eq. (2.5) to the fluid threshold of loose sand, for which interparticle forces are small, Bagnold (1941) obtained $A_\mathrm{ft} \approx 0.10$. Note that an expression similar to Eq. (2.5) can be used to describe the impact threshold $u_{*\mathrm{it}}$ (see Section 2.4), for which the proportionality constant $A_\mathrm{it} \approx 0.082$ (Bagnold 1937).

In order to derive an equation for $u_{*\mathrm{ft}}$ applicable to a wide range of Reynolds numbers and particle sizes, an understanding of the lift and interparticle forces is required. We discuss these forces in the next two sections, after which we review semi-empirical relations of the fluid threshold that account for these forces and thus have broader applicability than Eq. (2.5).

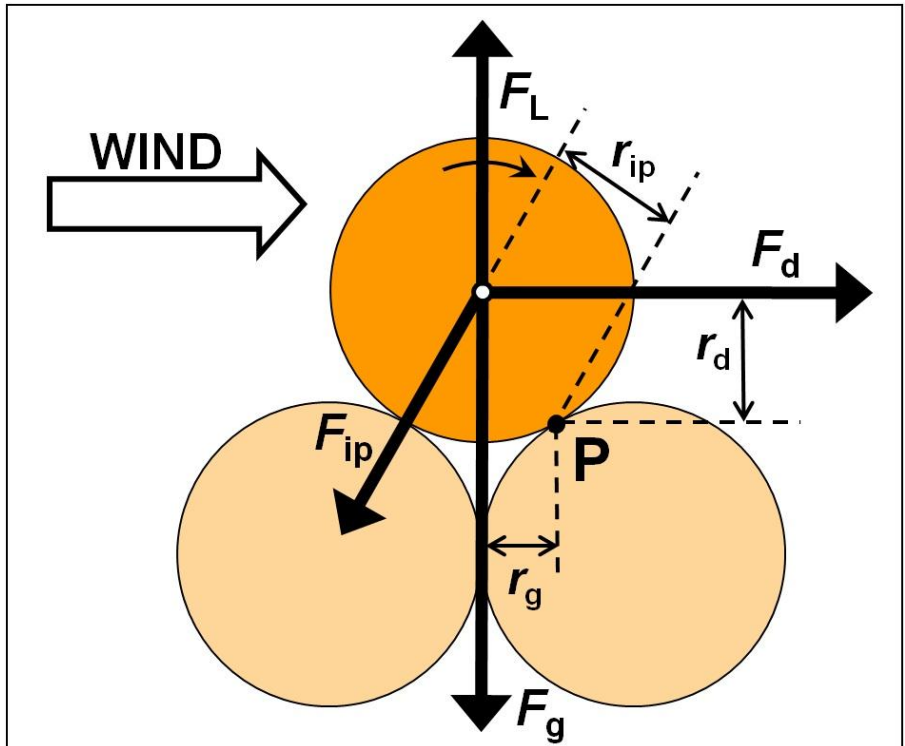

Figure 2.1. Schematic of the forces acting on a stationary sand particle resting on a bed of other particles (after Shao and Lu, 2000). Forces are denoted by thick arrows, and their moment arms relative to the pivoting point $P$ are indicated by thin arrows. When the moment of the aerodynamic lift and drag forces exceeds that of the gravitational and interparticle forces, the particle will be entrained into the flow by pivoting around $P$ in the indicated direction.

*2.1.1.1 The possible role of lift forces in initiating saltation*

The *Saffman lift force* is caused by the sharp gradient in the fluid velocity above the particle bed which, due to Bernoulli's principle, creates a lower pressure above the particle than below it (Saffman 1965). Measurements indicate that, at roughness Reynolds numbers (defined in Section 2.1.4.1) of $\sim 10^2 - 10^4$, the lift force is of the order of magnitude of the drag force (Einstein and El-Samni 1949, Chepil 1958, Apperley *et al.* 1968, Bagnold 1974, Brayshaw *et al.* 1983, Dwivedi *et al.* 2011). On the basis of these measurements, Greeley and Iversen (1985) and Carneiro *et al.* (2011) argued that the lift force plays an important role in initiating saltation. However, at lower roughness Reynolds numbers of $\sim 1 - 10$, characteristic for saltation in air, the theoretical treatments of Saffman (1965) and McLaughlin (1991) predict a lift force that is substantially smaller than the drag force. Measurements of the lift force at low Reynolds number are difficult to make, but available numerical simulations and measurements are nonetheless consistent with these theories (Loth 2008). The Saffman lift force might thus play a relatively minor role in the initiation of saltation, consistent with the negligible role it plays in determining the trajectories of saltating particles (see Section 2.1.2). Further research is required to settle this question.

In addition to the Saffman lift force, some experiments suggest that a lift force due to the pressure deficit at the core of dust devils and other vortices might aid in lifting particles (Greeley *et al.* 2003, Neakrase and Greeley, 2010a) (see Section 5.3.2).

*2.1.1.2 Interparticle forces*

When granular particles such as dust and sand are brought into contact, they are subjected to several kinds of cohesive interparticle forces, including van der Waals forces, water adsorption forces, and electrostatic forces (e.g., Castellanos 2005). Van der Waals forces are themselves a collection of forces that arise predominantly from interatomic and intermolecular interactions, for example from dipole – dipole interactions (Krupp 1967, Muller 1994). For macroscopic objects, the van der Waals forces scale linearly with particle size (Hamaker 1937). Electrostatic forces arise from net electric charges on neighboring particles (Krupp 1967). These charges can be generated through a large variety of mechanisms (Lowell and Rose-Innes 1980, Baytekin 2011), including through exchange of ion or electrons between contacting particles (McCarty and Whitesides 2008, Liu and Bard 2008, Forward *et al.* 2009, Kok and Lacks 2009). Finally, water adsorption forces are caused by the condensation of liquid water on particle surfaces, which creates attractive forces both through bonding of adjacent water films and through the formation of capillary bridges between neighboring particles, whose surface tension causes an attraction between these particles (see Section 4.1.1.1) (Herminghaus 2005, Nickling and McKenna Neuman 2009). Experiments using atomic force microscopy have found a strong increase of cohesive forces with relative humidity for hydrophilic materials, but not for hydrophobic materials (Jones *et al.* 2002). Mineralogy and surface contaminants affecting hydrophilicity can thus be expected to affect the dependence of cohesive forces on relative humidity.

Although the behavior of the different interparticle forces for aspherical and rough sand and dust is poorly understood (e.g., McKenna Neuman and Sanderson 2008), these forces can be expected to scale with the area of interaction. By combining this assumption with the Hertzian theory of elastic contact (Hertz 1882), Johnson, Kendall and Roberts (1971) developed a theory predicting that the force required to separate two spheres scales linearly with the particle size. This JKR theory has been verified by a range of experiments (e.g., Horn *et al.* 1986, Chaudhury and Whitesides 1992), including experiments that explicitly confirmed the linear dependence of the interparticle force on particle size (Ando and Ino, 1998, Heim *et al.* 1999). Therefore, absent a strong correlation of particle morphology or other surface properties with particle size, the interparticle force is expected to scale with the soil particle size (e.g., Shao and Lu 2000).

*2.1.1.3 Semi-empirical expressions for the fluid threshold*

Iversen and White (1982) used a combination of theory and curve-fitting to wind tunnel measurements (Iversen *et al.* 1976) to derive a semi-empirical expression for the saltation fluid threshold. Their expression is more general than that of Bagnold (Eq. (2.5)) because it includes the effects of lift and interparticle forces. Specifically, Iversen and White (1982) expressed the dimensionless threshold parameter $A_{ft}$ in Eq. (2.5) in a series of semi-empirical equations valid for different ranges of the friction Reynolds number $R_{*t}$ (see also Greeley and Iversen 1985):

$$A_{ft} = 0.2\sqrt{\frac{(1+0.006/\rho_p g D_p^{2.5})}{1+2.5R_{*t}}}, \text{ for } 0.03 \leq R_{*t} \leq 0.3,$$

$$A_{ft} = 0.129\sqrt{\frac{(1+0.006/\rho_p g D_p^{2.5})}{1.928 R_{*t}^{0.092}-1}}, \text{ for } 0.3 \leq R_{*t} \leq 10, \text{ and} \quad (2.6)$$

$$A_{ft} = 0.120\sqrt{(1+0.006/\rho_p g D_p^{2.5})}\{1-0.0858\exp[-0.0671(R_{*t}-10)]\}, \text{ for } R_{*t} \geq 10,$$

where

$$R_{*t} = \rho_a u_{*ft} D_p / \mu, \quad (2.7)$$

and where $\mu$ is the dynamic viscosity.

Although the semi-empirical expression of Iversen and White (1982) is in good agreement with a large number of wind tunnel measurements for varying conditions (e.g., Iversen *et al.*, 1976), Shao and Lu (2000) argued that Eq. (2.6) can be simplified substantially. In particular, they showed that the dependence of $A_{ft}$ on the Reynolds number appears limited, such that Eq. (2.6) can be simplified by assuming that $A_{ft}$ is independent of the Reynolds number. Shao and Lu (2000) furthermore assumed that the interparticle forces scale with the particle diameter, which is supported by both theory and experiments (Section 2.1.1.2). By balancing the entraining forces (aerodynamic drag and lift) against the stabilizing forces (gravity and interparticle forces), Shao and Lu then obtained a relatively straightforward and elegant expression for the saltation fluid threshold

$$u_{*ft} = A_N \sqrt{\frac{\rho_p - \rho_a}{\rho_a} g D_p + \frac{\gamma}{\rho_a D_p}}, \quad (2.8)$$

where the dimensionless parameter $A_N = 0.111$, which is close to the value originally obtained by Bagnold (Eq. (2.5)). The parameter $\gamma$ scales the strength of the interparticle forces; for dry, loose dust or sand on Earth, Shao and Lu (2000) recommended a value of $\gamma = 1.65\times10^{-4} - 5.00\times10^{-4}$ N/m. Kok and Renno (2006) obtained $\gamma = 2.9\times10^{-4}$ N/m, consistent with this proposed range, by fitting Eq. (2.8) to the threshold electric field required to lift loose dust and sand particles.

The above expressions of Iversen and White (1982) and Shao and Lu (2000) are compared to measurements of the fluid threshold in Figure 2.2. A critical feature of the threshold curves is the occurrence of a minimum in the fluid threshold for particle sizes of approximately ~75 – 100 μm. At this particle size, the interparticle forces are approximately equal in magnitude to the gravitational force, such that the lifting of smaller and larger particles is impeded by increases in the strength of the interparticle and gravitational forces, respectively, relative to the aerodynamic forces. The occurrence of this minimum in the fluid threshold is critical in understanding the physics of dust emission (Chapter 4). Indeed, it shows that saltation is initiated at wind speeds well below that required to aerodynamically lift dust aerosols of a typical size of ~1 - 10 μm. As a consequence, dust aerosols are only rarely lifted directly by wind (Loosmore and Hunt 2000). Instead, dust aerosols are predominantly emitted by the impacts of saltating particles on the soil surface (Gillette *et al.* 1974, Shao *et al.* 1993a). However, several plausible dust lifting mechanisms that do not directly involve saltation have been proposed to occur on Mars (see Section 4.2).

Note that the expressions for the saltation fluid threshold reviewed in this section are for ideal soils: loose, dry sand without roughness elements such as pebbles, rocks, or vegetation. As such, the applicability of these expressions is limited to dry sand dunes and beaches. In contrast, the emission of dust aerosols often occurs from soils containing roughness elements or soil moisture. Several corrections are thus required to calculate the fluid threshold for saltation (and thus dust emission) for these soils. The corrections for soil moisture content and for the presence of roughness elements are discussed in Sections 4.1.1.1 and 4.1.1.2, respectively.

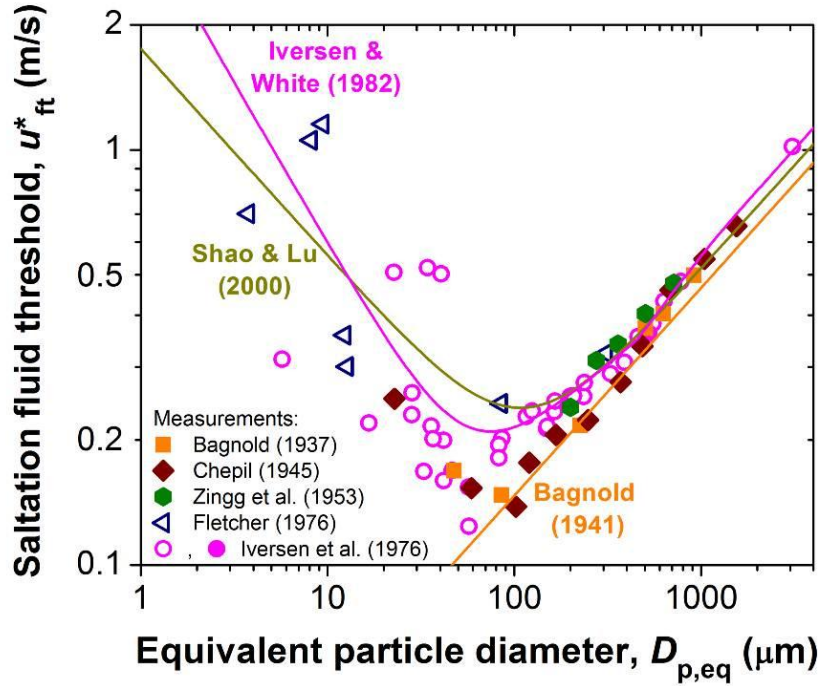

Figure 2.2. Semi-empirical expressions (colored lines) and measurements (symbols) of the threshold shear velocity required to initiate saltation for Earth ambient conditions. Measurements of the fluid threshold for sand and dust are denoted by filled symbols (Bagnold 1937, Chepil 1945, Zingg 1953, Iversen *et al.* 1976), whereas measurements of the fluid threshold for materials other than sand and dust are denoted by open symbols (Fletcher 1976, Iversen *et al.* 1976). The effect of the different density of these materials was accounted for by using the 'equivalent' particle diameter defined by Chepil (1951): $D_{p,eq} = D_p \rho_p / \rho_{p,sand}$, where $\rho_p$ is the particle density and $\rho_{p,sand}$ = 2650 kg/m$^3$. Note that this correction accounts for the effect of differences in material density on the fluid threshold (Eq. (2.5)), but not for the dependence of fluid forces on the particle size through the Reynolds number (which the analysis of Shao and Lu (2000) suggests might be limited), or the dependence of interparticle forces on the material type (e.g., Corn, 1961). An alternative scaling proposed by Iversen *et al.* (1976) accounts for these first two effects, but not the latter.

### 2.1.2 Particle trajectories

After saltation has been initiated, lifted sand particles undergo ballistic trajectories (Figure 1.1) that are determined primarily by the gravitational (2.2) and aerodynamic drag ($F_D$) forces. The acceleration of particles by the drag force transfers momentum from the fluid to the saltating particles and thus retards the wind profile in the saltation layer (Section 2.1.4). The drag force acting on a particle submerged in a fluid is given by

$$F_D = -\frac{\pi D_p^2}{8} \rho_a C_D v_R \mathbf{v_R}, \tag{2.9}$$

where $\mathbf{v_R} = \mathbf{v} - \mathbf{U}$ is the difference between the particle ($\mathbf{v}$) and wind ($\mathbf{U}$) velocities, and $v_R = |\mathbf{v_R}|$. The drag coefficient ($C_D$) is a function of the particle Reynolds number, $\mathrm{Re} = \rho_a v_R D_p / \mu$ (e.g., Morsi and Alexander 1972). The drag coefficient of natural sand is larger than that for spherical particles of the same volume, both because their irregular shape produces a larger effective surface area than a sphere and because regions of large curvature can lead to flow separation, thereby increasing the drag (Dietrich, 1982). Parameterizations for the drag coefficient of natural sand, based on measurements of the terminal fall speed of sand particles, are summarized in Camenen (2007).

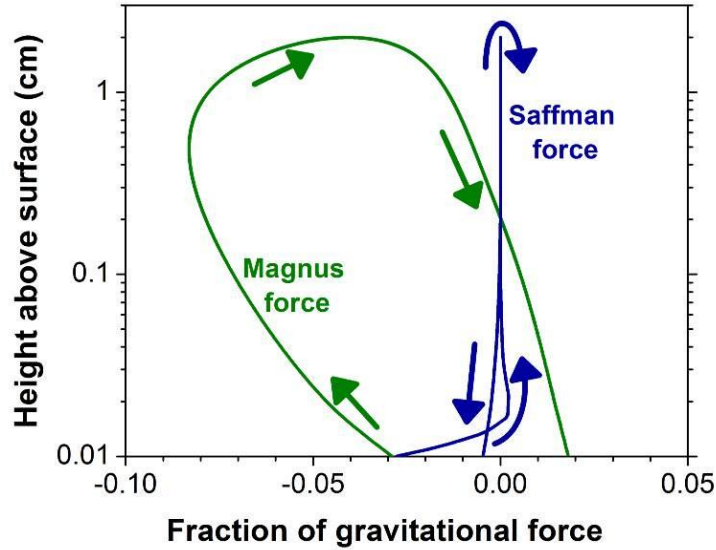

Figure 2.3. The Magnus force (green line) and the Saffman force (blue line) as a fraction of the gravitational force during the saltation trajectory of a 200 μm particle launched from the surface with a speed of 1 m/s at an angle of 40° from horizontal and a rotational speed of 400 rev/s. The arrows indicate the direction of particle motion. The particle trajectory and lift forces were calculated using Eqs. 1 – 6 in Kok and Renno (2009a), using the vertical profile of the horizontal wind speed of Eq. (2.18) with $u_* = 0.20$ m/s and $z_0 = D_p/30$. The asymmetries in the relative strength of both the Magnus and the Saffman force on the upward and the downward trajectories are due to changes in the velocity of the particle relative to the flow, which determines the strength and direction of these forces. The effects of turbulence were neglected.

In addition to the gravitational and aerodynamic drag forces, secondary aerodynamic forces on an airborne particle include aerodynamic lift forces due to particle spinning (the *Magnus force*; Rubinow and Keller 1961), the gradient in the wind speed (the *Saffman force* discussed in Section 2.1.1.1), and several other minor forces reviewed in Durán *et al.* (2011a). Various authors have suggested that these lift forces play a substantial role in saltation (White and Schulz 1977, White 1982, Zou *et al.* 2007, Xie *et al.* 2007, Huang *et al.* 2010). To further investigate this issue, Figure 2.3 shows the strength of the Magnus and Saffman forces relative to the gravitational force. Depending on the assumed spin rate, the Magnus force is of the order of 1% of the drag force and can affect individual saltation trajectories (White and Schulz 1977, Kok and Renno 2009a, Huang *et al.* 2010). In contrast, the Saffman force is several orders of magnitude smaller and has no substantial effect on particle trajectories.

Furthermore, some authors have argued that electrostatic forces can increase particle concentrations (Kok and Renno 2006, 2008), affect particle trajectories (Schmidt *et al.* 1998, Zheng *et al.* 2003) and the height of the saltation layer (Kok and Renno 2008), increase the saltation mass flux (Zheng *et al.* 2006, Kok and Renno 2008, Rasmussen *et al.* 2009), and possibly affect atmospheric chemistry on Mars (Atreya *et al.* 2006, Farrell *et al.* 2006, Kok and Renno 2009b, Ruf *et al.* 2009, Gurnett *et al.* 2010). Considering both the limited understanding of the processes that produce electrical charge separation during collisions of chemically similar insulators such as sand (Lowell and Truscott 1986, McCarty and Whitesides 2008, Kok and Lacks 2009, Lacks and Sankaran 2011), as well as the limited measurements of electric fields and particle charges in saltation (Schmidt *et al.* 1998, Zheng *et al.* 2003), further research will be required to establish whether the effect of electrostatic forces on saltation is indeed substantial. Recent reviews of the electrification of sand and dust and its effect on sand transport, erosion, and dust storms on Earth and Mars are given in Renno and Kok (2008) and Merrison (2012).

Several studies have also argued that mid-air collisions between saltators can affect particle trajectories and the mass flux at large shear velocities (Sorensen and McEwan 1996, Dong *et al.* 2002, 2005, Huang *et al.* 2007). Recent results suggest that the probability of a saltator colliding with another saltator during a single hop is on the order of 10-50 %, and increases with wind speed (Huang *et al.* 2007). As with electrostatic forces, further research is thus required to establish the potential effect of interparticle collisions on the characteristics of saltation.

*2.1.2.1 Turbulent fluctuations*

Most early numerical models of saltation neglected the effect of turbulence on particle trajectories because of the large computational cost (e.g., Ungar and Haff 1987, Anderson and Haff 1991). The steady increase in computational resources has allowed more recent numerical models to explicitly account for the effect of turbulence on particle trajectories (Almeida *et al.* 2006, Kok and Renno 2009a). The results of Kok and Renno (2009a) indicate that, for saltation on Earth, turbulence substantially affects the trajectories of particles smaller than ~200 μm (Figure 2.4). However, the characteristics of turbulence in the saltation layer remain uncertain, particularly because the effects of saltation on turbulence intensity and the Lagrangian timescale over which the fluctuating component of the fluid velocity is correlated (Van Dop *et al.* 1985) have been poorly studied. Indeed, Kok and Renno (2009a) reported discrepancies between the simulated and measured vertical flux profiles of ~100 – 200 μm particles, which could indicate inaccuracies in the treatment of turbulence in numerical saltation models. In particular, some wind tunnel measurements indicate that the wakes shed by saltating particles can increase the turbulence intensity (Taniere *et al.* 1997, Nishimura and Hunt 2000, Zhang *et al.* 2008, Li and McKenna Neuman 2012), which is not accounted for in most numerical saltation models.

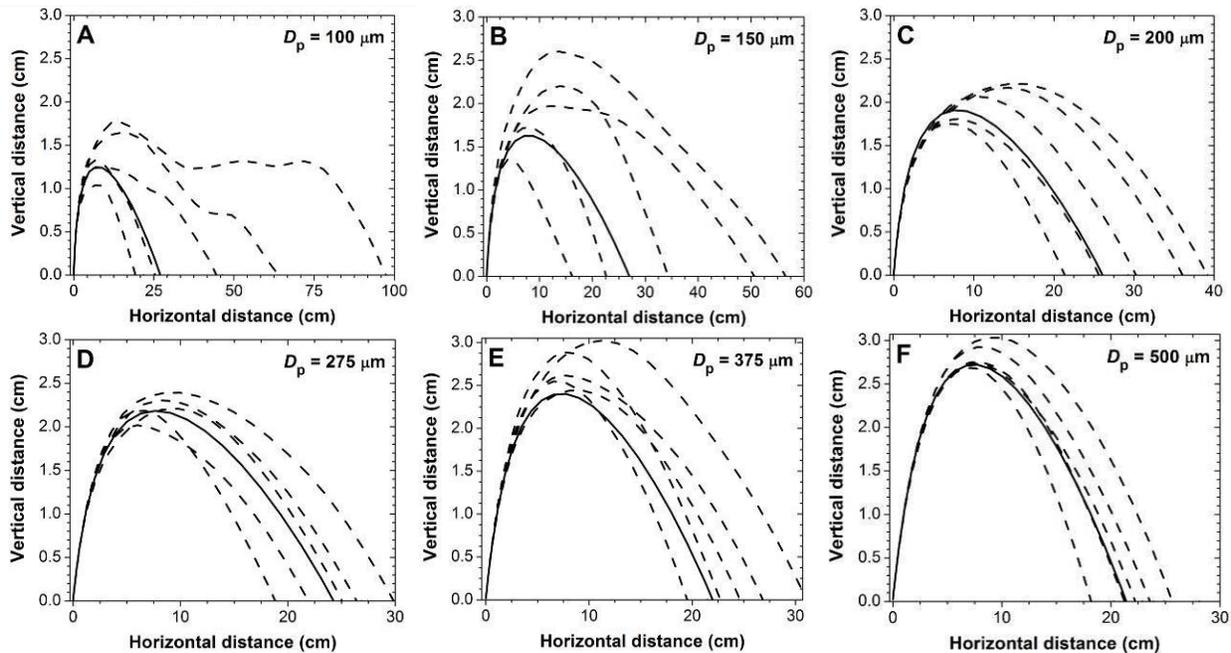

Figure 2.4. Simulated saltation trajectories for particles with diameters of (**a**) 100, (**b**) 150, (**c**) 200, (**d**) 275, (**e**) 375, and (**f**) 500 μm. The solid lines denote trajectories that do not include the effects of turbulence, whereas dashed lines denote five (stochastic) simulations that do include the effects of turbulence. Particles are launched from the surface with a speed of 1 m/s at an angle of 40° from horizontal and a rotational speed of 400 rev/s. The particle trajectories were calculated using the numerical saltation model COMSALT (Kok and Renno 2009a), for which the wind speed profile and turbulence characteristics were obtained by simulating saltation at $u_* = 0.4$ m/s for the soil size distribution reported in Namikas (2003).

*2.1.3 Saltator impacts onto the surface and the splashing of particles into saltation*
The ballistic trajectories of saltating particles are terminated by a collision with the surface. These particle impacts onto the soil surface (
Figure 2.5) are a critical process in saltation for two reasons. First, the splashing of surface particles by impacting particles is, for most conditions on Earth and Mars, the main source of new saltators after saltation has been initiated (see Sections 2.3.1 and 2.4). And second, since particles strike the soil nearly horizontally and rebound at angles of ~40° from horizontal (e.g., Rice *et al.* 1995), the impact on the soil surface partially converts the saltator's horizontal momentum gained through wind drag into vertical momentum. This conversion is critical to replenish the vertical momentum dissipated through fluid drag.

The impact of a saltating particle on the soil bed can thus produce a rebounding particle as well as one or more splashed particles. Analytical and numerical treatments of saltation need to account for the creation of these particles, but since the interaction of the impacting saltator with the soil bed is complex and stochastic, this process is resistant to an analytical solution (Crassous *et al.*, 2007). Instead, many laboratory (e.g., Rice *et al.* 1995, Beladjine *et al.* 2007) and numerical experiments (Anderson and Haff 1988, Oger *et al.* 2005, 2008) have been performed to better understand the *splash function* linking the characteristics of the impacting particle to those of the rebounding and splashed particles. Below, we summarize the insights into the splash function gained from these experiments.

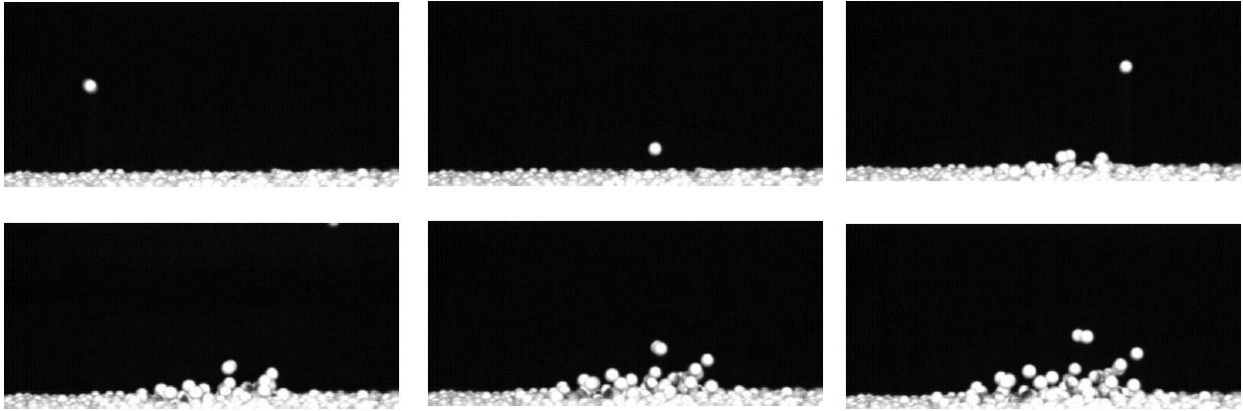

Figure 2.5. High-speed images of the splashing of surface particles by an impacting saltating particle; the time step between two successive images is 4 ms. (Reprinted with permission from Beladjine *et al.*, Physical Review E, 75, 061305, 2007. Copyright (2007) by the American Physical Society.)

*2.1.3.1 Characteristics of the rebounding particle*
The balance between saltators lost through failure to rebound and saltators gained through splash largely determines the characteristics of steady state saltation (Kok 2010a). Consequently, the probability that a particle does not rebound is a critical parameter in physically-based saltation models. Unfortunately, there have been very few studies of the probability that an impacting saltator fails to rebound, which contrasts with the many published studies on particle splash (e.g., Nalpanis *et al.* 1993, Beladjine *et al.* 2007). The most important exception is the numerical study of Anderson and Haff (1991), which suggested that the probability that a saltating particle will rebound upon impact can be approximated by

$$P_{reb} = B[1 - \exp(-\gamma v_{imp})], \qquad (2.10)$$

where $v_{imp}$ is the saltator's impact speed. Mitha *et al.* (1986) determined the parameter $B$ to be ~0.94 for experiments with 4 mm steel particles, whereas Anderson and Haff (1991) determined a similar value of $B \approx 0.95$ for ~250 μm sand particles. That is, even at large impact speeds not all saltating particles rebound from the surface, because some will dig into the particle bed (Rice *et al.* 1996). The parameter $\gamma$ has not been experimentally determined, although the numerical experiments of Anderson and Haff (1991) indicate that it is of order 2 s/m. This value is roughly consistent with the value of $\gamma = 1$ s/m that Kok and Renno (2009a) deduced from numerically reproducing the impact threshold, which is heavily dependent on Eq. (2.10) and the splash function. Note that the dimensionality of $\gamma$ suggests that it could depend on the particle diameter, the gravitational constant, the spring constant, and other relevant physical parameters.

If the saltator does rebound, the average *restitution coefficient* of the collision (i.e., the fraction of the impacting momentum retained by the rebounding particle) has been determined to lie in the range of ~0.5 – 0.6 for a loose sand bed by a wide range of numerical and laboratory experiments (Willetts and Rice 1985, 1986, Anderson and Haff 1988, 1991, McEwan and Willetts 1991, Nalpanis *et al.* 1993, Rice *et al.* 1995, 1996, Dong *et al.* 2002, Wang *et al.* 2008, Gordon and McKenna Neuman 2011). The mean restitution coefficient appears to be relatively invariant to changes in particle size (Rice *et al.*, 1995), but decreases with the impact angle (Beladjine *et al.* 2007, Oger *et al.* 2008). Moreover, recent measurements by Gordon and McKenna Neuman (2009) indicate that the restitution coefficient is a declining function of the saltator impact speed, which is consistent with both viscoelastic (Brilliantov *et al.* 1996) and plastic (Lu and Shao 1999) deformation of the soil bed during impact. However, experiments with plastic beads found that the restitution coefficient remains constant or decreases only slightly with impact speed (Rioual *et al.* 2000, Beladjine *et al.* 2007, Oger *et al.* 2008), such that further research is required to resolve this issue.

Measurements indicate that the mean angle from horizontal of the rebounding particle's velocity is around ~30 – 45° (McEwan and Willetts 1991, Nalpanis *et al.* 1993, Rice *et al.* 1995, Dong *et al.* 2002).

*2.1.3.2   Characteristics of splashed particles*
In steady state saltation, the loss of saltating particles to the soil bed through failure to rebound (Eq. (2.10)) must be balanced by the creation of new saltating particles through splash. Numerical simulations (Anderson and Haff 1988, 1991, Oger *et al.* 2005, 2008), laboratory experiments (Werner 1990, Rioual *et al.* 2000, Beladjine *et al.* 2007), and theory (Kok and Renno 2009a) indicate that the number of splashed particles (*N*) scales with the impacting momentum. That is,

$$N \overline{m_{\rm spl}} \overline{v_{\rm spl}} = \alpha m_{\rm imp} v_{\rm imp}, \tag{2.11}$$

where $\overline{m_{\rm spl}}$ and $\overline{v_{\rm spl}}$ are the average mass and speed of the particles splashed by the impacting saltator with mass $m_{\rm imp}$. The experiments of Rice *et al.* (1995) indicate that the fraction $\alpha$ of the average impacting momentum spent on splashing surface particles is of the order of ~15 % for a bed of loose sand particles. (Note that the ejection of dust aerosols from the soil, which is impeded by energetic interparticle bonds, probably scales with the impacting energy instead (Shao *et al.* 1993a, Kok and Renno 2009a).)

Since the number of splashed particles *N* is discrete, Eq. (2.11) produces two distinct regimes. For collisions with $N \sim 1$, an increase in the impacting particle momentum will result in an increase in the momentum of the single splashed particle, such that

$$\overline{v_{\rm spl}} = \alpha v_{\rm imp} m_{\rm imp} / \overline{m_{\rm spl}}. \qquad (N \sim 1) \tag{2.12}$$

This linear dependence of the speed of splashed particles on the impacting momentum is supported by the numerical simulations of Anderson and Haff (1988, see Figure 2.6). Conversely, for collisions with $N \gg 1$, momentum conservation and Eq. (2.11) require that the average speed of ejected particles becomes independent of the impacting momentum, as confirmed by the experiments of Werner (1987, 1990), Rioual *et al.* (2000), Beladjine *et al.* (2007), and Oger *et al.* (2008). That is,

$$\frac{\overline{v_{\rm spl}}}{\sqrt{gD_{\rm p}}} = \frac{\alpha}{a}, \qquad (N \gg 1) \tag{2.13}$$

where $\sqrt{gD_{\rm p}}$ nondimensionalizes the mean splashed particle speed, and the dimensionless parameter *a* is a proportionality constant that scales the number of splashed particles (Kok and Renno 2009a):

$$N = a \frac{m_{imp}}{m_{spl}} \frac{v_{imp}}{\sqrt{gD}}. \tag{2.14}$$

Measurements (McEwan and Willetts 1991, Rice *et al.* 1995, 1996) and results from the numerical saltation model COMSALT (Kok and Renno 2009a) indicate that *a* is ~ 0.01 – 0.03. The simplest way to capture the two limits of Eqs. (2.12) and (2.14) is then as follows

$$\frac{\overline{v_{spl}}}{\sqrt{gD}} = \frac{\alpha}{a}\left[1-\exp\left(-a\frac{m_{imp}}{m_{spl}}\frac{v_{imp}}{\sqrt{gD}}\right)\right], \tag{2.15}$$

which produces good agreement with numerical and laboratory experiments (Figure 2.6). The speed of splashed particles is thus generally about an order of magnitude less than that of the impacting particle (Rice *et al.*1995). Most splashed particles therefore move in low-energy reptation trajectories and quickly settle back to the soil bed. However, some splashed particles do gain enough momentum from the wind to participate in saltation and splash up more particles. This multiplication process produces a rapid increase in the particle concentration upon initiation of saltation (Section 2.2).

Measurements indicate that the distribution of splashed speeds at a given $\overline{v_{ej}}$ follows either an exponential or a lognormal distribution (Mitha *et al.*, 1986; Beladjine *et al.*, 2007; Ho et al. 2012; see Figure 5 in Kok and Renno 2009a for a compilation). Furthermore, laboratory and numerical experiments indicate that the mean angle at which particles are splashed is ~40-60 degrees from horizontal (Willetts and Rice 1985, 1986, 1989; Anderson and Haff 1988, 1991; Werner 1990; McEwan and Willetts, 1991; Rice *et al.* 1995, 1996, Gordon and McKenna Neuman 2011).

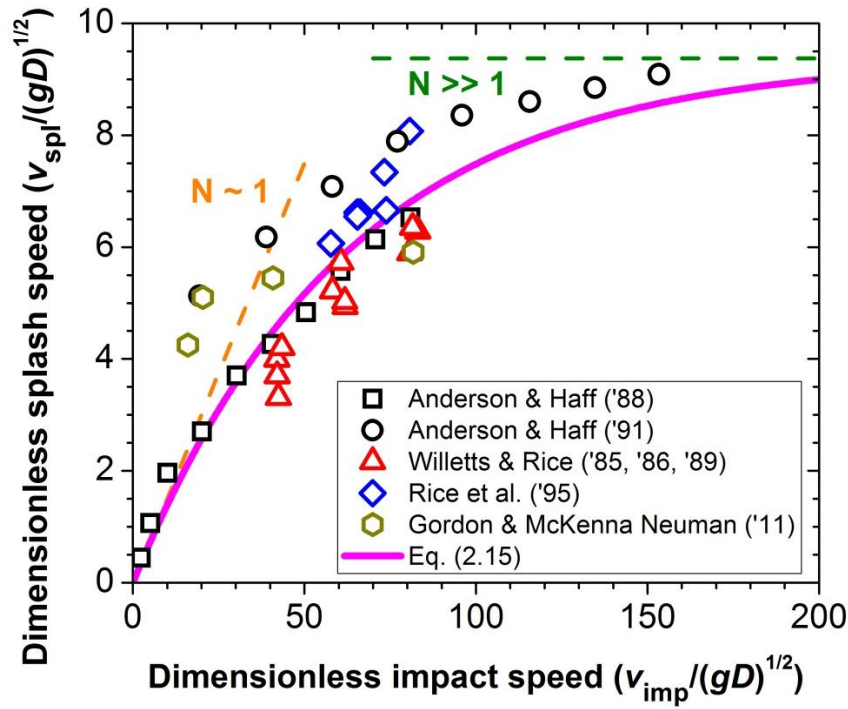

Figure 2.6. The average dimensionless speed of splashed surface particles ($\overline{v_{spl}}/\sqrt{gD}$) as a function of the dimensionless speed of the impacting particle ($v_{imp}/\sqrt{gD}$). Black and colored symbols respectively denote the results from numerical (Anderson and Haff 1988, 1991) and laboratory (Willetts and Rice 1985, 1986, 1989, Rice *et al.* 1995, 1996; Gordon and McKenna Neuman 2011) experiments with natural sand. The dashed orange and green lines show the limits at *N* ~ 1 (Eq. (2.12)) and *N* >> 1 (Eq. (2.13)), respectively, and the magenta line denotes the prediction of Eq. (2.15). These relations used $\alpha = 0.15$ and $a = 0.016$ (Rice *et al.* 1995, Kok and Renno 2009a).

*2.1.4 Modification of the wind profile by saltation*
Following our review of the initiation of saltation through particle lifting by the fluid (Section 2.1.1), the subsequent particle trajectories (Section 2.1.2), and the collision of saltating particles with the surface (Section 2.1.3), we now discuss the final critical process in saltation: the modification of the wind profile through momentum transfer to saltating particles. Indeed, it is the retardation of the wind profile through drag by saltating particles that ultimately limits the number of particles that can be saltating under given conditions.

*2.1.4.1 Wind profile in the absence of saltation*
The large length scale of the atmospheric boundary layer in which saltation occurs causes the Reynolds number of the flow to be correspondingly large, typically in excess of $10^6$ for both Earth and Mars, such that the flow in the boundary layer is turbulent. Since the horizontal fluid momentum higher up in the boundary layer exceeds that near the surface, eddies in the turbulent flow on average transport horizontal momentum downward through the fluid. Together with the much smaller contribution due to the viscous shearing of neighboring fluid layers, the resulting downward flux of horizontal momentum constitutes the fluid shear stress $\tau$. Because the horizontal fluid momentum is transported downward through the fluid until it is dissipated at the surface, $\tau$ is approximately constant with height above the surface for flat and homogeneous surfaces (Kaimal and Finnigan 1994).

The downward transport of fluid momentum through both viscosity and turbulent mixing equals (e.g., Stull 1988)

$$\tau = (K\rho_a + \mu)\frac{\partial \overline{U_x}(z)}{\partial z} \approx K\rho_a \frac{\partial \overline{U_x}(z)}{\partial z}, \qquad (2.16)$$

where we neglect the contribution from viscosity since the turbulent momentum flux exceeds the viscous momentum flux by several orders of magnitude for turbulent flows (e.g., White 2006). The eddy (or turbulent) viscosity $K$ quantifies the transport of momentum in turbulent flows in analogy with the exchange of fluid momentum through viscosity, and $\overline{U_x}(z)$ is the mean horizontal fluid velocity at a height $z$ above the surface. The size of turbulent eddies that transport the fluid momentum is limited by the distance from the surface, such that the eddy viscosity is usually assumed proportional to $z$ (Prandtl 1935, Stull 1988) as well as the shear velocity $u_*$ (defined in Eq. (2.4)),

$$K = \kappa u_* z, \qquad (2.17)$$

where $\kappa \approx 0.40$ is von Kármán's constant. (Note that recent results by Li *et al.* (2010) have called into question the presumed constancy of $\kappa$ during saltation.) Combining Eqs. (2.16) and (2.17) then yields

$$\overline{U_x}(z) = \frac{u_*}{\kappa}\ln\left(\frac{z}{z_0}\right), \qquad (2.18)$$

where $z_0$ is the aerodynamic surface roughness, which denotes the height at which the logarithmic profile of (2.18), when extrapolated to the surface, yields zero wind speed. The value of $z_0$ for different flow conditions is discussed in more detail below.

Eq. (2.18) is referred to as the logarithmic *law of the wall* and is widely used to relate the shear velocity $u_*$, and thus the wind shear stress $\tau$, to the vertical profile of the mean horizontal wind speed. However, there are several limitations to its use (Bauer et al. 1992, Namikas et al. 2003). First, Eq. (2.18) was derived assuming that the shear stress in the surface layer is constant with height. This is a realistic approximation for flat, homogeneous surfaces, but can be unrealistic for other conditions, such as for surface with non-uniform surface roughness or substantial elevation changes (e.g., Bauer et al. 1992). Second, as discussed in the next section, the drag by saltating particles reduces the horizontal momentum flux carried by the wind. Therefore, determining $u_*$ requires measurements of the wind speed above the saltation layer. Third, turbulence causes the instantaneous wind speed to substantially vary over time, such that Eq. (2.18) is only valid on timescales long enough for the fluid to access all relevant frequencies of the turbulence (Kaimal and Finnigan 1994, Van Boxel *et al.* 2004). Because of the large length scale of the atmospheric boundary layer (~1000 m), this time scale is of the order of ~10 minutes (Stull 1988, McEwan *et al.* 1993). And, finally, Eq. (2.18) is technically only valid when the stability of the atmosphere is neutral (that is, the vertical temperature profile follows the adiabatic lapse rate). Unstable conditions produce a lower wind speed at given values of $z$ and $u_*$, whereas stable conditions produce a higher wind speed. However, the corrections required to account for the stability of the atmosphere in the wind speed profile of Eq. (2.18) are small close to the surface (Kaimal and Finnigan 1994), such that this equation can be considered sufficient for most studies of saltation.

An exception to the predominantly turbulent flow in the atmospheric boundary layer occurs near the surface, where surface friction can cause viscous forces to dominate over inertial forces, resulting in laminar flow. For a smooth surface, the thickness of this viscous (or laminar) sublayer is approximately (White 2006)

$$\delta_{\text{vis}} \cong \frac{5\mu}{\rho_a u_*}, \tag{2.19}$$

where $\mu$ is the dynamic viscosity. The thickness of the viscous sublayer at the fluid threshold is ~0.4 mm on Earth and ~2 mm on Mars. Consequently, sand particles protrude into the viscous sublayer, thereby enhancing turbulent transport by shedding wakes and partially dissipating the viscous sublayer. The thickness of the viscous sublayer thus depends on the size of these roughness elements, as denoted by the roughness Reynolds number (White 2006),

$$\text{Re}_r = \frac{\rho_a k_s u_*}{\mu}, \tag{2.20}$$

where $k_s$ is the Nikuradse roughness (Nikuradse, 1933). For a homogeneously arranged bed of monodisperse spherical particles we have that $k_s \approx D_p$ (Bagnold 1938), whereas for a more realistic irregular surface of mixed sand $k_s$ equals two to five times the median particle size (Thompson and Campbell 1979, Greeley and Iversen 1985, Church *et al.* 1990). For $\text{Re}_r > ~60$, the turbulent mixing generated by the roughness elements is sufficient to destroy the viscous sublayer and the flow is termed aerodynamically rough (Nikuradse 1933). Consequently, the aerodynamic surface roughness $z_0$ (defined by Eq. (2.18)) is related in a straightforward manner to the Nikuradse roughness (White 2006),

$$z_0 = k_s/30. \quad (\text{Re}_r > 60) \tag{2.21}$$

In contrast, for $\text{Re}_r < ~4$ the roughness elements are too small to substantially perturb the viscous sublayer, and the flow is termed aerodynamically smooth (Nikuradse 1933). In this case, the aerodynamic surface roughness is thus proportional to the thickness of the viscous sublayer (Eq. (2.19)) and equals (Greeley and Iversen, 1985)

$$z_0 = \frac{\mu}{9\rho_a u_*}. \quad (\text{Re}_r < 4) \tag{2.22}$$

Aeolian saltation on Earth takes place for roughness Reynolds numbers of ~1 – 100 and thus usually occurs in the transition zone between the smooth and rough aerodynamic regimes (Dong *et al.* 2001). Since the roughness in the transition regime does not differ much from that in the aerodynamically rough regime (see figure 11 in Nikuradse 1933), most studies have used Eq. (2.21) to approximate the surface roughness (e.g., Anderson and Haff 1991, Sorensen 1991, 2004, Kok and Renno 2009a, Jenkins *et al.* 2010). In contrast, the much lower air density on Mars causes saltation there to usually take place in the smooth aerodynamic regime, resulting in values of $z_0$ substantially larger than predicted by Eq. (2.21). However, most studies of martian saltation have still used this equation for $z_0$ (e.g., White 1979, Almeida *et al.* 2008, Kok 2010a, b) because, once saltation is initiated, the motion of the saltating particles through the viscous sublayer dramatically increases the vertical mixing of horizontal fluid momentum (e.g., Owen 1964), which likely eliminates the viscous sublayer (White 1979).

*2.1.4.2 Modification of the wind profile through drag by saltating particles*

In the presence of saltation, the logarithmic wind profile of Eq. (2.18) is modified by the transfer of momentum between the fluid and saltating particles. Some of the downward horizontal momentum flux in the saltation layer is thus carried by saltating particles, with the total downward momentum flux remaining constant. That is (e.g., Raupach 1991),

$$\tau = \tau_a(z) + \tau_p(z), \quad (2.23)$$

with the particle momentum flux $\tau_p$ given by (Shao, 2008)

$$\tau_p(z) = \phi(z)\Delta v_x(z), \quad (2.24)$$

and where the fluid momentum flux in the saltation layer $\tau_a(z)$ reduces to $\tau$ for $z$ above the saltation layer, $\phi(z)$ is the mass flux of particles passing the height $z$ in either the upward or downward direction, and $\Delta v_x(z)$ is the difference between the average velocity of descending and ascending particles. Numerical models find that $\tau_p$ decays approximately exponentially from a maximum value near the surface (Figure 2.7). Pähtz *et al.* (2012) derive the dependence of the e-folding length of the particle shear stress on physical parameters such as particle size and the wind shear velocity.

In analogy with Eq. (2.4), the shear velocity within the saltation layer equals

$$u_{*a}(z) = \sqrt{\tau_a(z)/\rho_a}, \quad (2.25)$$

which similarly reduces to $u_*$ (the shear velocity above the saltation layer) for $z$ above the saltation layer. Combining Eq. (2.25) with the analogs of Eqs. (2.16) and (2.17) for the saltation layer then yields

$$\frac{\partial \overline{U_x}(z)}{\partial z} = \frac{1}{\kappa z}\sqrt{u_*^2 - \tau_p(z)/\rho_a}. \quad (2.26)$$

Although Eq. (2.26) does not have a straightforward analytical solution (Sorensen 2004, Durán and Herrmann 2006a, Pähtz *et al.* 2012), its implementation in numerical saltation models allows a straightforward calculation of the wind profile (see Section 2.3.2.4 and, e.g., Werner 1990). Moreover, by assuming a plausible function for the vertical profile of the particle shear stress, such as an exponential profile (Figure 2.7), analytical treatments have been able to derive approximate expressions of the wind profile (Sorensen 2004, Li *et al.* 2004) and the aerodynamic roughness length in saltation (Raupach *et al.* 1991, Durán and Herrmann 2006a, Pähtz *et al.* 2012; see Section 2.3.2.4).

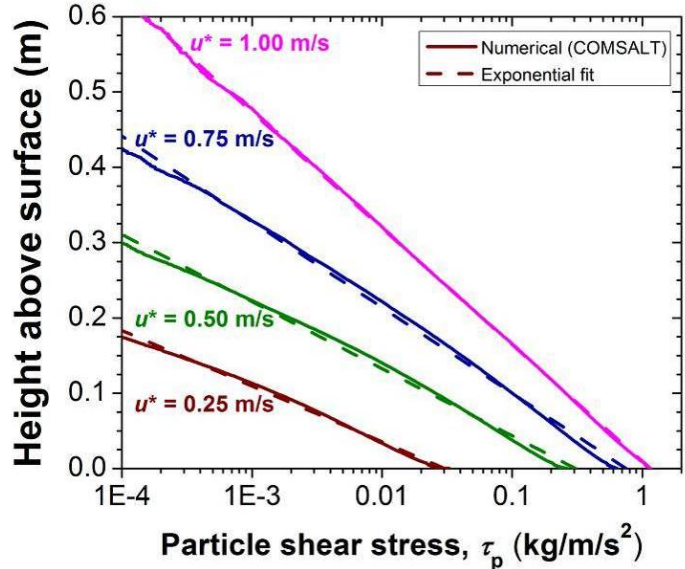

Figure 2.7. The particle shear stress as a function of height above the surface as simulated by the numerical saltation model COMSALT (Kok and Renno 2009a) for 250 μm particles. The dashed lines represent exponential fits to the numerical results.

*2.2 The path of saltation to steady state*

The interplay of the four main saltation processes reviewed in Section 2.1 determines both the path to and the characteristics of steady state saltation. After the saltation fluid threshold has been exceeded (Section 2.1.1), particles lifted from the surface are quickly accelerated by the wind into ballistic trajectories (Section 2.1.2) and, after several hops, can have gathered sufficient momentum to splash surface particles (Section 2.1.3). These newly ejected particles are themselves accelerated by wind and eject more particles when impacting the surface, causing an exponential increase in the horizontal saltation flux in the initial stages of saltation (Anderson and Haff 1991, McEwan and Willetts 1993, Andreotti *et al.* 2010, Durán *et al.* 2011a). This rapid increase in the particle concentration produces a corresponding increase in the drag of saltating particles on the fluid, thereby retarding the wind speed (Section 2.1.4). This in turn reduces the speed of saltating particles, such that a steady state is reached when the speed of saltating particles is reduced to a value at which there is a single particle leaving the soil surface for each particle impacting it (Ungar and Haff 1987). Due to the finite response time of saltating particle speeds to the wind speed, the horizontal saltation flux can 'overshoot' the eventual steady state mass flux (Anderson and Haff 1988, 1991; Shao and Raupach 1992; McEwan and Willetts 1993), after which the momentum fluxes of the fluid and saltating particles reach an equilibrium.

The distance required for saltation to reach steady state in the manner reviewed above is characterized by the *saturation length* (e.g., Sauermann *et al.* 2001). Its value depends on several length scales in saltation, such as the length of a typical saltation hop, the length needed to accelerate a particle to the fluid speed, and the length required for the drag by saltating particles to retard the wind speed (e.g., Andreotti *et al.* 2010). As reviewed in more detail in Durán *et al.* (2011a), these finite length scales cause the saltation flux to require between ~1 (Andreotti *et al.* 2010) to ~10 – 20 meters (Shao and Raupach 1992, Dong *et al.* 2004) of horizontal distance to saturate. The cause of the large range of the saturation length measured in the literature is not well understood, but is possibly due to differences in the soil size distribution and the shear velocity (Durán *et al.* 2011a). As we discuss in Section 3.1.1, the saturation length is critical for understanding the typical length scales of dunes.

In addition to the saturation length, there is another characteristic length scale over which the horizontal saltation flux increases to a steady state: the *fetch distance* (Gillette *et al.* 1996). The corresponding *fetch effect* arises because the atmospheric boundary layer flow adjusts to the increased roughness of the surface layer produced by saltation (discussed in more detail in Section 2.3.2.4). The increased surface roughness acts as a greater sink of horizontal fluid momentum, which increases the downward flux of fluid momentum, thereby increasing the wind shear velocity for a given free stream wind speed in the atmospheric boundary layer. This process acts as a positive feedback on saltation and is termed the *Owen effect* after the theoretical paper of Owen (1964) that identified it. Field studies indicate that the fetch distance for a flat field is of the order of ~100 meters (Gillette *et al.* 1996). After this initial increase over the fetch distance, the near-surface shear stress relaxes to a lower equilibrium value. This results in a decrease of sand flux with distance further downwind (McEwan and Willetts 1993), as recently verified by field measurements of sand flux in a dune field (Jerolmack *et al.* 2012).

*2.3 Steady state saltation*

Saltation is in steady state when its primary characteristics, such as the horizontal mass flux and the concentration of saltating particles, are approximately constant with time and distance. Since wind speed can undergo substantial turbulent fluctuations, this is rarely true on timescales longer than minutes or often even seconds, causing saltation on Earth to be highly intermittent (Stout and Zobeck, 1997). In fact, a substantial fraction of sand transport occurs in *aeolian streamers* or *sand snakes* (Figure 2.8), which are probably produced by individual eddies of high-speed air (Baas and Sherman 2005). These streamers have typical widths of ~0.2 meters, thereby producing strong variability on short time and length scales (Baas and Sherman 2005, Baas 2006). However, numerical models and field measurements indicate that the saltation mass flux responds to changes in wind speed with a characteristic time scale of a second (Anderson and Haff 1988, 1991, McEwan and Willetts 1991, 1993, Butterfield 1991, Jackson *et al.* 1997, Ma and Zheng 2011). Consequently, it seems plausible that saltation is close to steady state for most conditions (Durán *et al.*, 2011a). This hypothesis is experimentally supported by (i) the finding that particle speeds near the surface do not depend on $u_*$ (see Namikas (2003) and Section 2.3.2.1), (ii) the occurrence of the *Bagnold focus* (Figure 2.9), and (iii) the relative insensitivity of the saltation layer height to changes in $u_*$ (Section 2.3.2.2). All these observations are consistent with theories and numerical models of steady-state saltation. Nonetheless, more measurements are needed to explicitly test to what extent natural saltation is in steady state.

Note that this section reviews the steady state characteristics of *transport limited* saltation, for which the amount of saltating sand is limited by the availability of wind momentum to transport the sand. This contrasts with *supply limited* saltation, for which the amount of saltating sand is limited by the availability of loose soil particles that can participate in saltation, which can occur for crusted or wet soils (e.g. Rice *et al.* 1996, Gomes *et al.* 2003). The characteristics of supply limited saltation are reviewed in Nickling and McKenna Neuman (2009).

The particle concentration in transport limited saltation is in steady state when there is exactly one particle leaving the soil bed for each particle impacting it (Ungar and Haff 1987, Andreotti 2004, Kok 2010a). An equivalent constraint is that for each saltating particle lost to the soil bed due to failure to rebound upon impact (Section 2.1.3.1), another particle must be lifted from the soil bed and brought into saltation by either splash or aerodynamic entrainment (Shao and Li 1999, Doorschot and Lehning 2002). In order to understand and predict the characteristics of steady state saltation we thus need to determine whether particles are lifted predominantly by fluid drag or through splashing by impacting saltating particles. We do so in the next section.

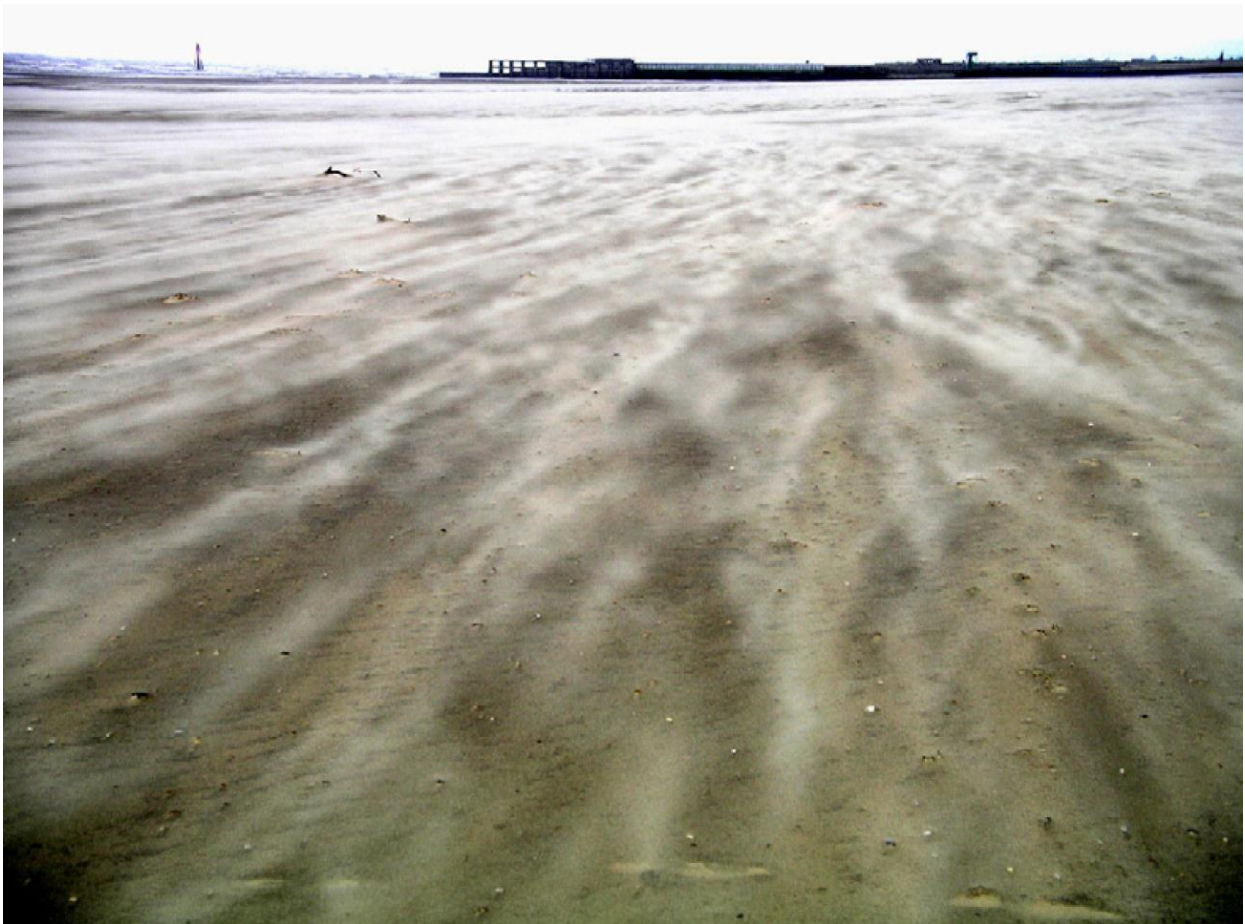

Figure 2.8. Aeolian streamers moving toward the observer. Reprinted from Baas (2008), with permission from Elsevier.

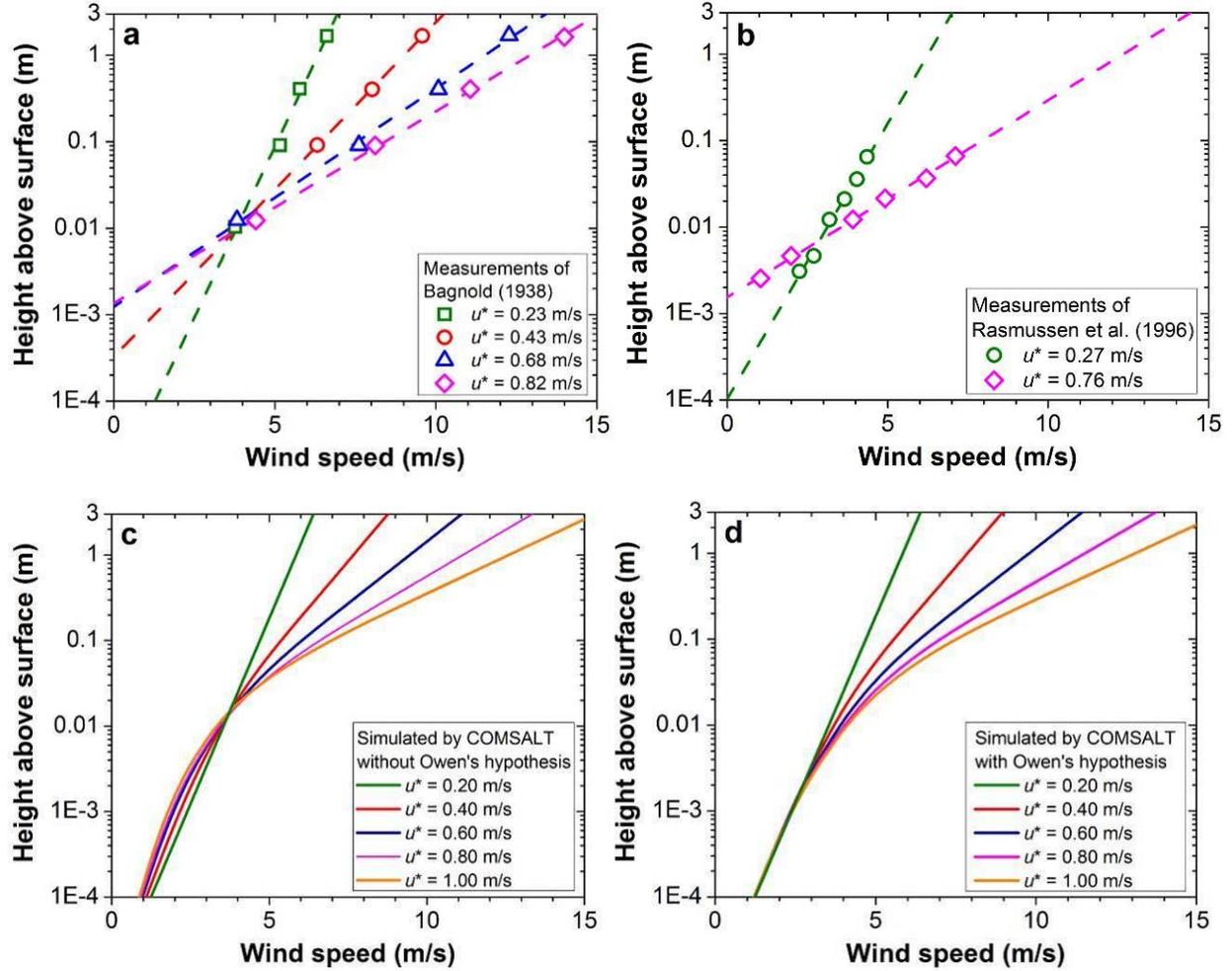

Figure 2.9. Measurements of the wind profile during saltation of ~250 μm sand by Bagnold (1938) and Rasmussen *et al.* (1996) show the occurrence of a 'focus' of wind profiles close to the surface (panels **(a)** and **(b)**). This *Bagnold focus* can be reproduced by saltation models that account for the occurrence of splash as the main particle entrainment mechanism (panel **(c)**; also see e.g. Werner 1990), but not by models that use Owen's hypothesis that fluid lifting is the main particle entrainment mechanism (**d**).

### 2.3.1 Is particle lifting in steady state due to aerodynamic or splash entrainment?

In his influential theoretical paper, Owen (1964) argued that the near horizontal impact of saltators on the soil surface would be ineffective at mobilizing surface particles, such that particles must be predominantly lifted by aerodynamic forces in steady state saltation. Owen further reasoned that, as a consequence, the fluid shear stress at the surface must equal a value "just sufficient to ensure that the surface grains are in a mobile state," which he took as the impact threshold ($\tau_{it}$). Owen argued that, if the surface shear stress falls below the impact threshold, fewer particles are entrained by wind. This in turn reduces the transfer of momentum from the fluid to saltating particles, thereby increasing the surface shear stress back to its threshold value. Conversely, if the surface shear stress exceeds the threshold value, more particles are entrained, again restoring the surface shear stress to its critical value. That is, Owen hypothesized that

$$\tau_a(0) \approx \tau_{it} \text{, or, equivalently} \tag{2.27}$$

$$u_{*\text{sfc}} \equiv u_{*a}(0) \approx u_{*it}. \tag{2.28}$$

These equations can greatly simplify analytical and numerical studies of saltation and have thus been widely adopted (e.g., Raupach 1991, Marticorena and Bergametti 1995, Sauermann *et al.* 2001, Kok and Renno 2008, Huang *et al.* 2010, Ho *et al.* 2011). Unfortunately, as we review below, there is strong evidence that Owen's hypothesis that the bed fluid shear stress remains at the impact threshold is incorrect and that its use can produce incorrect results.

Owen's hypothesis rests on the assumption that splash plays a minor role in entraining surface particles into the fluid flow. However, this assumption is inconsistent with the occurrence of saltation below the fluid threshold (Bagnold 1941), which indicates that splash is more efficient than direct fluid drag in transferring momentum to surface particles. Moreover, a large number of wind tunnel experiments (e.g., Willetts and Rice 1985, 1986, 1989, Nalpanis *et al.* 1993, Rice *et al.* 1995, 1996, Gordon and McKenna Neuman 2009, 2011) show that particles are splashed at impact speeds typical of saltation (~1 m/s for loose sand on Earth; see Section 2.3.2.1). Implementing the results of wind tunnel measurements of splash into numerical models causes the fluid shear stress at the surface to decrease with the wind shear velocity $u_*$ to values well below the impact threshold (Figure 2.10), contradicting Owen's hypothesis of Eq. (2.28).

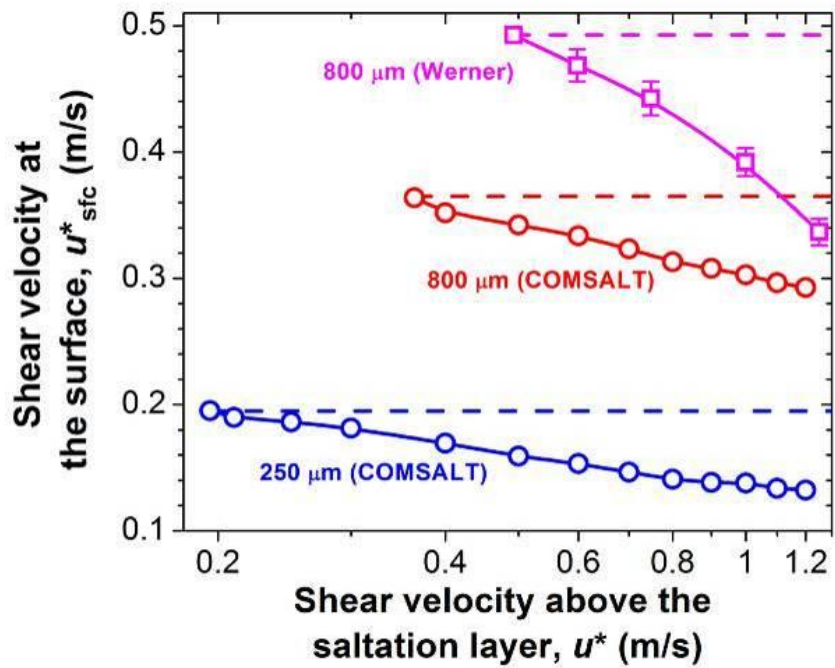

Figure 2.10. Results from the physically-based numerical saltation models of Werner (1990) and Kok and Renno (2009a) indicate that the shear velocity at the surface (solid line and symbols) decreases with $u_*$. These findings contradict Owen's hypothesis that the surface shear stress stays at the impact threshold (dashed lines). The physical reason for the decrease of $u_{*\mathrm{sfc}}$ with $u_*$ is discussed in Section 2.3.2.4.

Further evidence of the incorrectness of Owen's hypothesis is provided by measured vertical profiles of the wind speed. Indeed, wind profiles simulated with Eq. (2.28) cannot reproduce the focusing of wind profiles for different values of $u_*$ at a height of ~1 cm (Figure 2.9). This well-known feature of wind profiles in the presence of saltation is known as the *Bagnold focus* (Bagnold 1936). In contrast, numerical models that do not use Eq. (2.28), and instead include a parameterization for the splashing of surface particles do reproduce the Bagnold focus (Ungar and Haff 1987, Werner 1990, Kok and Renno 2009a, Durán *et al.* 2011a, Ma and Zheng 2011, Carneiro *et al.* 2011).

Owen's hypothesis is thus inconsistent with laboratory measurements of splash, simulations of the wind profile by physically-based numerical models (Figure 2.10), the occurrence of the Bagnold focus (Figure 2.9), and, as we discuss in Section 2.3.2.1 below, measurements of the particle speed in saltation. We thus conclude that Owen's hypothesis is incorrect, which has two important implications. First, the surface shear stress does not remain at the impact threshold, but instead decreases with $u_*$ (Figure 2.10). And, second, particle entrainment in steady state is dominated by splash, not by direct fluid lifting (e.g., Anderson and Haff 1988, 1991, Werner 1990). In the next few sections, we will show that the dominance of splash entrainment in steady state provides powerful constraints on the characteristics of steady state saltation.

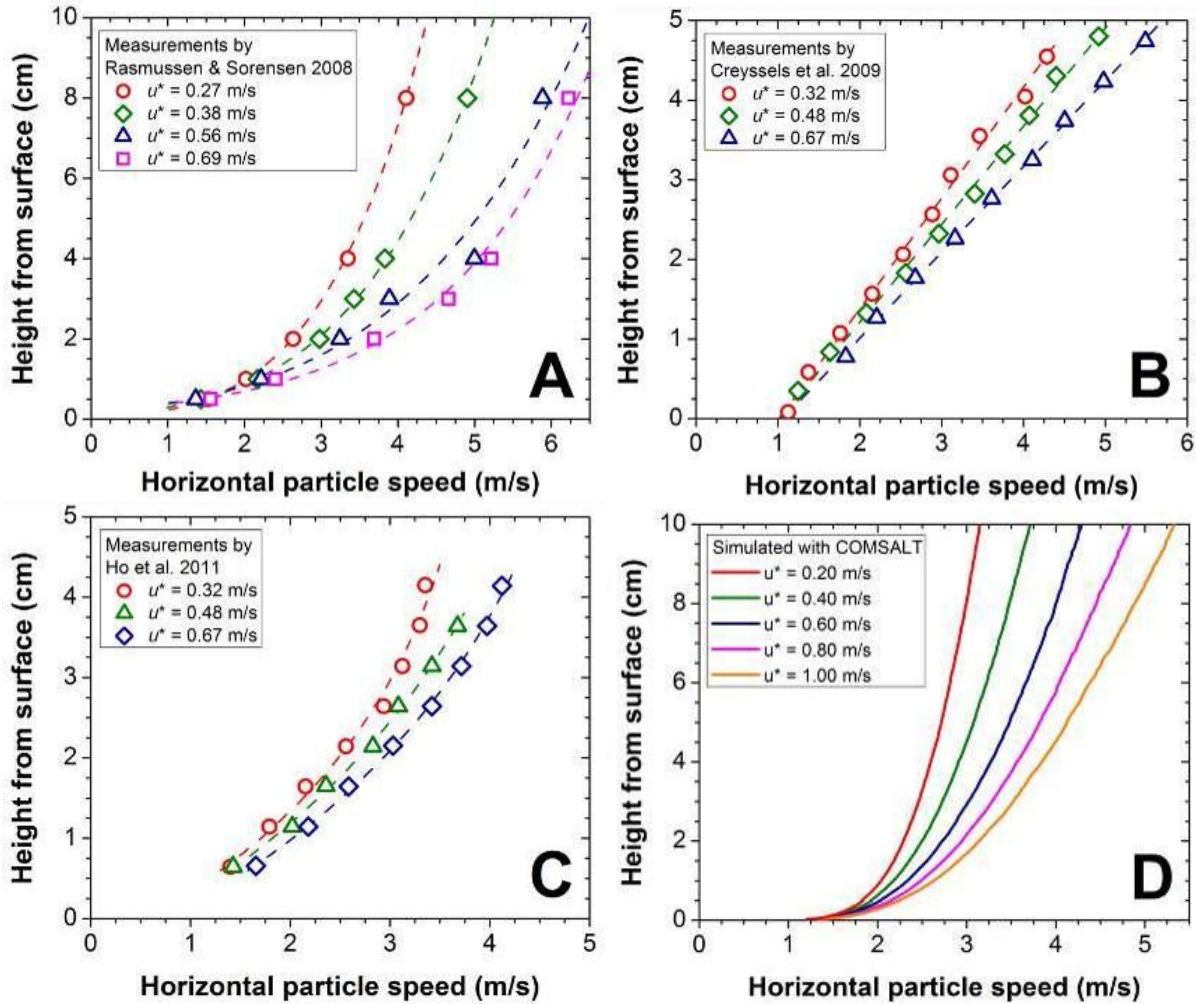

Figure 2.11. Vertical profiles of the horizontal particle speed in steady state saltation from the wind-tunnel measurements of (**a**) Rasmussen and Sorensen (2008), (**b**) Creyssels *et al.* (2009), and (**c**) Ho *et al.* (2011) show that the particle speed converges to a common value near the surface (see also Figure 2.12). This is reproduced by numerical saltation models (panel (**d**) and figure 23 in Durán *et al.*, 2011). All results are for particle sizes of ~250 μm.

## 2.3.2 *Characteristics of steady state saltation*

Since particle entrainment in steady state saltation is dominated by splash, the number of saltating particles lost to the soil bed through failure to rebound (Section 2.1.3.1) must be balanced by the creation of new saltating particles through splash. That is, the mean *replacement capacity* (Werner 1990) of a saltator impact must equal unity. As is evident from Eqs. (2.10) and (2.14), the steady state impact speed at which the replacement capacity equals unity is not dependent on $u_*$ (Ungar and Haff 1987, Andreotti 2004, Kok 2010a, Durán *et al.* 2011a). Consequencely, the mean particle speed at the surface must remain constant with $u_*$, which contradicts assumptions made in the seminal works of Bagnold (1941) and Owen (1964), but is strongly supported by measurements (discussed in the next section).

The independence of surface particle speeds with $u_*$ is a powerful constraint that can be used to understand and predict many of the characteristics of steady state saltation. Below, we discuss the particle speed, the height of the saltation layer, the horizontal flux of saltating particles, the wind profile, and the size distribution of saltating particles in steady–state saltation.

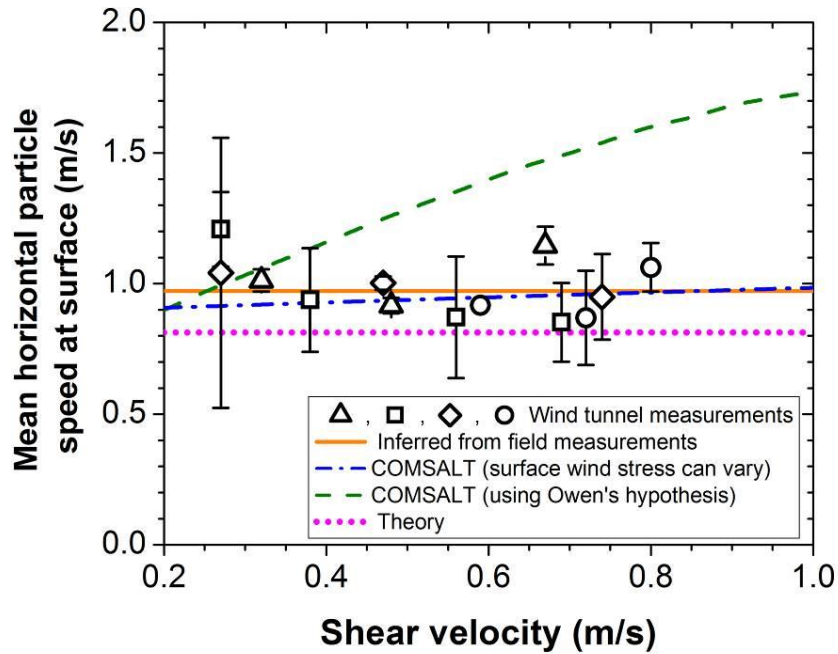

Figure 2.12. Wind tunnel measurements of the speed of ~250 – 300 µm saltating particles (symbols) indicate that the mean horizontal speed at the surface stays constant with $u_*$. Similarly, Namikas (2003) inferred from his field measurements that the speed with which saltating particles are launched from the surface is independent of $u_*$ (solid orange line). Also included are COMSALT simulations with and without Owen's hypothesis (dashed green and dash-dotted blue lines, respectively), and the prediction of the theory of Kok (2010a) (dotted magenta line). (The launch speed and angle of ~0.70 m/s and ~35° inferred by Namikas (2003) were converted to a mean horizontal surface speed by using that the rebound speed is $\sim v_{\text{imp}}/2$, and that the impact and launch angles are ~12° and ~35° degrees (e.g., Rice *et al.* 1995). Estimates of the surface particle speed in the wind tunnel measurements of Rasmussen and Sorensen (squares and diamonds respectively denote measurements with 242 and 320 µm sand), Creyssels *et al.* (triangles, 242 µm sand), and Ho *et al.* (circles, 230 µm sand) were obtained by linearly extrapolating horizontal particle speed measurements within 2 mm of the surface. Error bars were derived from the uncertainty in the fitting parameters.)

*2.3.2.1 Particle speed*
As discussed above, the dominance of splash entrainment in steady state saltation requires the mean particle speed at the surface to be independent of $u_*$. Since the mean speed of particles high in the saltation layer can be expected to increase with $u_*$, the vertical profiles of particle speed must converge towards a common value near the surface. A range of recent wind tunnel measurements (Rasmussen and Sorensen 2008, Creyssels *et al.* 2009, Ho *et al.* 2011), field measurements (Namikas 2003), and numerical saltation models (Kok and Renno 2009a, Durán *et al.* 2011a) confirm both these results (Figure 2.11 and Figure 2.12). Note that the independence of surface particle speeds with $u_*$ cannot be reproduced by numerical saltation models using Owen's hypothesis (Figure 2.12).

Although the mean speed of particles at the surface thus remains approximately constant with $u_*$, the probability distribution of particles speeds at the surface does not. Indeed, increases in $u_*$ produce increases in the wind speed above the Bagnold focus but decreases in wind speed below the Bagnold focus (Figure 2.9). Consequently, the speed of energetic particles moving mostly above the Bagnold focus increases with $u_*$, whereas the speed of less energetic particles moving mostly below the Bagnold focus decreases (Figure 2.13). In other words, the probability distribution of particle speeds at the surface will broaden. In particular, the tail of very large impact speeds increases strongly with $u_*$, which is offset by a decrease in the probability of lower particle speeds (Figure 2.13).

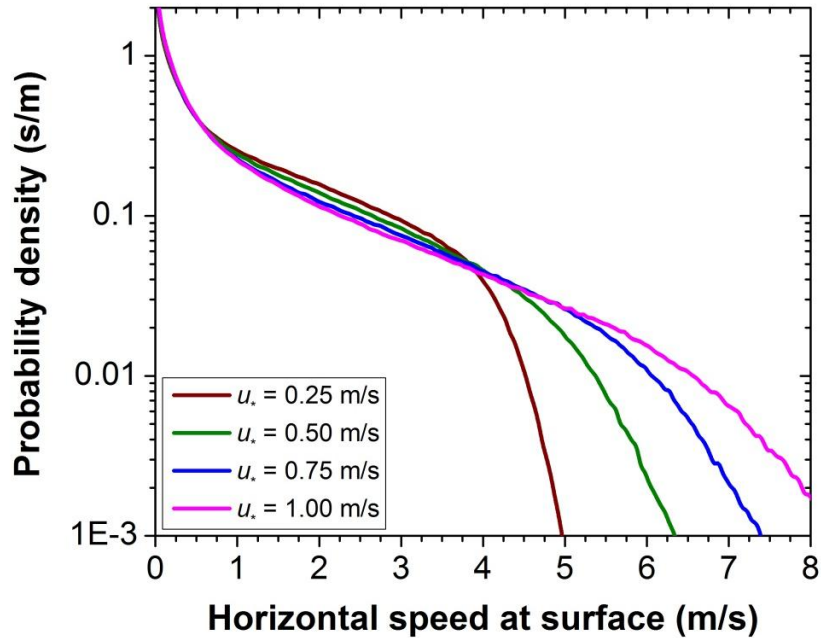

Figure 2.13. Probability density function of the horizontal particle speed at the surface of a bed of 250 μm particles, simulated with COMSALT. Note that the high-energy tail increases sharply with the shear velocity.

### 2.3.2.2  Height of the saltation layer

Since particle trajectories vary widely (e.g., Anderson and Hallet 1986, Namikas 2003), there is no universal measure of the height of the saltation layer. However, since vertical profiles of the horizontal mass flux are commonly measured in wind tunnel and field experiments (e.g., Greeley *et al.* 1996, Namikas 2003, Dong *et al.* 2007), convenient ways to define the saltation layer height include the e-folding length in the vertical profile of the horizontal mass flux (e.g. Kawamura 1951, Farrell and Sherman 2006) and the height below which 50 % of the mass flux occurs (Kok and Renno 2008). The height of the saltation layer is thus determined by (i) the distribution of speeds with which saltators leave the surface, which sets the vertical concentration profile of saltators, and (ii) the mean horizontal saltator speed at each height. As such, measurements of the saltation layer height can constrain theoretical and numerical predictions of important saltation properties such as the mass flux and particle impact speed.

Early theoretical predictions of the saltation layer height $z_{\text{salt}}$ assumed that the speed with which particles leave the surface scales with $u_*$ (Bagnold 1941, Owen 1964), resulting in

$$z_{\text{salt}} = c_z u_*^2 / 2g, \qquad (2.29)$$

with $c_z$ of the order of 1. However, as discussed above, advances in theory, numerical saltation models, and measurements have shown that the average speed with which saltating particles are launched from the surface is approximately constant with $u_*$ (Figure 2.12), such that Eq. (2.29) is likely incorrect (Pähtz *et al.* 2012). And indeed, field measurements of the saltation layer height (Greeley *et al.* 1996, Namikas 2003) find no evidence of the sharp increase with $u_*$ predicted by Eq. (2.29). Instead, the field measurements of Greeley *et al.* (1996) and Namikas (2003) suggest that the height of the saltation layer remains approximately constant with $u_*$ (Figure 2.14), whereas the recent field measurements of Dong et al. (2012) suggest a slight increase in the saltation layer with wind speed (see their Figure 7). This latter result is also what is expected from theory and predicted by models: although the particle speed at the surface stays constant, numerical simulations and wind-tunnel measurements find that the particle speed does increase with $u_*$ above the surface (Figure 2.11). This means that the mass flux higher up in the saltation layer increases relative to that in lower layers, producing a slight increase in the saltation layer height with $u_*$ (blue line in Figure 2.14).

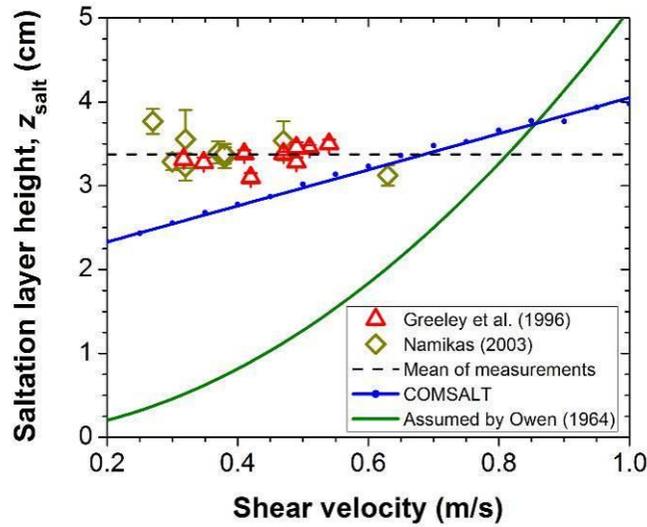

Figure 2.14. Field measurements (symbols) indicate that the height of the saltation layer (defined here as the height below which 50 % of the mass flux occurs) stays approximately constant with wind speed, whereas classical theory (Bagnold 1941, Owen 1964) predicts a sharp increase (solid green line denotes Eq. (2.29)), which is thus incorrect. The numerical saltation model COMSALT (Kok and Renno 2009a) predicts a more gradual increase in the saltation layer height with $u_*$, consistent with Dong et al. (2012) and the increase in above-surface particle speeds found in wind tunnel experiments (Figure 2.11).

#### 2.3.2.3 *Saltation mass flux*
The saltation mass flux $Q$, obtained by vertically integrating the horizontal flux of saltating particles, represents the total sand movement in saltation and is thus a critical measure for wind erosion and dune formation studies. Consequently, much effort has been devoted to formulating equations that effectively predict the saltation mass flux (e.g., Bagnold 1941, Kawamura 1951, Owen 1964, Lettau and Lettau 1978, Sorensen 2004).

The saltation mass flux can be derived from the momentum balance in the saltation layer, that is, that the sum of the horizontal momentum fluxes due to particles ($\tau_p$) and the fluid ($\tau_a$) equals the fluid momentum flux at the top of the saltation layer ($\tau$) (Eq. (2.23)). Applying this relation at the surface yields
$$\tau_p = \rho_a \left( u_*^2 - u_{*\text{sfc}}^2 \right). \tag{2.30}$$

The saltation mass flux is thus generated by the absorption of the excess horizontal wind momentum flux of Eq. (2.30) by saltating particles. The contribution $q$ of a typical saltating particle hop to the mass flux is the lengthwise mass transport per unit time,

$$q = mL/t_{\text{hop}} \tag{2.31}$$

where $m$ is the particle mass, $L$ is the particle's hop length, and $t_{\text{hop}}$ is the duration of the hop. This hop extracts an amount of momentum $p$ per unit time from the wind,

$$p = m(v_{\text{imp}} - v_{\text{lo}})/t_{\text{hop}} \tag{2.32}$$

where $v_{\text{imp}}$ and $v_{\text{lo}}$ are respectively the horizontal particle speed upon impact and lift-off from the surface. The steady state saltation mass flux $Q$ is then the available horizontal wind momentum flux $\tau_p$ available to accelerate particles (Eq. (2.30), multiplied by the mass flux $q$ (Eq. (2.31)) generated by a unit momentum flux $p$ (Eq. (2.32)). That is,

$$Q = \tau_p \frac{q}{p} = \rho_a \left(u_*^2 - u_{*\text{sfc}}^2\right)\frac{L}{\Delta v}, \tag{2.33}$$

where $\Delta v$ is the average difference between the particle's impact and lift-off speeds.

Different assumptions about the dependence of $L$, $\Delta v$, and $u_{*\text{sfc}}$ on $u_*$ have resulted in different equations relating $Q$ to $u_*$, as summarized in Table 2.1. In particular, $u_{*\text{sfc}}$ is usually approximated with $u_{*\text{it}}$, which, as we noted in Section 2.3.1, is incorrect and produces errors when implemented in analytical or numerical saltation models (Figure 2.9 and Figure 2.12). However, this approximation is more reasonable in this instance because when $u_*$ is close to $u_{*\text{it}}$, we have that $u_{*\text{it}} \approx u_{*\text{sfc}}$ (Figure 2.10), whereas when $u_* \gg u_{*\text{it}}$, both $u_*^2 - u_{*\text{it}}^2$ and $u_*^2 - u_{*\text{sfc}}^2$ approximate $u_*^2$.

Another assumption commonly made to simplify Eq. (2.33) is that the particle speed in saltation scales with $u_*$, resulting in $L/\Delta v \propto u_*$ (Bagnold 1941). Using this assumption results in a scaling of $Q$ with $u_*^3$, as for example proposed by Bagnold (1941), Kawamura (1951), and Owen (1964). However, more recent studies have questioned the linear scaling of particle speeds with $u_*$ and thus the scaling of $Q$ with $u_*^3$ (Ungar and Haff 1987, Andreotti 2004, Kok 2010a, 2010b, Durán et al. 2011a, Ho et al. 2011, Pähtz et al. 2012). As discussed in Sections 2.3.1 and 2.3.2.1, these studies are supported by recent experimental, numerical, and theoretical results showing that particle speeds in steady state, transport limited saltation are not proportional to $u_*$ (e.g., Figure 2.11 and Figure 2.12). Instead, the mean particle speed at the surface is independent of $u_*$, whereas the mean particle speed above the surface increases gradually with $u_*$ (as well as with height). The speed with which particles are launched from the surface (and thus $\Delta v$) is thus independent of $u_*$, and the saltation hop length is only a weak function of $u_*$ (see Figure 2.12 and Namikas 2003, Rasmussen and Sorensen 2008, Kok 2010a, Ho et al. 2011).

We can obtain a more physically realistic relation for the mass flux from Eq. (2.33) by neglecting the weak scaling of $L$ with $u_*$ (e.g., Namikas 2003) and using that particle speeds scale with wind speed in the saltation layer, which in turn scales with $u_{*\text{it}}$. This yields $L/\Delta v = C_{\text{DK}} u_{*\text{it}}/g$, such that the mass flux can be approximated by (Kok and Renno 2007, Durán et al. 2011a)

$$Q_{\text{DK}} = C_{\text{DK}} \frac{\rho_a}{g} u_{*\text{it}} \left(u_*^2 - u_{*\text{it}}^2\right). \tag{2.34}$$

Experiments and numerical simulations suggest that, for ~250 μm sand, we have that $L \approx 0.1$ m (Namikas 2003), $\Delta v \approx 1$ m/s (Figure 2.12), and $u*_{\text{it}} \approx 0.2$ (e.g. Bagnold 1937), such that the parameter $C_{\text{DK}} \approx 5$. We now use Eq. (2.34) to non-dimensionalize the mass flux, which provides a convenient way to compare different mass flux relations. That is, we take the dimensionless mass flux as

$$\hat{Q} = \frac{gQ}{\rho_a u_{*it}\left(u_*^2 - u_{*it}^2\right)}. \tag{2.35}$$

Figure 2.15 shows the dimensionless mass flux predicted by the relations in Table 2.1. Although the scatter in the experimental data is substantial, most mass flux data sets seem to support the scaling of Eqs. (2.34) and (2.35).

Table 2.1. List of the most commonly used saltation mass flux relations.

| Mass flux equation | Comments | Study |
| --- | --- | --- |
| $Q_{\text{Bagnold}} = C_B \sqrt{\dfrac{D_p}{D_{250}}} \dfrac{\rho_a}{g} u_*^3$ | $C_B$ = 1.5, 1.8, or 2.8 for uniform, naturally graded, and poorly sorted sand, respectively. | Bagnold (1941) |
| $Q_{\text{Kawamura}} = C_K \dfrac{\rho_a}{g} u_*^3 \left(1 - \dfrac{u_{*it}^2}{u_*^2}\right)\left(1 + \dfrac{u_{*it}}{u_*}\right)$ | $C_K$ = 2.78 (Kawamura 1951) or 2.61 (White 1979). The origin of this relation is often confused to be White (1979); see Namikas and Sherman (1997). | Kawamura (1951) |
| $Q_{\text{Owen}} = \dfrac{\rho_a}{g} u_*^3 \left(0.25 + \dfrac{v_t}{3u_*}\right)\left(1 - \dfrac{u_{*it}^2}{u_*^2}\right)$ | $v_t$ is a particle's terminal fall speed. | Owen (1964) |
| $Q_{\text{Lettau}} = C_L \sqrt{\dfrac{D_p}{D_{250}}} \dfrac{\rho_a}{g} u_*^3 \left(1 - u_{*it}/u_*\right)$ | $C_L$ = 6.7. | Lettau and Lettau (1978) |
| $Q_{\text{UH}} = C_{\text{UH}} \rho_a \sqrt{\dfrac{D_p}{g}} u_*^2 \left(1 - \dfrac{u_{*\text{sfc}}^2}{u_*^2}\right)$ | Ungar and Haff (1987) did not estimate a value of $C_{\text{UH}}$. | Ungar and Haff (1987) |
| $Q_{\text{Sorensen}} = \dfrac{\rho_a}{g} u_*^3 \left(1 - u_{*it}^2/u_*^2\right)\left(\alpha + \gamma u_{*it}/u_* + \beta u_{*it}^2/u_*^2\right)$ | $\alpha$, $\beta$, and $\gamma$ are parameters that characterize the dimensions of a typical saltation hop. | Sorensen (2004) |
| $Q_{\text{DK}} = C_{\text{DK}} \dfrac{\rho_a}{g} u_{*it} u_*^2 \left(1 - \dfrac{u_{*it}^2}{u_*^2}\right)$ | $C_{\text{DK}} \approx 5$. | Proposed here and in Durán *et al.* (2011a) |

$D_{250}$ is a reference diameter of 250 μm.

### 2.3.2.4 *Wind profile and aerodynamic roughness length during saltation*

The extraction of wind momentum by saltating particles produces a steady state wind profile that accelerates saltating particles to an impact speed that, on average, results in a single particle leaving the soil bed for each particle impacting it (Section 2.3.1). However, as $u_*$ increases, the wind speed higher up in the saltation layer will increase as well (Figure 2.9), thereby increasing the saltator impact speed. But since the mean saltator impact speed must remain constant with $u_*$ (Figure 2.12), this increase of the wind speed higher in the saltation layer must be compensated by a decrease of the wind speed lower in the saltation layer. Consequently, when wind profiles at different values of $u_*$ are plotted on the same graph, the wind profiles intersect in the *Bagnold focus* at a height of ~1 cm above the surface for saltation over loose sand (Bagnold 1936). The wind speed below the focus, as well as the wind speed gradient and thus $u_{*a}$, actually *decreases* with $u_*$ (Figure 2.9, Figure 2.10, and Figure 2.16).

The absorption of wind momentum by saltating particles acts as roughness elements that increase the aerodynamic roughness length sensed by the flow above the saltation layer (Owen, 1964). The wind speed above the saltation layer is thus determined by this increase of the aerodynamic roughness length (Pähtz *et al.* 2012). Several models have been proposed to relate $z_{0s}$, the aerodynamic roughness length in saltation, to $u_*$. These include the empirical Charnock (1955) and modified Charnock (Sherman 1992, Sherman and Farrell 2008) models, and the Raupach (1991) model, which is based on Owen's hypothesis (Eq. (2.28)). More recent physically-based models of the aerodynamic roughness length include Durán and Herrmann (2006a) and the Pähtz *et al.* (2012) model, which uses the balances of momentum and energy in the saltation layer to derive a physically-based expression for $z_{0s}$. The various relations are listed in Table 2.2, and comparisons of some these relations with measurements are given in Sherman and Farrell (2008).

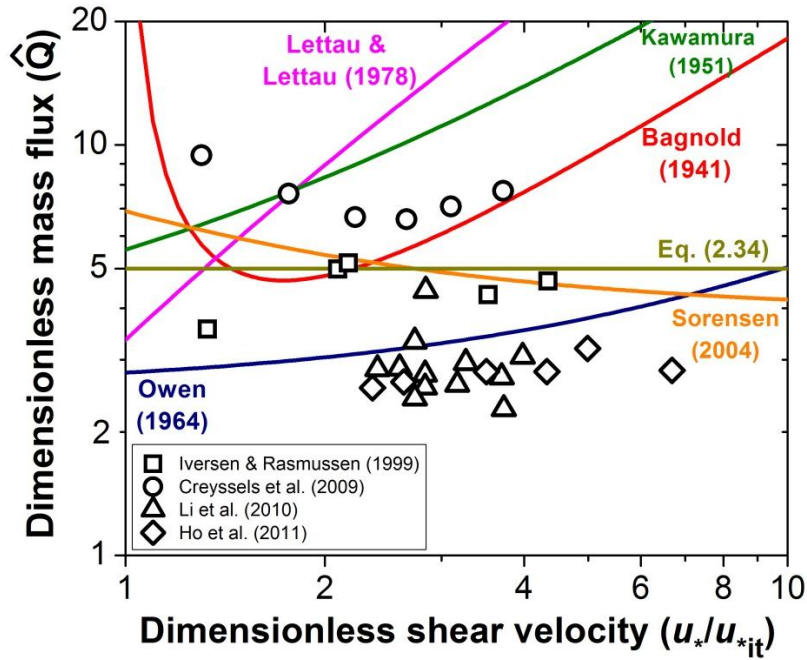

Figure 2.15. The dimensionless saltation mass flux predicted by the relations listed in Table 2.1 (colored lines) and measured in recent wind tunnel (Iversen and Rasmussen 1999, Creyssels *et al.* 2009, Ho *et al.* 2011) and field (Li *et al.* 2010) experiments for sand with a diameter of ~250 μm. For the Sorensen relation, we used $\alpha = 0$, $\beta = 3.9$, and $\gamma = 3.0$ from figure 3 in Sorensen (2004). The apparent overestimation of the mass flux by theoretical relations is at least partially due to sand collectors having an efficiency of only ~50-90 % (Shao *et al.* 1993b, Greeley *et al.* 1996, Rasmussen and Mikkelsen 1998).

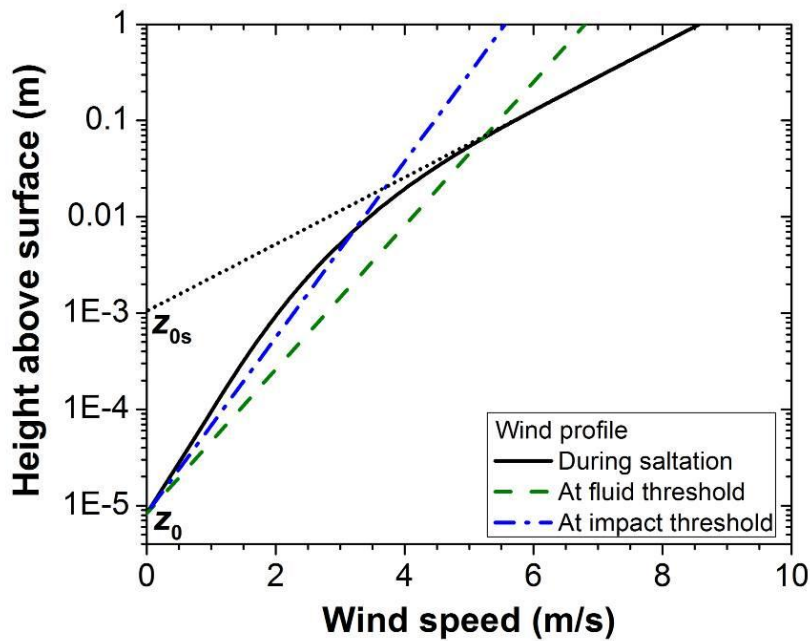

Figure 2.16. The wind speed profile at the fluid threshold (green line), the impact threshold (blue line), and during steady state saltation (black line) as simulated by COMSALT (Kok and Renno, 2009a). The aerodynamic roughness length $z_{0s}$ during saltation is also indicated.

Table 2.2. Relations for the aerodynamic roughness length in saltation. The relations of Durán and Herrmann (2006a, see their Eqs. 20 - 24) and Pähtz *et al.* (2012, see their Eqs. 67a-e) are not included here due to their complexity.

| **Relation** | **Comments** | **Study** |
|---|---|---|
| $z_{0S} = C_c u_*^2 / g$ | $C_c \approx 0.010$ for wind tunnels and $\approx 0.085$ for natural saltation (Sherman and Farrell, 2008) | Charnock (1955) |
| $z_{0S} = \left(\dfrac{Au_*^2}{2g}\right)^{1-r} z_0^r$ | $r = u_{*it}/u_*$; $A \approx 0.2 - 0.4$ (Raupach 1991, Gillette *et al.* 1998); a similar relationship was derived by Owen (1964). | Raupach (1991) |
| $z_{0S} = z_0 + C_m \dfrac{(u_* - u_{*it})^2}{g}$ | $C_c \approx 0.012$ for wind tunnels and $\approx 0.13$ for natural saltation (Sherman and Farrell, 2008) | Sherman (1992) |

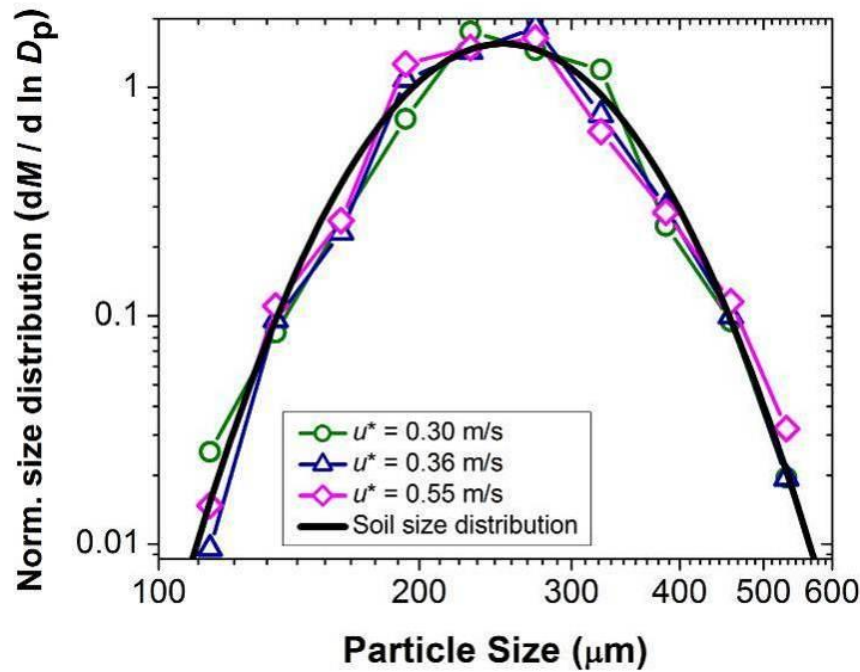

Figure 2.17. The normalized size distribution of saltating particles from the field measurements of Namikas (2003, 2006) at $u_* = 0.30$ m/s, 0.36 m/s, and 0.55 m/s are similar to the soil size distribution (solid line). Saltating particle size distributions were obtained by integrating the horizontal distribution of the saltation mass flux given in Figure 3 of Namikas (2006).

*2.3.2.5 Size distribution of saltating particles*

As discussed in Section 2.3.1, after saltation has been initiated by the aerodynamic entrainment of surface particles, subsequent lifting of surface particles occurs predominantly through splash. The fluid threshold therefore has little influence on the physical properties of steady state saltation. In particular, the size distribution of saltating particles is not determined by the dependence of the fluid threshold on particle size, as many previous studies have assumed (e.g., Marticorena and Bergametti 1995, Shao *et al.* 1996), but is rather determined by (i) the probability that a particle of a given size is splashed into the fluid stream, and (ii) the time the particle spends in saltation before settling back to the soil bed. The experiments of Rice *et al.* (1995) indicate that particles with a larger cross-sectional area have a correspondingly higher change of being splashed by an impacting saltator. Conversely, larger particle have short lifetimes because their high inertia causes them to attain lower speeds when splashed (Eq. (2.15)), leading them to quickly settle back to the soil surface (Kok and Renno 2009a). These two competing effects appear to result in a size distribution of saltating particles that is similar to the soil size distribution (Figure 2.17).

The size distribution of saltating particles is thus primarily determined by the speed (or momentum) of impacting saltators. Since the impact speed of saltators remains constant with $u_*$ (Figure 2.12), the size distribution of saltating particles should also be relatively invariant to changes in $u_*$, as indeed indicated by field measurements (Figure 2.17).

Table 2.3. Comparison of the characteristics of saltation on Earth, Mars, Venus, and Titan.

| Planetary body | Gravitational constant (g) | Air density (kg/m$^3$) | Dynamic viscosity (kg/m/s) | Particle composition | Particle density (kg/m$^3$) | Typical particle size (μm) | Threshold shear velocity (m/s)[a] | Ratio of impact to fluid threshold[b] | Typical saltation height (cm)[c] | Typical saltation length (cm)[c] |
|---|---|---|---|---|---|---|---|---|---|---|
| Earth | 1 | 1.2 | 1.8×10$^{-5}$ | Quartz | 2650 | 150 – 250 | ~0.2 | ~0.8 | ~3 | ~30 |
| Mars | 0.378 | 0.02 | 1.2×10$^{-5}$ | Basalt | 3000 | 100 – 600 | ~1.5 | ~0.1 | ~10 | ~100 |
| Venus | 0.904 | 66 | 3.2×10$^{-5}$ | Basalt | 3000 | Unknown | ~0.02 | >1 | ~0.2[d] | ~1[d] |
| Titan | 0.138 | 5.1 | 6.3×10$^{-6}$ | Tholin/ice | ~1000 | Unknown | ~0.04[e] | >1 | ~0.8 | ~8 |

[a] After Iversen and White (1982)
[b] Calculated with the saltation model COMSALT using the procedures outlined in Kok and Renno (2009a) and Kok (2010b). The impact threshold is defined here as the minimum shear velocity at which saltation can be sustained by splash in the absence of direct fluid entrainment and can thus exceed the fluid threshold for saltation in dense fluids such as on Venus and Titan.
[c] Typical saltation heights and lengths are defined as the mean height and length of trajectories of 250 μm particles that have rebounded from the surface at least once, and are calculated with the saltation model COMSALT for a shear velocity equaling twice the minimum value for which saltation can be sustained (i.e., $2u_{*it}$ on Earth and Mars, and $2u_{*ft}$ on Venus and Titan). The lift-off speed of surface particles entrained by aerodynamic forces, which dominates over splash on Titan and Venus, was calculated using Eq. (26) and (28) in Hu and Hui (1996).
[d] Trajectory calculations for Venus are consistent with the calculations of White (1981) and the experiments of Greeley *et al.* (1984).
[e] The calculated fluid threshold is consistent with the threshold of $u_{*ft} \approx 0.036$ m/s inferred by Tokano (2010) from the orientation of Titan's equatorial dunes.

*2.4 Saltation on Mars, Venus, and Titan*

As discussed in the introduction, aeolian processes are not limited to Earth, but have also shaped the surfaces of Mars, Venus, and Titan by creating dunes, ripples, and a plethora of erosional features (e.g., Greeley and Iversen 1985). The characteristics of saltation on these planetary bodies differ from those on Earth and depend on factors such as gravity, air density, viscosity, and the properties of the granular material being transported. Whereas terrestrial sand is composed primarily of silicon dioxide quartz with a density of 2650 kg/m$^3$, both martian (Yen *et al.* 2005, Golombek *et al.* 2006b) and Venusian (Greeley *et al.*, 1991; Basilevsky and Head, 2003) sand is most likely predominantly basalt, which has a density of approximately 3000 kg/m$^3$ (Johnson and Olhoeft, 1984). The granular material transported in saltation on Titan is rather different from that on Earth, Mars, and Venus, and likely consists of a mixture of ice and tholins (organic heteropolymers formed by ultraviolet irradiation of organic compounds; Imanaka *et al.* 2004, McCord *et al.* 2006), which has a density substantially lighter than basalt and quartz.

Large variations in gravity, air density, viscosity, and material density produce correspondingly large differences in the typical characteristics of saltation on the different planets (Table 2.3 and Figure 2.18). In addition to large differences in typical particle trajectories, caused primarily by differences in the gravitational constant and the terminal fall speed (Figure 2.19), the fluid and impact thresholds differ substantially between the planetary bodies. Indeed, the ratio of the impact and fluid thresholds ranges from ~0.1 for Mars (Kok 2010a, b) to ~0.8 for Earth (Bagnold, 1941) to >1 for Venus and Titan. This implies that the physics of saltation on Titan and Venus is fundamentally different from that on Earth and Mars: whereas the lifting of surface particles is dominated by splash on Earth and Mars, it is dominated by direct fluid lifting on Venus and Titan. Indeed, the terminal fall speed of granular particles on Venus and Titan is less than the typical speed of ~1 m/s required to splash surface particles (Figure 2.19).

Below, we review the characteristics of saltation on Mars, Venus, and Titan in more detail.

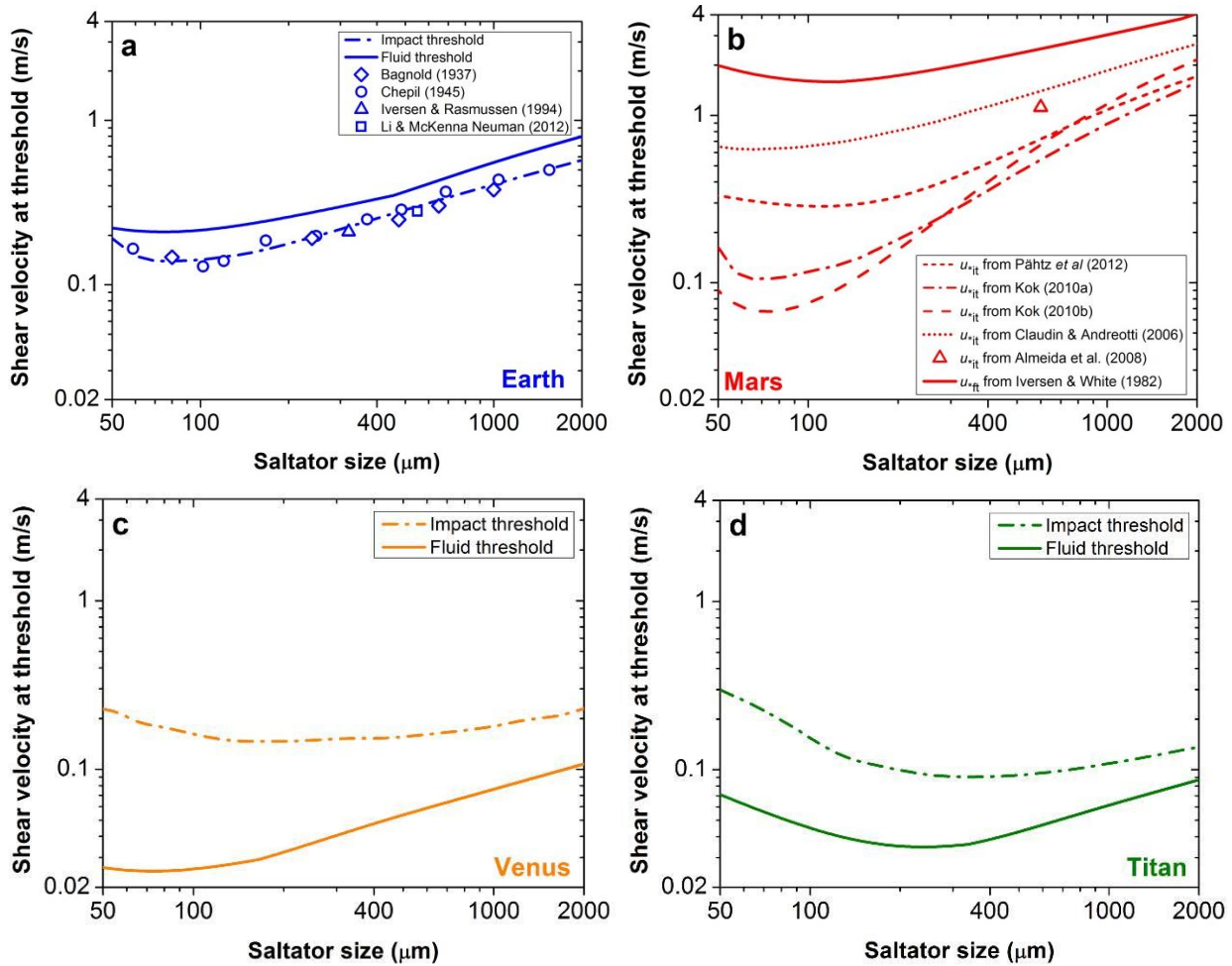

Figure 2.18. Predictions of the fluid and impact thresholds on (**a**) Earth, (**b**) Mars, (**c**) Venus, and (**d**) Titan. The fluid thresholds (solid lines) are calculated using Eq. (2.6) (Iversen and White 1982, Greeley and Iversen 1985), whereas the impact thresholds (dash-dotted lines) are calculated with the numerical model COMSALT as the minimum shear velocity at which saltation can be sustained by splash (Kok and Renno 2009a, Kok 2010b).Wind tunnel measurements of the impact threshold were included for Earth conditions (symbols), whereas model predictions by Claudin and Andreotti (2006), Almeida *et al.* (2008), Kok (2010a), and Pähtz *et al.* (2012) are included for Mars conditions. The effects of interparticle forces on the impact threshold were accounted for as described in Kok (2010a), with the scaling constant of the interparticle force for Titan and Venus assumed equal to that for Earth.

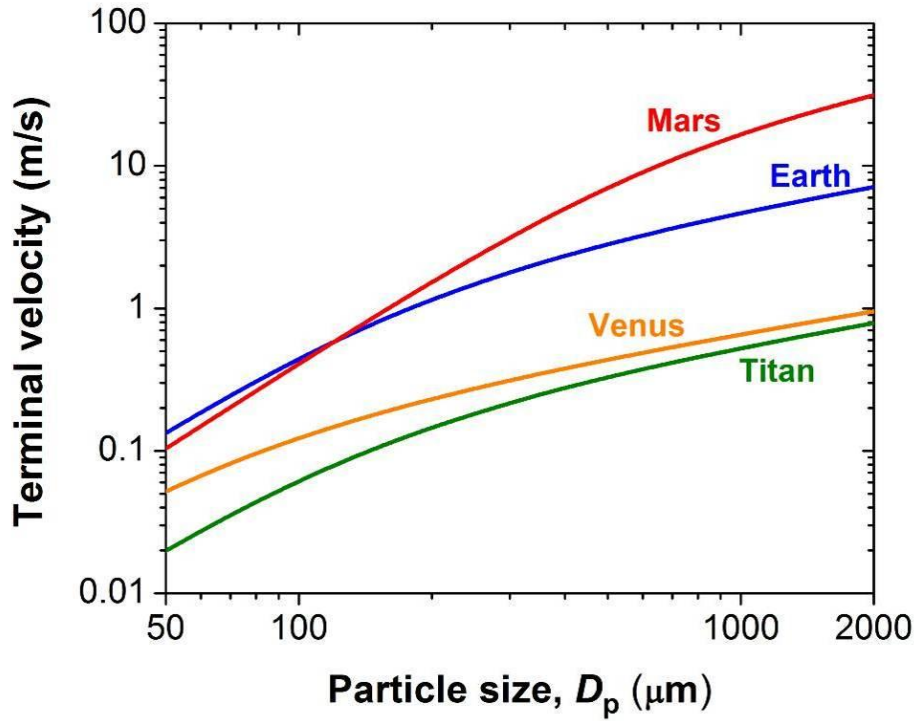

Figure 2.19. The terminal velocity of natural sand (Cheng 1997) on Earth, Mars, Venus, and Titan. Note that the low terminal velocity on Venus and Titan limits saltation particle speeds, thereby causing splashing to be inefficient.

*2.4.1 Saltation on Mars*

The characteristics of martian saltation have been widely studied (e.g., Sagan and Bagnold 1975, White *et al.* 1976, Greeley 2002, Almeida *et al.* 2008, Kok 2010a, b) as a consequence of the observed abundance of dunes, ripples, and wind-abraded landforms on the martian surface (e.g. Greeley and Iversen 1985, Bridges *et al.* 1999, 2010, Sullivan *et al.* 2005, 2008, Greeley *et al.* 2006). Moreover, dust devils and dust storms are common on Mars (Thomas and Gierasch 1985, Cantor *et al.* 2001) and saltation is a likely mechanism of raising dust in these phenomena (Greeley 2002, Kok 2010b), although other plausible dust emission mechanism have recently been suggested (Merrison *et al.* 2007, Sullivan *et al.* 2008, Wurm *et al.* 2008) (see Section 4.2).

The typical size of saltating particles on Mars is still under discussion. Edgett and Christensen (1991) determined an average grain diameter of $500 \pm 100$ $\mu$m for Mars' dunes by relating the remotely sensed thermal inertia of martian dunes to the porosity of the granular material, and thus the grain diameter. Larger and smaller values were suggested in subsequent years (Presley and Christensen 1997, Fenton *et al.* 2003). However, Claudin and Andreotti (2006) more recently found $d = 87 \pm 15$ μm from high resolution images of ripples' sand obtained *in situ* by rovers, although is it unclear what the applicability of this result at a single location on Mars is to the typical particle size of martian dunes.

*2.4.1.1 The puzzle of active martian sand transport*

A foremost open question is how commonly saltation occurs on Mars. Sporadic lander measurements suggest that winds in the light martian atmosphere rarely exceed the saltation fluid threshold (Zurek *et al.* 1992; Sullivan *et al.* 2000; Holstein-Rathlou *et al.* 2010), which is consistent with simulations of both mesoscale (Fenton *et al.* 2005, Chojnacki et al. 2011) and global circulation models (Haberle *et al.* 2003). These findings are difficult to reconcile with recent reports of active sand transport at many locations on Mars (Fenton *et al.* 2005, 2006, Greeley *et al.* 2006, Sullivan *et al.* 2008, Bourke *et al.* 2008, Geissler *et al.* 2010, Silvestro *et al.* 2010, 2011, Chojnacki *et al.* 2011, Hansen *et al.* 2011, Bridges *et al.* 2012a, Kok 2012), which in at least some cases results in sand fluxes that are similar to those on slow-moving dunes on Earth (Bridges et al. 2012b).

Part of the explanation for these seemingly contradictory findings might be that small-scale topography and convection generate strong localized winds, which cannot be accurately simulated by the coarse resolution of martian atmospheric circulation models (Fenton and Michaels 2010). Moreover, once such strong localized winds initiate saltation, recent results indicate that it can be sustained by wind speeds reduced by up to a factor of ~10 (Kok 2010a, b). That is, the martian impact threshold is only ~10 % of the fluid threshold (see Figure 2.18b), thus allowing saltation to occur at much lower wind speeds than previously thought possible. The ratio of impact to fluid thresholds on Mars is lower than on Earth (Table 2.3) because the lower gravity and air density allows particles to travel higher and longer trajectories on Mars, causing them to be accelerated by wind for a longer duration during a single hop than on Earth (Kok 2010a). This effect combines with the lower atmospheric density on Mars to produce an impact threshold that is comparable to that on Earth (Figure 2.18). However, the lower atmospheric density causes the martian fluid threshold to be an order of magnitude larger than on Earth (Eq. (2.5)). Consequently, the ratio of the impact to fluid thresholds is much smaller on Mars than it is on Earth (Table 2.3).

*2.4.1.2 Typical saltation trajectories on Mars*
The trajectories of saltating particles on Mars are largely determined by the typical speed with which they are launched from the surface. Interestingly, the mean saltator launch speed during steady state saltation over a bed of loose sand is predicted to be very similar on Mars and Earth (Kok 2010a, b). The reason for this is that the efficiency of the splashing process is essentially independent of the air density (Durán *et al.* 2011a) and only weakly dependent on particle size and gravity. However, the lower gravity and vertical air drag cause martian saltators to undergo trajectories that are almost an order of magnitude longer and higher than those on Earth (see Table 2.3 and Figure 3b in Kok, 2010b). These results contrast with the substantially larger saltation trajectory estimates of Almeida *et al.* (2008), who did not explicitly account for splash. Since steady state on Mars is probably determined by splash entrainment, neglecting splash can yield unphysical results, such as an increase in the saltator impact speed with $u_*$ (e.g., Figure 5 in Almeida *et al.* 2008). The saltation trajectory estimates of Almeida *et al.* (2008) can nonetheless be considered plausible for *supply limited saltation* (Nickling and McKenna Neuman, 2009), for which saltator impact speeds generally do increase with $u_*$ (Houser and Nickling 2001, Ho *et al.* 2011).

*2.4.2 Saltation on Venus and Titan*
As discussed in the introduction, evidence of aeolian saltation transport has been observed not only on Mars, but also on Venus and Titan. The characteristics of saltation on Venus and Titan are rather different from those on Earth and Mars because the much greater atmospheric density dissipates the vertical motion of saltating particles (White 1981, Williams and Greeley 1994, Lorenz *et al.* 1995). This reduces the typical height to which particles bounce to less than a centimeter (Table 2.3). Moreover, the dissipation of vertical momentum (Figure 2.19) makes saltating particles unable to access the larger wind speeds higher in the saltation layer. Particles therefore cannot readily pick up large speeds as they do on Earth and Mars, such that splash on Titan and Venus is inefficient and particle lifting is done primarily by direct fluid lifting. Saltation on Titan and Venus thus cannot be sustained below the fluid threshold (Table 2.3) and is likely more similar to saltation in water (Bagnold 1973, Claudin and Andreotti 2006) than saltation in air on Earth and Mars.

## 3. Sand dunes and ripples

The previous chapter reviewed the physics of wind-blown sand (saltation), while neglecting any changes to the sand bed. However, a flat sand bed exposed to a wind strong enough to set grains into motion is unstable. That is, saltation over an initially flat sand bed results in the generation of two kinds of bedforms with distinct length-scales: *ripples*, a few centimeters tall, and *dunes*, which are typically 5–10 meters tall but can reach heights of a hundred meters. Dunes occur frequently as isolated objects moving on a firm ground (such as barchan dunes in a corridor) but also as part of compounds evolving on a dense sand bed (see Section 3.1.5). Ripples appear most commonly on the surface of dunes as chains of small undulations that orient transversely to the wind trend (see Section 3.1.2). The physics governing the formation of ripples and dunes has been studied since the pioneering field works by Bagnold (1941). Indeed, many insights have been gained during the last few decades from computer modeling. With the help of computer simulations, it has become possible to reproduce in a few hours dune processes that take place within decades or centuries.

In this chapter our aim is to highlight, from the point of view of physics, the progress achieved in understanding the formation and evolution of ripples and dunes on Earth, Mars, Venus, and Titan. A brief introduction to the physics of ripples and dunes, including relevant insights into dune dynamics gained from field and experimental investigations, is given in Section 3.1. We then describe recent progress achieved from numerical modeling in Section 3.2. Finally, in Section 3.3 we discuss aeolian bedforms on Mars, Venus and Titan.

### 3.1 The physics of sand dunes and ripples

Anyone who walked on a dune in a desert or on a windy beach has noted that the surface is armored with small undulations — sand ripples. Saltation is the primary transport mode responsible for the formation of both dunes and ripples. However, these bedforms emerge from totally different types of sand-wave instabilities, which will be discussed next in Sections 3.1.1 and 3.1.2. Many decades of field and experimental works have pushed forward our understanding about the factors controlling the scale and morphology of aeolian bedforms, which are discussed in Sections 3.1.3 – 3.1.5. Dune dynamics are the subject of Section 3.1.6.

#### 3.1.1 Dunes

The instability leading to dunes is of hydrodynamic origin. The presence of a bump or small hill on the soil induces a modification in the logarithmic profile of the wind (Section 2.1.4.1). There is an excess pressure at the upwind front of the bump, which acts as an obstacle to the wind. This pressure provides a force that deflects upward the air flow approaching the bump from upwind. At the crest, a negative pressure perturbation keeps the flow attached to the surface, in such a way that the flow streamlines get closer to each other at the upper portion of the bump (Figure 3.1). The gradient of wind velocity is thus larger at the crest than in the valleys, whereas the perturbation in the flow profile extends up to a height that is of the order of the length of the bump (Fourrière *et al*. 2010). The compression of the streamlines due to the decrease of atmospheric pressure above the bump with an increase of the shear stress at the bump's top is a linear perturbation. This perturbation is symmetric with respect to the downstream crest position (see e.g. Kroy *et al*. 2002). Indeed, the turbulent nature of the flow over the dune leads to nonlinear hydrodynamic effects in the shear stress perturbation (Stam 1997, van Boxel *et al*. 1999, Kroy *et al*. 2002, Andreotti *et al*. 2002a, Fourrière *et al*. 2010). In particular, the inertia of the turbulent velocity fluctuations introduces a resistance to the flow deflection over the bump that is not symmetric even for symmetric bumps: it counteracts the upward deflection of the streamlines at the streamward side and their downturn at the lee side (Kroy *et al*. 2002). Due to this nonlinear part of the hydrodynamic perturbation, the maximum basal shear stress is not located at the crest. Rather, it is shifted *upwind* with respect to the profile of the terrain (Figure 3.1). In the hypothetical situation where the sand flux responds without delay to wind velocity, maximum erosion would always occur upstream of the crest, such that sand would be deposited on the bump, thus leading to dune growth.

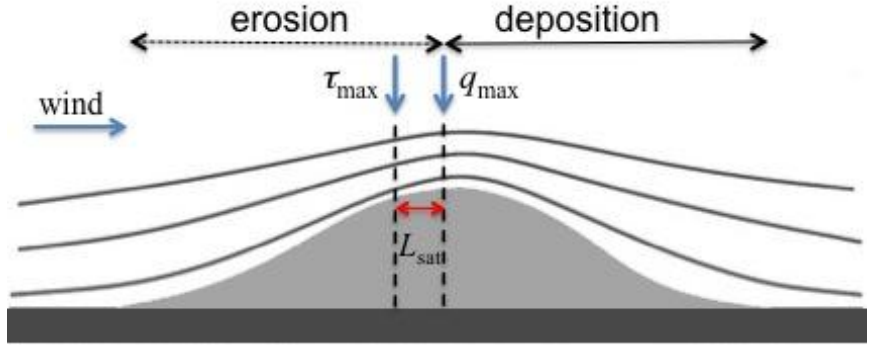

Figure 3.1 Hydrodynamic instability leading to sand dunes. Schematic diagram showing streamlines of the wind flow over a bump of Gaussian profile in the direction longitudinal to the wind (solid surface). $\tau_{max}$ and $q_{max}$ indicate the positions of maximum shear stress and sand flux, respectively. The saturation length, $L_{sat}$, gives the horizontal distance of relaxation of the flux ($q$) towards its saturated value ($q_s$) due to the variation in the wind flow along the bump's profile. The bump grows if the maximum flux is reached upwind of the bump's crest. After Kroy *et al.* (2002) and Fourrière *et al.* (2010).

However, the saltation flux takes a finite time — or, equivalently, a saturation length ($L_{sat}$) — to react to a variation of wind speed (Section 2.2). The bump grows only if its wavelength is large enough such that the flux is maximal upwind of the bump's crest. The saturation length thus defines a minimum length-scale for the growth of the bump and dune formation. Based on field observations and water tank experiments on dune formation, Hersen *et al.* (2002) concluded that $L_{sat}$ approximately scales with the quantity,

$l_{drag} = d \, \rho_{fluid}/\rho_{grains}$ , (3.1)

where $\rho_{fluid}$ and $\rho_{grains}$ are densities of the fluid and of the sediments, respectively, and $d$ is the particle diameter. The flux saturation length $L_{sat}$ is the only relevant length-scale in the physics of dunes: it determines the minimum size below which the bump is eroded and a dune cannot form. The physics governing the saturation length is still matter of debate (see Section 3.2.1.2 and 3.2.1.4).

The coupling between erosion-deposition and the hydrodynamics due to the nonlinear perturbation described above results in an asymmetric dune profile with gentle windward slope and steep lee side. As the dune grows and the surface curvature becomes very large, the negative pressure perturbation at the crest does not suffice to keep the flow attached to the topography and flow separation occurs: at the lee of the dune a zone of recirculating flow develops (Figure 3.2). Further, when the slope at the lee exceeds the angle of repose of the sand (~ 34°) the surface relaxes through avalanches and a slip face forms. Dune motion thus consists of saltating grains climbing up the windward side, accumulating on the crest and thereafter rolling down the slip face being deposited at the bottom on the lee side of the dune.

Since the flow over dunes is fully turbulent with eddies at all length-scales, it has no characteristic length-scale. The pressure perturbation around the bump is essentially a function of the surface shape, such that one expects the flow pattern to depend essentially on the dune shape but not on its size. Field measurements on barchan dunes (Sauermann *et al.* 2000, Andreotti *et al.* 2002a) show that, for large enough dunes, the shape of dunes is scale invariant. Scale invariance of dune shape is broken as dune size becomes close to the minimal size – barchans close to the minimal size display less steep profiles than large enough barchans do and incipient slip face (Sauermann *et al.* 2000, Kroy *et al.* 2002). The breaking of scale invariance is thus a consequence of the flux saturation length, which dictates the onset for the growth of the bump into an asymmetric dune shape with a steep lee side and a separation bubble, as explained above. $L_{sat}$ is thus an essential ingredient for morphodynamic models of sedimentary landscapes, i.e. models that incorporate a physical description of the interaction between the fluid and sediment bed leading to bedform formation (see Section 3.2).

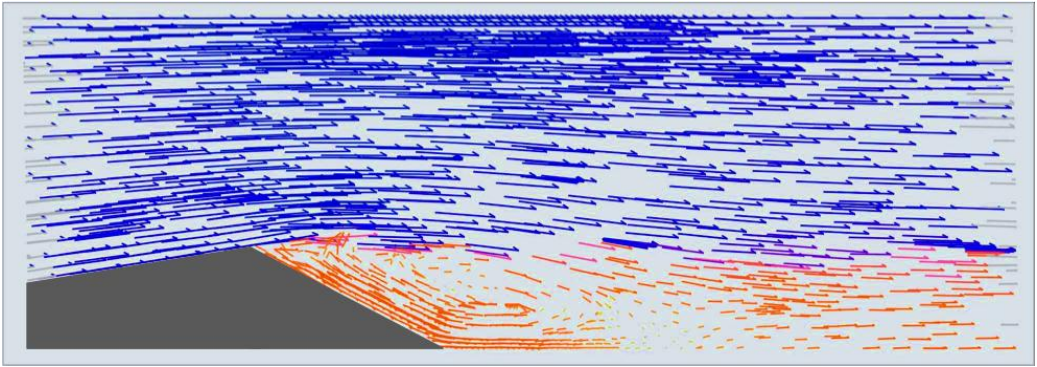

Figure 3.2 Calculation of the flow over a barchan dune using CFD (after Herrmann *et al.* 2005). The image shows the velocity vectors, colored by velocity magnitude (increasing from red to blue), over the cut along the symmetry plane of a barchan dune (in grey). The wind blows from left to right. The separation of the flow at the brink of the dune and the large eddy in the wake of dune are clearly visible.

*3.1.2 Ripples*

Ripples are smaller than the saturation length, and in this manner they cannot develop from the hydrodynamic instability described above. Instead, ripples stem from a "screening" instability of the sand surface exposed to the impact of saltating particles (Bagnold 1941, Anderson 1987a, Fourrière *et al.* 2010). Saltating grains colliding obliquely onto a sand bed generate small depressions on the surface, which then develop a chain of rough, small-scale undulations of asymmetric profile — the windward side of the perturbations is less steep than the lee side (Figure 3.3). Each collision may result in the ejection of many reptating particles that move forward several grain diameters, while the impacting grain may rebound and land again at much larger distances. Indeed, the upwind side of each roughness element receives more impacts of saltating grains than downwind surface areas, the so-called "shadow zones" (Figure 3.3). Ejected particles that reach the shadow zones are less likely to be impacted again by saltating grains compared to particles on upwind surface areas, and in this manner sand accumulates in the shadow zones leading to the growth of the perturbations. The role of saltating particles for ripple formation is to furnish energy to the motion of low-energy particles through reptation, in effect the transport mechanism of ripples. The instability leading to aeolian ripples is thus controlled by grain impact, rather than by hydrodynamics, which is the case for dunes.

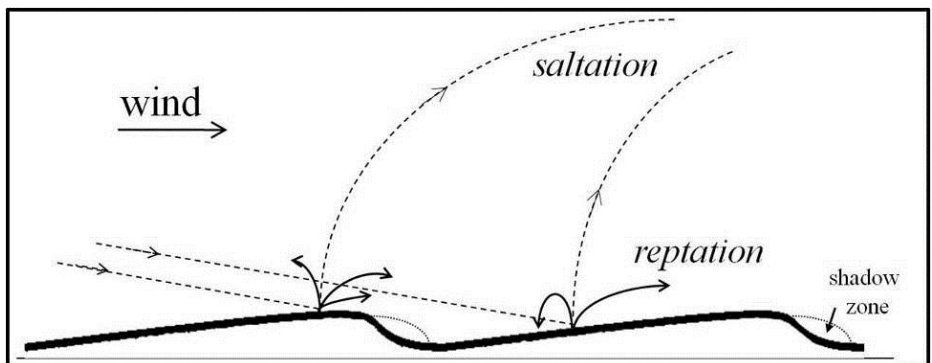

Figure 3.3 Sketch of the transport modes leading to ripple formation (after Prigozhin 1999). Saltating grains (dashed lines) impacting onto the sand bed eject several particles which enter reptation (solid lines).

Bagnold (1941) and Sharp (1963) first documented systematic field observations on ripples' morphology and migration speeds. Wind tunnel studies on the formative processes of ripples, conducted by Sepällä and Lindè (1978), showed an increase in ripples' wavelength in time until a "fully-developed state" was reached without further growth. This saturated state was studied in detail in wind tunnel experiments by Andreotti *et al.* (2006), who found a scaling for the final wavelength of ripples with the grain diameter multiplied by ($u_*/u_{*\text{it}} -1$) (but see Manukyan and Prigozhin 2009, Section 3.2.4.2). In fact, the time-scale of ripple formation is much smaller than the one of dunes. Therefore, ripples oriented transversely to the local wind direction on the surface of dunes provide an example of bedforms in the fully-developed state, and can be used for indirectly inferring local wind trends over the terrain (Andreotti *et al.* 2006).

### 3.1.3 *Aeolian versus underwater bedforms*
Experiments of dune formation under water provide a means to study the long-term dynamics of sand dunes (see Section 3.1.6.5). Much effort has been dedicated to understanding the different mechanisms of transport and bedform dynamics under water (Fourrière *et al.* 2010).

Impact ripples (Figure 3.3) cannot form under water where the efficiency of saltating particles in producing a splash of reptating grains is negligible compared to the aeolian case (Section 2.4). The instability leading to underwater bedforms is of the same type as the one generating aeolian dunes (hydrodynamic); the relevant length scale is the flux saturation length (Hersen *et al.* 2002). Following Fourrière *et al.* (2010), underwater bedforms may be called "ripples" if their wavelength is much smaller than the water depth, and "dunes" otherwise. According to this definition, dunes in a water stream are those bedforms that result from amalgamation of ripples (Fourrière *et al.* 2010). Subaqueous dunes have thus a similar origin as giant aeolian dunes, which grow progressively by merging of smaller dunes (Lancaster 1988, Andreotti *et al.* 2009). We note that, due to the large atmospheric densities of Venus and Titan, sediment transport on these planetary bodies is more akin to underwater transport than to sand transport in air (see Section 2.4.2). As shown experimentally, Venusian microdunes behave similarly to aeolian dunes or subaqueous ripples, rather than to aeolian impact ripples (see Section 3.3.3).

### 3.1.4 *Factors limiting dune size*
One essential factor governing the long-term growth of dunes is the amalgamation of dunes (see Section 3.2.1). Dune coalescence may lead to giant dunes several hundred meters in height, which often exhibit a more complex morphology than their smaller companions and commonly host superimposed bedforms of different types (Section 3.1.5.2).

Indeed, the size of giant aeolian dunes is limited by the height of the atmospheric boundary layer (where sand transport takes place) in the same manner as the size of subaqueous dunes is bounded by the water depth (Andreotti *et al.* 2009, Fourrière *et al.* 2010). The topography of the sand terrain excites surface waves on the interface between the inner layer and the free surface (which in the aeolian case is the upper thermal inversion layer). The effect of these waves is to squeeze the flow streamlines downstream of the bump's crest (instead of upstream as in Figure 3.1), thus providing a stabilizing mechanism (Andreotti *et al.* 2009). When dune wavelength (or interdune spacing in the case of a dune field) becomes comparable to the thickness of the atmospheric boundary layer (which amounts to a few kilometers), this stabilizing effect becomes sufficiently large to prevent further dune growth (Andreotti *et al.* 2009). In rivers, the radiation of internal waves into the inner layer may not be as relevant as in the aeolian case, and underwater "mega-dunes" of wavelength much larger than the water depth have been observed (possibly in connection with polydispersity of sand; Fourrière *et al.* 2010).

In fact, dunes of wavelength bounding the atmospheric boundary layer thickness are not typical of Earth deserts (Bagnold 1941, Pye and Tsoar 1990, Lancaster 1995). Collisions between dunes, as well as variations in wind direction by small angles, can also act as destabilizing factors for large dunes (Schwämmle and Herrmann 2003, Elbelrhiti *et al.* 2005, 2008, Hugenholtz and Barchyn 2012; see Section 3.1.6.3). Such processes can lead to ejection of significant amounts of mass from large bedforms, for instance in corridors of barchan dunes extending over hundreds of kilometers, thus hindering dune growth and helping to maintain the dune size well below the atmospheric boundary layer thickness.

### 3.1.5 *Factors controlling bedform morphology*
The main factors determining the shape of aeolian bedforms are the wind directionality and the amount of sand available for transport (Wasson and Hyde 1983). When the wind blows from approximately the same direction most of the time and there is enough sand to completely cover the ground, *transverse bedforms*, aligned perpendicularly to the net transport direction, develop on the sand bed. Ripples on a dune provide a typical example of transverse bedforms. In areas where the amount of sand on the ground is scarce, unidirectional winds lead to *barchan dunes*. Barchans have a crescent-shaped form with two limbs (also called horns or arms) that point in the migration direction. In many areas, the wind trend varies seasonally between at least two main directions (Fryberger and Dean 1979) leading to longitudinal *seif dunes*, which have a characteristic meandering shape and align parallel to the vector resultant sand transport direction (Tsoar 1983, Tsoar 1984, Livingstone 1989, Bristow *et al.* 2000). When multiple wind directions become relevant, dunes become depositional centers thus developing a star-like shape that displays multiple arms radiating from a central peak. These *star dunes* can reach up to several hundred meters in height (Lancaster 1989a, b). The aforementioned dune types, which are the four main dune types occurring in nature (Wasson and Hyde 1983, Pye and Tsoar 1990, Lancaster 1995), are illustrated in the sketches in Figs. 3.4 a–d (top).

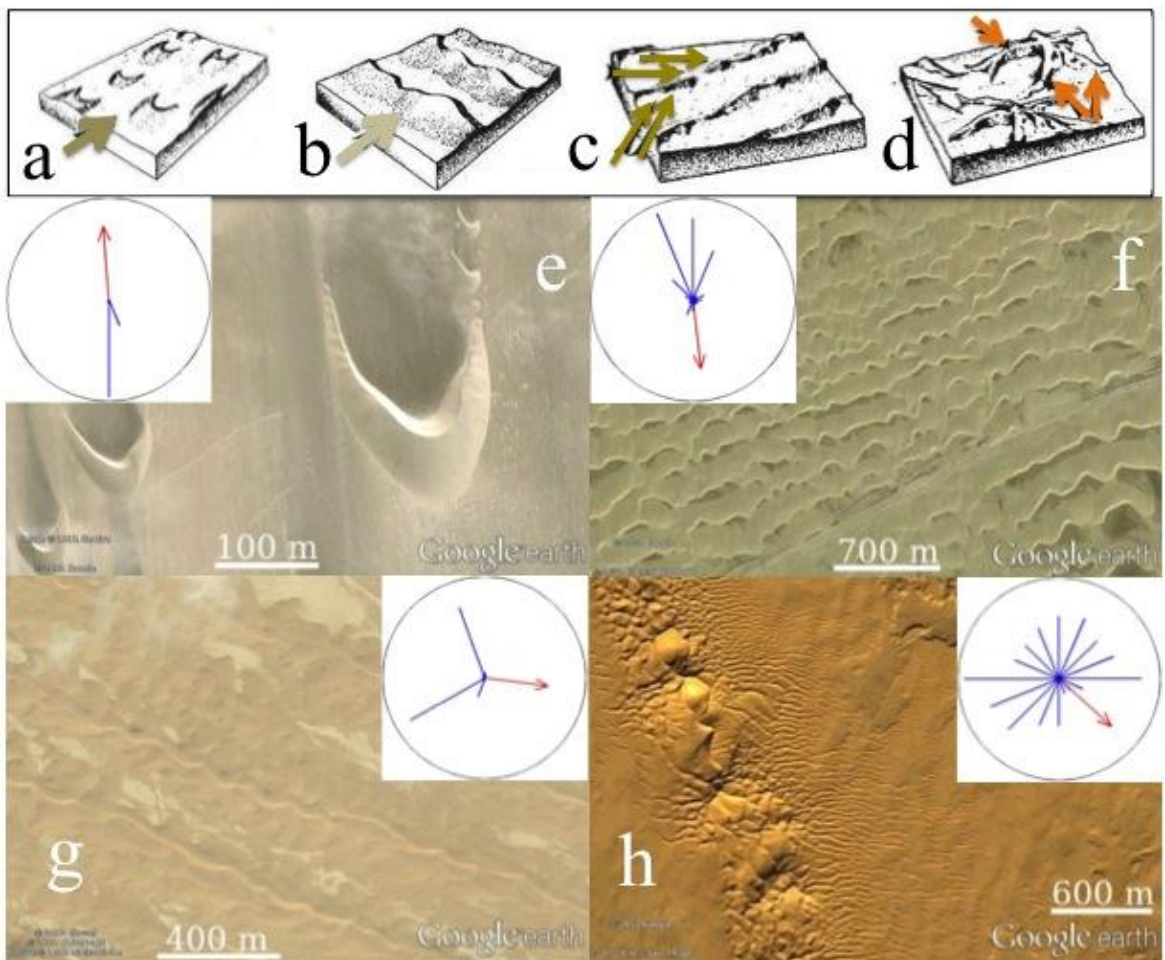

Figure 3.4 The four main types of dunes occurring in nature. In the box on top we see sketches of barchans (a), transverse dunes (b), longitudinal dunes (c) and star dunes (d). Arrows indicate the prevailing wind directions typical for the formation of each dune type. Modified after Greeley and Iversen (1985). In the bottom we see satellite images of: (e) barchans in Peru, near 15°07 S, 75°15 W; (f) transverse dunes in Bahrein, 25°49N, 49°55E; (g) longitudinal seif dunes at Bir Lahfan, Sinai, near 30°52N, 33°52E, and (h) star dunes in Algeria, near 30°35 N, 2°51E. Images credit: Google Earth. The corresponding sand roses are also shown for each dune field. The length of each blue vector is proportional to the $DP_i$ value in the corresponding direction, while the red arrow gives the vector resultant sand transport direction and is proportional to the RDP value. The values of RDP/DP are (e) 1009/1034; (f) 307/510; (g) 112/199, and (h) 95/686.

Figures 3.4e-h show aerial photographs of barchan, transverse, seif and star dunes with the corresponding "sand rose" for each dune field. The vector length of each direction (*i*) of the sand rose gives the potential rate of sand transport from that direction, or the *drift potential (DP)*,

$$DP_i = \sum U^2 [U - U_t] f_U, \tag{3.2}$$

where $U$ is the wind velocity (in knots) at a height of 10 m, $U_t = 12$ knots is the threshold wind velocity and $f_U$ is the fraction of time the wind was above $U_t$ (Fryberger and Dean 1979, Tsoar 2001). In fact, DP is related to the (potential) bulk sand flux while $U$ and $U_t$ are proportional to $u_*$ and $u_{*t}$, respectively. Thus, Eq. (3.2) for the potential rate of sand transport is consistent with the sand flux equation by Lettau and Lettau (1978) (see Table 2.1). The resultant drift potential (RDP) is obtained by calculating the vector sum of the drift potential for each of the directions of the sand rose. Wind directionality can then be quantified in terms of the ratio $\beta$ = RDP/DP, where DP is the sum of the magnitude of the drift potential values for all directions. A value of $\beta$ close to unity means unidirectional wind regimes, while $\beta$ close to 0 means multimodal wind systems.

*3.1.5.1 Experimental investigation of ripples in directionally varying flows*
Rubin and Hunter (1987) designed an experiment to simulate aeolian ripple formation under bimodal winds. A sand-covered turntable subject to a nearly steady wind of constant direction was periodically rotated by a divergence angle $\theta_w$. The table was held at one position for a time $T_{w1}$, then turned to the other position, where it remained a time $T_{w2}$, then turned back to the first position, where the process was repeated in a cyclic manner. Ripples always oriented transversely (longitudinally) to the net transport trend when $\theta_w$ was smaller (larger) than $\theta_L = 90°$. A "mixed state" was observed for $\theta_w$ within the range $90° < \theta_w < 112°$. "Oblique bedforms" occurred for obtuse divergence angle when the transport ratio $R = T_{w1}/T_{w2}$ differed from unity (Rubin and Hunter, 1987). The experiments showed that bedform alignment is such that it yields the maximum gross bedform-normal transport — which is defined as the amount of sand transported normal to the bedform axis (Rubin and Hunter 1987). Gross transport means that transport from opposite directions do not cancel, instead both contribute to build bedform surface. The prediction $\theta_L = 90°$ has been confirmed in the field (McKee and Tibbitts 1964, Tsoar 1983, 1989, Rubin and Hunter 1985, Bristow *et al.* 2007, Rubin *et al.* 2008, Dong *et al.* 2010), in water-tank experiments (Rubin and Ikeda 1990, Reffet *et al.* 2010), and in numerical simulations of aeolian dunes (Werner 1995, Reffet *et al.* 2010, Parteli *et al.* 2009, see Sections 3.2.1.4 and 3.3.2.2).

*3.1.5.2 The complex shape of giant dunes*
Very large dunes commonly display superimposed dunes of a different morphology – such dune compounds are said to be *complex* dunes (McKee 1979). Examples of complex dunes are the giant longitudinal dunes in Namibia, which host smaller superimposed transverse bedforms migrating obliquely to the main dune trend (Lancaster 1988, Andreotti *et al.* 2009). The origin of complex dunes is that bedforms of different sizes respond to wind regimes associated with different time-scales (see Section 3.1.6.1).

*3.1.5.3 The role of stabilizing agents for bedform morphology*
Some natural agents can serve as sand stabilizers, thus greatly affecting bedform morphology. For instance, in vegetated areas U-shaped parabolic dunes constitute the predominant dune morphology (Tsoar and Blumberg 2002; Section 3.2.3.1). Straight-crested, downwind-elongating linear dunes (where "linear" refers to dunes that are much longer than they are wide; see e.g. Rubin and Hesp 2009) can also form due to the action of vegetation, though the formative processes of these dunes are still poorly understood (Pye and Tsoar, 1990). Some authors hypothesized that sand induration or cementation by salts or moisture in the dune could also lead to linear dunes elongating longitudinally to a single wind direction (Schatz *et al.* 2006, Rubin and Hesp 2009; see Section 3.3.2.3). Furthermore, the presence of an exposed water table may affect local rates of erosion and deposition and thus play a relevant role for dune morphology and mobility (Kocurek *et al.* 1992, Luna *et al.* 2012).

*3.1.6 Dune dynamics*

The large length scales involved in dune processes make the study of aeolian dunes in a wind-tunnel impossible. Consequently, for decades, field studies have provided much of our knowledge on the physics dictating the morphodynamics of aeolian dunes (Livingstone *et al.* 2007).

*3.1.6.1 Scaling relations for barchan dunes and their consequences*

The best-known type of dune is the barchan, which is also of particular interest due to its high mobility: barchans can migrate between 30 and 100 meters in a single year (Bagnold 1941, Pye and Tsoar 1990). Many authors reported linear relations between barchan's width ($W$), length ($L$) and height ($H$) and also a scaling of dune migration velocity ($v$) with the inverse of dune size (Bagnold 1941, Finkel 1959, Long and Sharp 1964, Hastenrath 1967, Embabi and Ashour 1993, Hesp and Hastings 1998, Sauermann *et al.* 2000, Hersen *et al.* 2004, Elbelrhiti *et al.* 2005).

The migration velocity of a barchan dune can be approximately calculated for a dune in steady state by applying mass conservation at the dune's crest (Bagnold 1941). The dune migration velocity scales as $v_m \sim cQ_0/W$, where $c \approx 50$ (Durán *et al.* 2010). Since the barchan's volume $V$ scales with $W^3$ with a constant of proportionality of about 0.017 (Durán *et al.* 2010), one can write $v_m \approx kQ_0/V^{1/3}$, with $k \approx 12.8$. The *turnover time* ($T_m \sim W/v_m$) of a dune can be understood as the time scale needed for a sand grain, once buried at the lee, to reappear at the windward dune's foot (Allen 1974, Lancaster 1988, Andreotti *et al.* 2002a). $T_m$ follows the scaling relation,

$$T_m \sim aW^2/Q_0, \tag{3.3}$$

where $Q_0$ is the bulk sand flux and $a \approx 0.02$ from field observations (Hersen *et al.* 2004). So the smaller the dune, the shorter its turnover time, and thus the faster it readapts to a change in flow conditions.

Indeed, the coexistence of dunes of different types associated with distinct sizes in an environment of complex wind regime (e.g. small transverse dunes occuring superimposed on or nearby much larger star or longitudinal dunes; see Lancaster 1988) can be understood on the basis of Eq. (3.3): sufficiently small bedforms can orient transversely to average wind directions that prevail for a short time-scale $T_w$ comparable to the turnover time of these small dunes (10 to 100 years); in contrast, the shape of much larger dunes provides a proxy for changes in wind regime occurring within time-scales similar to their much longer turnover time (1000 to 100000 years). Consequently, the aforementioned scaling relations have crucial implication for understanding the different shapes of dunes occurring in Earth and Mars deserts and their connection to local variations in wind regimes occurring at a range of time-scales.

*3.1.6.2 Long-term evolution of dunes as proxy for climatic changes*

As explained previously, dune dynamics respond to variations in wind regime, availability of mobile sediments and different factors that are influenced by local climate, such as vegetation cover. The shape of mobile dunes prevailing in a given field may change over time in response to changes in the local environmental conditions, with new dunes emerging superimposed on more ancient bedforms (Levin *et al.* 2007). Indeed, it is possible to infer information about past climates from the shape of fossil dunes (Lancaster *et al.* 2002). Also, knowledge of how dune dynamics react to climatic changes has potential application to predicting the remobilization of currently inactive sand systems due to global warming (Thomas *et al.* 2005), or to infer information about the history of the Martian climate (Fenton and Hayward 2010, Gardin *et al.* 2012).

*3.1.6.3 Collisions and wind trend fluctuations as factors for dune size selection*

Changes in wind regimes not only affect the shape of dunes but can also serve as a control mechanism for dune size in a dune field. Large barchans display in general small superimposed dunes ("surface-wave instabilities") induced by variations in wind trend (Elbelrhiti *et al.* 2005). When subject to a storm wind blowing from a direction that makes a small angle with the primary wind, the dune's surface is unstable.

On the windward side, small transverse bedforms with size of the order of the minimal dune occur, which then migrate along the dune surface finally escaping through the limbs. This mechanism reduces the mass of large dunes, thus hindering the formation of a single giant dune through merging of all dunes in the field (Elbelrhiti *et al.* 2005).

A further factor relevant for dune size selection is the occurrence of collisions. Due to the scaling of dune velocity with the inverse of dune width, if a small barchan is migrating behind a larger one, a collision will occur. Collisions between barchans in a corridor may generate small "baby"-barchans escaping through the limbs of large dunes (Schwämmle and Herrmann 2003, Endo *et al.* 2004, Hersen 2005, Durán *et al.* 2005) and could therefore also contribute to regulating the dune size. The respective roles of dune collisions and wind trend fluctuations for the stability of dune fields are currently a matter of substantial debate (Hersen and Douady 2005, Elbelrhiti *et al.* 2008, Durán *et al.* 2009, Diniega *et al.* 2010, Elbelrhiti and Douady 2011, Durán *et al.* 2011b).

*3.1.6.4 Further insights from field investigations*

Excellent reviews on field investigations of dunes can also be found elsewhere (Pye and Tsoar 1990, Lancaster 1995, Livingstone *et al.* 2007, among others). Field observations provide the basis for *conceptual models*, as for example the model by Tsoar and Blumberg (2002) for parabolic dunes (Section 3.2.3.1) or the model by Lancaster (1989a) for star dunes. The latter author suggested that a dune shaped by two opposite wind directions (i.e. a transverse dune of reversing type) may transform into a star dune with four arms owing to leeside "secondary flow patterns", which lead to along-axis sand transport from the sides towards the center of the dune. First, the secondary winds build two longitudinal arms, one at the windward side and the other at the lee, whereupon the growth of the emerging arms is accentuated by a third oblique wind direction. Conceptual models have also been proposed for the common occurrence of asymmetric barchan shapes, where one of the barchan limbs extends downwind, eventually developing into a seif dune (Bagnold 1941, Tsoar 1983, Bourke 2010). Experiments or numerical simulations, discussed next, might be employed to test such models.

*3.1.6.5 Experiments under water as an approach to investigate (aeolian) dune dynamics*

Water tank experiments provide an alternative approach to study the long time scale processes of dune dynamics (Rubin and Ikeda 1990, Hersen *et al.* 2002, Endo *et al.* 2004). The major difference between subaqueous and aeolian dunes lies in their scale. Owing to the difference between the fluid densities of water and air (see Eq. (3.1)), underwater dunes are almost 1000 times smaller than their aeolian counterparts. Hersen *et al.* (2002) designed an experimental set-up to produce unimodal water streams strong enough to put glass beads, with size and density similar to those of desert sand grains, into saltation. The experiments produced subaqueous barchan-like bedforms displaying nearly the same morphological relations as those of aeolian dunes, thus suggesting that dune formation should not be strictly dependent upon the details of the transport mechanism (Hersen *et al.* 2002). By using water tank experiments, it has been possible to reproduce in the laboratory different dune morphologies appearing under varying flow conditions (Rubin and Ikeda 1990, Hersen 2005, Reffet *et al.* 2010) as well as to investigate in a systematic manner the collision between barchans (Endo *et al.* 2004) and the formation of a barchan belt emerging from a sand source (Katsuki *et al.* 2005). Water tank experiments have been also employed as a means to test hypotheses on the origin of some extraterrestrial dune shapes (Taniguchi and Endo 2007; see also Section 3.3.2.2).

*3.2 Numerical modeling*

In this Section, our aim is to present a summary of the progress achieved in our understanding of ripple and dune formation obtained from modeling. Two main modeling approaches can be adopted: morphodynamic (continuum) modeling or cellular automaton (CA) modeling. In the morphodynamic modeling approach, a mathematical expression is sought for the local mass transport of sediment over the sand bed (Anderson 1987a, Sauermann *et al.* 2001). Then, the rate of change of local bed elevation ($h$) can be obtained from the divergence of mass flux ($q$) of sediments using the Exner equation,

$$\frac{\partial h}{\partial t} = -\frac{1}{\rho_b} \nabla \cdot \boldsymbol{q}, \qquad (3.4)$$

where $\rho_b$ is the bulk density of the sand. In contrast, in a CA model for aeolian transport, physical processes such as erosion or deposition of sand are implemented by defining time-dependent stochastic interaction rules between nearest-neighbors of a lattice. The topography of the sand terrain evolves through the individual transport of so-called "sand slabs" according to prescribed rules (Werner 1995). Below we present a short description of the main morphodynamic and cellular automaton models for dunes and ripples and highlight the main achievements in the understanding of bedform dynamics obtained from modeling. At the end of the Section, a table is presented (Table 3.1), which displays a summary of the main numerical models for dunes and ripples discussed in the subsections which follow.

### 3.2.1  Morphodynamic (also continuum) model for aeolian dunes

Attempts to develop a continuum model for aeolian dunes started in the eighties (Wippermann and Gross 1986, Zeman and Jensen 1988, Stam 1997, van Dijk *et al.* 1999). Although none of these models was able to reproduce the shape of a migrating barchan, they provided the basis for the first successful continuum dune model by Sauermann *et al.* (2001) and Kroy *et al.* (2002), which has been further improved by many authors during the last decade. In the following, the main ingredients for modeling aeolian dunes using the continuum approach are discussed. At the same time, insights gained from continuum modeling into dune dynamics are highlighted.

#### 3.2.1.1  Wind field over dunes

A surface with dunes introduces a perturbation in the shear stress profile over a flat ground. Computational fluid dynamics (CFD) simulations have been employed to solve the Navier-Stokes equations of the turbulent wind over barchans and transverse dunes (Wiggs 2001, Parsons *et al.* 2004, Herrmann *et al.* 2005, Schatz and Herrmann 2006, Livingstone *et al.* 2007). Such an approach has been helpful in elucidating the patterns of recirculating flow at the dune lee (Figure 3.2) as well as the modification of the flow due to the presence of an array of many dunes. However, such calculations are too computationally expensive to be utilized in the framework of a continuum three-dimensional dune model.

A more suitable, less computationally expensive approach consists of using an analytical model to calculate the average turbulent wind shear stress field over the sand terrain. Such a model has been developed in the course of several years (Jackson and Hunt, 1975, Weng *et al.* 1991), and has been widely used in dune models (Stam 1997, Kroy *et al.* 2002, Andreotti *et al.* 2002b). It computes the Fourier-transformed components of the shear stress perturbation $(\hat{\vec{\tau}})$ in the directions longitudinal and transverse to the wind, whereupon the (perturbed) shear stress $\tau$ can be calculated,

$$\vec{\tau} = |\vec{\tau}_0| \left( \vec{\tau}_0 / |\vec{\tau}_0| + \hat{\vec{\tau}} \right), \qquad (3.5)$$

where $\vec{\tau}_0$ is the undisturbed shear stress over a flat ground. Note that the shear stress formulation is a linear solution of the Reynolds averaged Navier-Stokes equations. A consequence of this is that the shear stress perturbation is scale invariant: it only depends on the shape of the dune and not its size.

This analytical model works only for smooth heaps and gentle slopes. It does not apply for a dune with a slip face, since it does not account for flow separation occurring at sharp edges and steep slopes. Zeman and Jensen (1988) therefore suggested a heuristic approach by introducing a *separation bubble* at the dune lee, which comprises the area of recirculating flow. For the purpose of calculating the flow at the windward side of the dune, the wind model can be solved for the envelope,

$$h_s(x,y) = \max\{h(x,y), s(x,y)\}, \qquad (3.6)$$

comprising the separation streamlines ($s(x,y)$) and the dune surface ($h(x,y)$) (Figure 3.5). Then, at the lee, the shear stress within the separation bubble, where the resulting net transport essentially vanishes, can be set as zero.

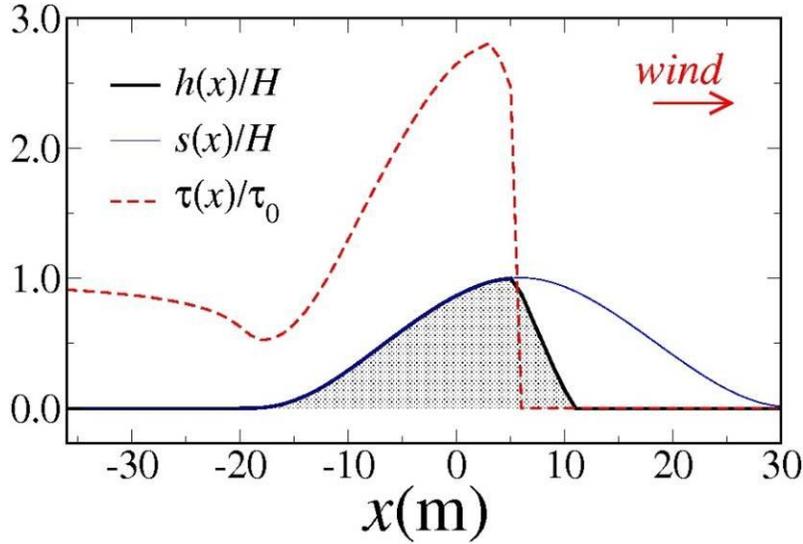

Figure 3.5 Average shear stress and separation bubble calculated over the profile of an isolated transverse dune. The shear stress profile, $\tau(x)$, appears rescaled with the upwind shear stress ($\tau_0$), while both the dune profile and the separation bubble, $h(x)$ and $s(x)$, respectively, are normalized using the dune height ($H$). $s(x)$ is a third order polynomial computed according to a phenomenological model (Kroy *et al.* 2002, Durán et al. 2010, see also Paarlberg et al. 2007).

*3.2.1.2  Continuum model for saltation flux*
In addition to the shear stress distribution, a continuum dune model requires an equation for the mass flux, $q$(x,y), which is then combined with Eq. (3.4) to give the bed evolution. At this stage, many authors (Wippermann and Gross 1986, Zeman and Jensen 1988, Stam 1997) attempted to use an equation for the saturated flux ($q_s$), such as the one by Lettau and Lettau (1978), c.f. Section 2.3.2.3. However, the flux upwind of a barchan moving on the bedrock is well below the saturated value (Fryberger *et al.* 1984, Wiggs *et al.* 1996). The simulation of a barchan using an equation for saturated flux therefore leads to an unphysical result: the decrease in wind strength close to the dune at the upwind causes deposition at the windward foot (which is not observed in the field; Wiggs *et al.* 1996). The dune's foot then stretches and the dune flattens. This problem was illustrated by van Dijk *et al.* (1999), who added to the saturated flux equation of Lettau and Lettau (1969) a phenomenological "adaptation length" to account for the distance over which sediment transport adapts to a new equilibrium condition due to a spatial change in transport conditions.

The continuum model by Sauermann *et al.* (2001) accounts for a physical modeling of the saturation transients of the flux. The cloud of saltating particles is regarded as a thin fluid-like layer moving on top of the immobile bed. The flux is described by calculating the average density and velocity of grains jumping with a "mean saltation length", whereas it is not distinguished between reptating and saltating particles. The spatial evolution of the average height-integrated mass flux over the sand bed is computed by solving the following equation, i.e. the well-known logistic equation for population growth,

$$\vec{\nabla}\cdot\vec{q} = \left[1 - |\vec{q}|/q_s\right]\cdot|\vec{q}|/L_{\text{sat}}, \tag{3.7}$$

where the saturated flux $q_s$ and the saturation length $L_{\text{sat}}$, which respectively scale with $[\tau-\tau_t]$ and $[\tau/\tau_t-1]$, encode the information on grain diameter, gravity, fluid viscosity and densities of fluid and particles (Sauermann *et al.* 2001, Durán and Herrmann 2006a, Parteli and Herrmann 2007b). It was shown that the minimal dune width $W_{\min}$ in the simulations is around $12L_{\text{sat}}$ (Parteli *et al.* 2007a).

So the scaling of the saturation length in the model of Sauermann et al. (2001) includes a dependence on the wind shear velocity which is not included in the scaling $L_{\text{sat}} \approx 2l_{\text{drag}}$ (see Section 3.1.1). Experiments show that the distance measured from the edge of a flat sand patch up to the horizontal position downwind where the flux is nearly saturated scales with the shear stress as predicted in the model of Sauermann et al. (2001) for $L_{sat}$ (Andreotti *et al*. 2010). However, $L_{sat}$ appears to be nearly independent of $\tau$ when $q$ is close to $q_s$. Hersen *et al*. (2002) proposed a simplified version of Eq. (3.7), where the RHS of this equation is replaced by $(q_s - q) / L_{sat}$, with $L_{sat}$ independent of the shear stress. There is an ongoing debate about the modeling of the saturation length, and so we refer to ongoing research papers for a more detailed discussion (Andreotti and Claudin 2007, Parteli *et al*. 2007b, Fourrière *et al*. 2010).

*3.2.1.3   Avalanches*
The continuum model then computes the bed evolution by inserting Eq. (3.7) into the mass conservation equation (Eq. (3.4)) (Kroy *et al*. 2002). Wherever the local slope exceeds the angle of repose of the sand ($\theta_r$), the surface must relax through avalanches in the direction of the steepest descent. Avalanches can be considered to occur instantaneously, since their time scale is much smaller than the time scale of dune evolution. Durán *et al*. (2010) presented a model of relatively fast convergence for computing avalanches at the dune lee.

*3.2.1.4   Model validation and contribution to understanding dune dynamics*
The continuum dune model by Kroy *et al*. (2002) was applied successfully by Sauermann *et al*. (2003) to predict the average shear stress and flux profiles over the symmetry axis of a real barchan dune in northeastern Brazil. It also reproduced quantitatively the three-dimensional shape of a barchan in Morocco (Figure 3.6).

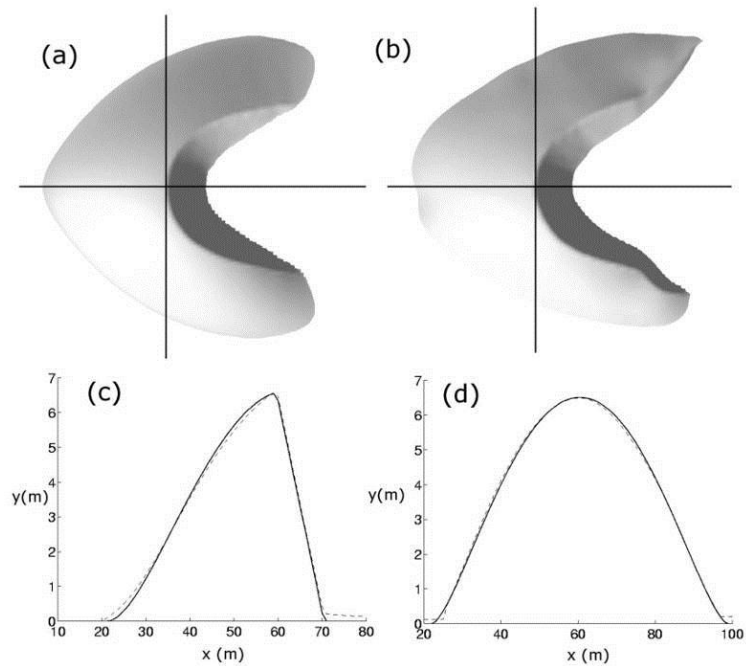

Figure 3.6 (a) Top view of a barchan dune produced with the model by Kroy *et al.* (2002); (b) a real barchan in Morocco; (c) longitudinal, and (d), transverse profiles at the respective crests of the simulated (continuous line) and real (dashed line) barchans of panels (a) and (b). The profile of the dune was measured as described in Sauermann *et al.* (2000). Reprinted with permission from O. Durán and H. J. Herrmann, *Modelling of saturated sand flux*, Journal of Statistical Mechanics: Theory and Experiment P07011 (2006). Copyright (2006) by IOP Publishing Ltd.

Also, important insights into the formation and dynamics of dunes and dune fields were gained using the continuum model.

In particular, the simulations showed that collisions between dunes lead to different dynamics depending on the volume ratio of the small and big dunes, $r = V_{small}/V_{big}$. The dunes merge if $r < 7\%$. If $r >$ 25%, then sand from the larger dune downwind is caught in the wake of the smaller dune behind, which then increases in size, eventually becoming larger (and therefore slower) than its downwind companion. Because the dynamics resembles a small dune crossing over the larger one (Besler 1997), it has been ill-named "solitary wave-like" behavior of dunes (Schwämmle and Herrmann 2003, Andreotti *et al*. 2002b, Durán *et al*. 2005), whereas the simulation showed that in reality what happens is that large and small dunes simply interchange their roles through the net release of sand from the large dune to the small one (Schwämmle and Herrmann 2003). For *r* between 14% and 25%, the collision results in the ejection of small "baby barchans" from the limbs of the larger dune downwind, whereas a further dune is realeased from the central part for 7% < *r* < 14% (Schwämmle and Herrmann 2003, Durán *et al.* 2005). Indeed, the ejection of small dunes resulting from collisions between dunes has implication for maintaining an equilibrium size distribution in barchan corridors (Durán *et al*. 2009, 2011b, see discussion in Sections 3.1.4 and 3.1.6.3).

Furthermore, by using an extension of the model to study dune formation under bimodal winds (Parteli *et al*. 2009) it was possible to reproduce many exotic dune shapes that occur on Mars (Section 3.3.2). The simulations could also help to understand the meandering of longitudinal seif dunes – which depends, as shown in Fig. 3.7, on the duration ($T_w$) of each of both prevailing winds relative to the dune turnover time ($T_m$).

Moreover, application of the model to investigate the genesis and evolution of barchan dune fields from a sand source has yielded important insights into the origin of dunes under realistic conditions (Durán *et al*. 2010). It was shown that, when a flat sand sheet is subjected to a saturated sand influx, it becomes a source for transverse dunes which emerge from sand-wave instabilities (see Section 3.1.1) on the sand surface. The transverse dunes become unstable after reaching the bedrock (Reffet *et al*. 2010, Parteli *et al*. 2011, Niiya *et al*. 2012, Melo *et al*. 2012) and decay into chains of barchans (Fig. 3.8). The genesis of barchan dune fields has also been studied with inclusion of vegetation (Luna *et al*. 2009, 2011, see also Section 3.2.3) and an exposed water table (Luna *et al*. 2012).

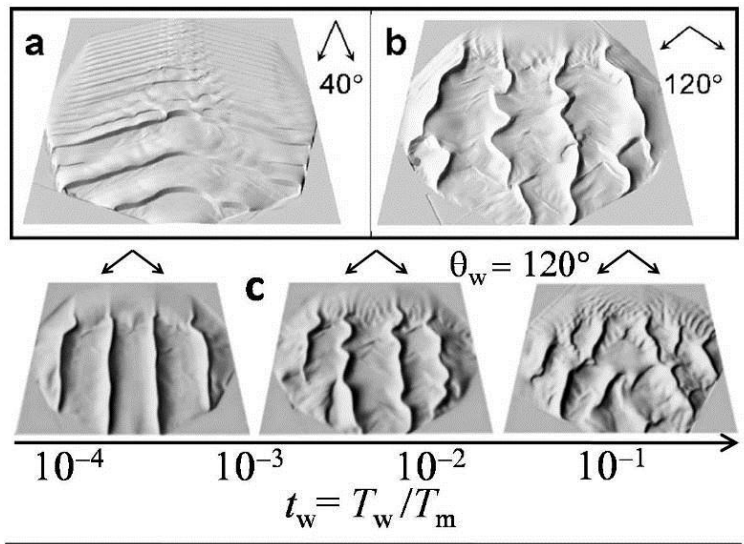

Figure 3.7 Seif dunes formed under wind regimes with different angles between the two main wind directions (panels (**a**) and (**b**)), and for different values of the ratio $t_w$ of the wind duration and dune turnover times (panel (**c**)). Longitudinal alignment occurs for obtuse divergence angles. Straight dunes form when $t_w < 0.1\%$; meandering (seif dunes) occur for $0.1\% < t_w < 10\%$; beyond this range, transverse dunes with periodically reversing orientation are formed (Parteli *et al*. 2009). Reproduced with permission from E. J. R. Parteli *et al*., Proceedings of the National Academy of Sciences of the USA 106, 22085-89 (2009). Copyright (2009) by the National Academy of Sciences of the USA.

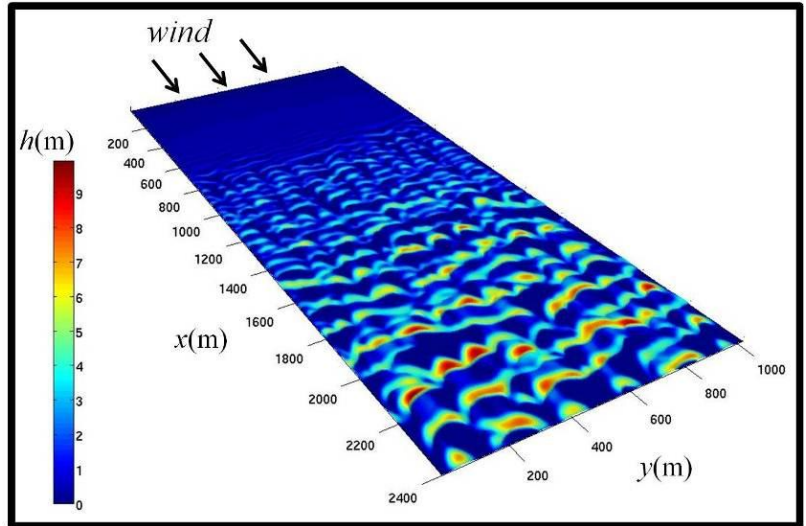

Figure 3.8 Numerical simulation of the genesis of a dune field (Durán *et al.* 2010, Luna *et al.* 2011). The flat sand surface upwind is subjected to a wind of constant direction. The instabilities occurring on the bed evolve into small transverse dunes, which destabilize on the bedrock, decaying into barchans (Parteli *et al.* 2011). The average dune height increases in size downwind due to the strong upwind flux (Luna *et al.* 2011).

While the continuum model can potentially serve as a helpful tool in the quantitative assessment of planetary dunes, some important limitations must be emphasized. One major issue concerns the modeling of the saturation length, $L_{\text{sat}}$. The saturation length of the dune model does account for inertia of the transported sediments, which limits its application to small values of wind shear stress (Parteli *et al.* 2007b). Indeed, due to the scaling of $L_{\text{sat}}$ with $(\tau-\tau_t)^{-1}$, a rapid increase of the minimal dune size with decreasing values of wind shear stress close to $\tau_t$ is predicted. While measurements confirming this prediction are scarce (Andreotti *et al.* 2010), the physical processes leading to flux saturation are currently the matter of substantial debate (Andreotti and Claudin 2007, Parteli *et al.* 2007b, Durán *et al.* 2011b). Moreover, the dependence of the threshold friction speed of the wind on the local slope is not accounted for in the model (Andreotti and Claudin 2007, Parteli *et al.* 2007b). One further issue concerns the model for the separation bubble (Figure 3.5). For isolated dunes, setting the wind shear at the lee to zero might have minor consequences for the barchan shape (Kroy *et al.* 2002). However, complex three-dimensional flow patterns at the lee could be critical for the dynamics of closely-spaced dunes (Parteli *et al.* 2006) or the elongation of seif dunes (Bristow *et al.* 2000) and should be considered in future modeling.

*3.2.2 Discrete (CA) models for aeolian dunes*

Werner (1995) introduced the first cellular-automaton model for aeolian dunes. This model successfully reproduced, in three dimensions, all four main types of dunes encountered in nature: barchans, transverse dunes, longitudinal and star dunes. In Werner's model, sand is transported in the wind direction as *sand slabs*, as depicted in Figure 3.9. The model can be summarized as follows.

a) *Transport algorithm* — A sand slab on the surface is chosen at random and moves downwind to a new lattice site $l$ (typically equal to 5) sites away (Werner 1995). The slab is then deposited at the site of arrival with a probability $p_d$, if there is no sand there, or $p_s$, if the site has at least one sand slab, where $p_d < p_s$ in consistence with the observation that grains rebound on the bedrock with greater likelihood than on a sand surface (Gordon and McKenna Neuman, 2009). If no deposition occurs, then the slab is moved again $l$ sites in the direction of the wind until deposition occurs. Next, another slab on the surface is chosen at random and the process is executed again. As shown by Nield and Baas (2008a), the wind strength can be mimicked through adjusting the mean path length $l/p_d$, where an increase/decrease in the probability of deposition $p_d$ means a lower/higher wind power.

b)  *Avalanches* — If the slope becomes larger than 34º (a) due to deposition, then the slab is moved down successively in the direction of the steepest descent, until all slopes are below the critical value; (b) due to erosion, then neighbouring slabs are moved downslope one lattice site, starting from the eroded site up the steepest gradient.
c)  Finally, a shadow zone (equivalent to the *separation bubble*, Figure 3.5) is introduced downwind of the surface covering the area enclosed by an angle of 15°. Slabs entering this shadow zone are deposited there.

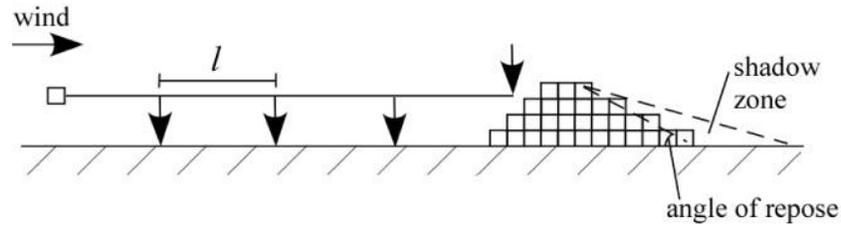

Figure 3.9 Sketch showing the main elements of the model by Werner (1995). Reprinted with permission from J. D. Pelletier *et al.*, Earth Surface Processes and Landforms 34, 1245-54 (2009). Copyright (2009) John Wiley & Sons Inc.

Werner's model has been applied as a tool for investigating the long-term evolution of dune fields (Werner and Kocurek 1997, 1999, Kocurek and Ewing 2005, Eastwood *et al.* 2011), and has been also improved by many authors. For instance, Momiji *et al.* (2000) added the condition of no erosion within shadow zones, since net transport within the separation bubble is negligible (McDonald and Anderson 1995). Also, a phenomenological non-linear equation for the transport length (*l*) was incorporated in order to account for the wind speed-up effect on the dune's windward side, c.f. Figure 3.5 (Momiji *et al.* 2000, Bishop *et al.* 2002). These improvements led to dunes with more realistic, asymmetric cross-section profile as compared to the triangular-shaped dunes of the original model (Werner 1995).

*3.2.2.1  Toward a physically based model for dunes*
In comparison to morphodynamic modeling CA algorithms may offer simpler and computationally less expensive numerical tools for simulating dune dynamics. One drawback of CA models is the lack of a physical modeling of the sand flux and the wind profile, which is essential for the quantitative assessment of dunes. Length and time scales are not explicitly established in CA transport rules (Werner 1995), and it is also not possible to perform a stability analysis of dunes using this type of modeling (Narteau *et al.* 2009). Recently, attempts were made to develop more realistic CA models through including quantitative descriptions of the average turbulent wind profile or saltation flux. For example, the model by Zheng *et al.* (2009) considered a wind friction velocity ($u_*$) that increased linearly with the surface height, while in the model by Pelletier (2009), Werner's transport algorithm was coupled to a generalized version of the boundary layer flow model by Jackson and Hunt (1975). Narteau *et al.* (2009) introduced a three-dimensional model in which the Navier-Stokes equations for the turbulent fluid were solved using a lattice gas cellular automaton model (LGCA). This model was coupled to a probabilistic sand transport model that considered the erosion rate proportional to the excess shear stress, $\tau - \tau_t$, thus accounting for the exchange of momentum between particles and fluid that lead to flux saturation. The simulations produced secondary (superimposed) bedforms on the surface of large dunes (Narteau *et al.* 2009, Zhang *et al.* 2010), thus indicating the minimal dune size in units of elementary cells. Therefore, it was possible to set real length associated with the size of a unit cell from the (real) size of minimal dunes on Earth deserts. As explained by Narteau *et al.* (2009), a similar procedure could be used in order to simulate dunes in other environments using the model.

*3.2.2.2 Other models*

Besides Werner's algorithm, different phenomenological models with deterministic transport rules were proposed (Nishimori *et al*. 1998, Katsuki *et al*. 2005a). The model by Nishimori *et al*. (1998), which included gravity-driven creep flux and lateral diffusion, also produced different dune geometries consistent with the imposed wind regime (c.f. Section 3.1.3). Katsuki *et al.* (2005a) introduced a simpler model including empirical deterministic rules for saltation flux on the windward side of dunes and avalanches at the lee. This model was applied to study collisions between barchans (Katsuki *et al.* 2011) and the genesis of a barchan belt from a sand source (Katsuki *et al.* 2005b). Katsuki and Kikuchi (2011) found that a field of barchans forms only if the sand source is subject to a large influx, which is in qualitative agreement with the findings of Durán *et al.* (2010), c.f. Section 3.2.1.4.

*3.2.2.3 Dunes as interacting particles*

Insights gained from continuum and CA models for dunes can provide basis for a different class of models where dunes are modeled as "particles" interacting according to phenomenological rules (Diniega *et al.* 2010). Such models can be potentially useful to assess the long time-scale processes involved in the evolution of a field of barchans (Lima *et al.* 2002, Durán *et al.* 2011b) or transverse dunes (Parteli and Herrmann 2003, Lee *et al.* 2005).

*3.2.3 Dune stabilization*

Some natural agents can act as stabilizers of dune sand on Earth and other planetary bodies. On Earth the prominent sand stabilizer (and the one most studied in dune modeling) is vegetation. In this Section our aim is to describe the progress achieved from modeling in our understanding of the dynamics of dunes with vegetation. We discuss stabilizing agents of dune sand on Mars and Titan in Sections 3.3.2.3 and 3.3.4, respectively.

*3.2.3.1 Vegetation*

*A conceptual model for the barchan-parabolic transition* —Plants locally reduce the velocity of the wind blowing over dunes, thus inhibiting sand erosion and enhancing deposition (see Section 4.1.1.2). The first places colonized by vegetation are those where erosion or deposition is small, i.e. the dune's arms, crest and the surrounding terrain. Indeed, erosion is an important factor for the dying of vegetation, since sand removal denudes the plants' roots and exposes deep layers to evaporation (Tsoar and Blumberg 2002). Since erosion prevents vegetation from growing on the windward side, the central part of a barchan or transverse dune can still migrate while two marginal ridges are left behind. A barchan that enters a stabilizing vegetated area thus undergoes a transition to a U-shaped parabolic dune (Figure 3.10): its shape "inverts", after which the vegetated limbs point opposite to the migration direction. When the rates of erosion or deposition at the crest become sufficiently small, vegetation finally stabilizes also the central part and the dune is rendered inactive (Tsoar and Blumberg 2002).

*Numerical simulations* — The first successful modeling of the barchan-parabolic transition (Tsoar and Blumberg 2002) was achieved by Durán and Herrmann (2006b), who extended the model by Kroy *et al.* (2002) to account for vegetation growth. The simulations indicated that the relevant parameter for the dune shape is the so-called fixation index (Durán and Herrmann, 2006b),

$$\theta \equiv Q_0 / (V^{1/3} V_v), \qquad (3.8)$$

where $Q_0$ is the saturated bulk sand flux, $V$ is the dune volume and $V_v$ is the characteristic growth rate of the vegetation cover. The fixation index gives the ratio between the erosion rate, which is proportional to $Q_0/V^{1/3}$, and the vegetation growth rate, $V_v$. Durán and Herrmann (2006b) found that parabolic dunes form when $\theta < \theta_c$ ($\approx 0.5$), while barchans appear for larger $\theta$ (Durán and Herrmann 2006b). This prediction roughly matches field observations (Reitz *et al.* 2010), and good quantitative agreement is found between the shape of real and simulated parabolic dunes (Durán *et al.*, 2008). Using the same model, Luna *et al.*

(2011) found that a sand beach evolves into a vegetated sand barrier (or "foredune", c.f. Hesp, 1989), rather than into a field of barchans, when $\theta < 0.1$.

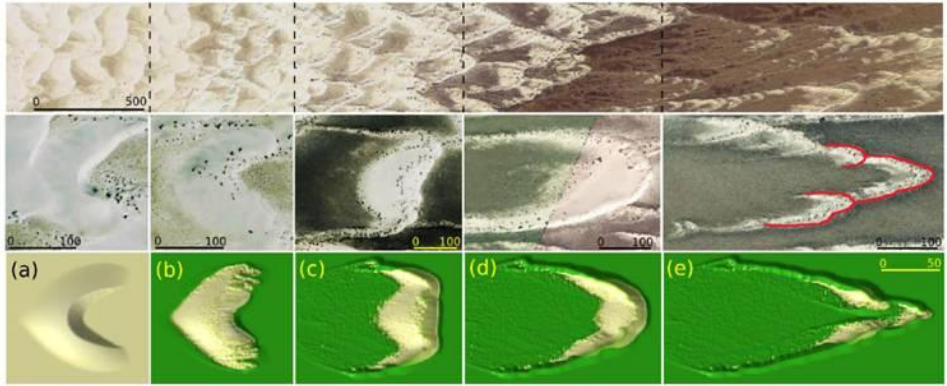

Figure 3.10 Deactivation of sand dunes due to vegetation growth. Images of dunes at White Sands, New Mexico (top and center) and numerical simulations of the transition barchan-parabolic dune (bottom, a-e) by Durán and Herrmann (2006b) using fixation index $\theta = 0.22$. Areas with vegetation appear dark and green in the images and in the simulations, respectively. Reproduced with permission from O. Durán and H. J. Herrmann, Physical Review Letters 97, 188001 (2006). Copyright (2006) by the American Physical Society.

In an earlier (and simpler) model, Nishimori and Tanaka (2001) extended the algorithm of Nishimori *et al.* (1998) to include a phenomenological vegetation cover density with growth rate depending on local erosion and deposition rates. Geometries resembling barchans (parabolic dunes) were obtained for strong (weak) winds, i.e. for long (short) hop length of the slabs (Nishimori and Tanaka 2001), essentially the information encoded in Eq. (3.8). Also, Baas (2002) extended Werner's model (Section 3.2.2) by adding a vegetation cover density, which linearly altered the probability of erosion at a given cell. Again, the vegetation growth rate depended on the local rates of erosion and deposition (Durán and Herrmann 2006b, Nishimori and Tanaka 2001). The barchan-parabolic transition was achieved by allowing different vegetation species (each associated with a maximum growth velocity and response rate to changes in erosion or deposition) to grow concurrently: pioneer species slowed down the flanks (first) for the stabilizer species to subsequently immobilize the edges into depositional ridges (Baas and Nield, 2007).

Modeling has been used to investigate the evolution of large sand systems in response to variations in vegetation growth due to climatic changes (Nield and Baas, 2008b) or altitude-induced changes in soil fertility (Pelletier *et al.* 2009). In some places, barchans and parabolic dunes can even coexist under the same climatic conditions. This can be understood by noting that dunes with different sizes have different fixation indices (Eq. (3.8)). Yihzaq *et al.* (2007) proposed a simple mathematical model that elucidates this bistability phenomenon. For moderate wind power condition, both barchans and parabolic dunes can form, while the former (latter) prevail for high (low) wind power (Yihzaq *et al.* 2007).

### 3.2.4 *Models for aeolian ripples in single-grain-sized beds*
Significant understanding of the physics of aeolian ripples was achieved through the work of Anderson (1987a). Anderson's revolutionary continuum model is in contrast with Bagnold's picture of a rhythmic barrage of grains saltating from one ripple to another (Bagnold 1941, Ellwood *et al.* 1975; see also modeling using this picture, e.g. Nishimori and Ouchi 1993, Kurtze *et al.* 2000, Miao *et al.* 2001, Kang and Guo 2004). It is based on the assumption that *reptation*, rather than saltation, is the transport mechanism through which ripples form and move. Saltating grains only provide the energy necessary for initiating and maintaining reptation.

A flat sand bed with a small periodic perturbation in the wind direction (*x*) is exposed to a horizontally uniform saltation flux,

$$q_{\text{sal}} = mN\ell,\tag{3.9}$$

where $m$ is the particle mass, $N$ is the impact rate and $\ell$ is the mean saltation trajectory length. A flux $q_{rep}$ of reptating particles is produced by the impacts. After entrainment, a reptating grain travels a distance $a$ (with probability density $p(a)da$ and average $\langle a \rangle$) before landing. The average mass flux of reptating grains is written as,

$$q_{rep} = mNn\langle a \rangle, \tag{3.10}$$

where $n$ is the average number of particles ejected upon impact. Importantly, the stoss side is more exposed to grain impacts than the lee area (Figure 3.3). By including this asymmetry of impact rate in the model, Anderson (1987a) arrived at an expression for the mass flux ($q(x)$) per unit width across the plane perpendicular to the flow, which is valid in the limit of small perturbations,

$$q(x) = q_{sal} + q_{rep} + mNn \cot \alpha \int_0^\infty [h(x) - h(x-a)] p(a) da, \tag{3.11}$$

where $\alpha$ is the impact angle, defined positively when the descent is inclined downwind (Anderson 1987a). The flux is thus largest at those areas where the bed slopes most steeply upwind, and is smallest in the lee areas (shadow zones), thereby causing the ripple to migrate upwind. A linear stability analysis of the model yields a preferred wavelength which is about six times the mean reptation length (Anderson 1987a), in agreement with the initial wavelength observed in wind tunnel experiments on ripple formation (Sepällä and Lindè 1978).

Anderson's model for initial ripple growth served as the basis for later models (summarized below), which were developed to study the long-term evolution of ripples.

### 3.2.4.1 *Cellular automaton models of aeolian ripples*
Forrest and Haff (1992) introduced a discrete cellular automaton simulation following the trajectory of each moving particle. The initial ripples grew from the spatial asymmetry in the flux of impacting grains (screening instability) in agreement with Anderson (1987a). In the nonlinear regime of the bed evolution, *merging* of ripples automatically filtered out ripples of significantly different sizes (e.g., Anderson 1990). Forrest and Haff (1992) suggested that bedform merging should play a major role for attaining a fully-developed state of ripples with constant size and spacing after a long time.

This hypothesis was explored in a toy-model introduced by Werner and Gillespie (1993), in which ripples were represented as one-dimensional entities ("worms") of different lengths moving around a discretized ring of $N$ segments. The head of each "ripple" moved one segment clockwise with probability inversely proportional to its length, thereby absorbing one segment of the "downwind ripple". Without any account for the detailed physics of ripples, the "worm" model (as well as its two-dimensional version) produced a logarithmic increase of "ripple" size with time (Werner and Gillespie 1993). Landry and Werner (1994) then introduced a three-dimensional model including a slope-dependent reptation length (Werner and Haff 1988) as well as rolling: after landing, the particle moved down the steepest descent (with a prescribed "pseudomomentum") until halting on a stable pocket on the granular bed. Bedform spacing grew at a decreasing rate with time, however without indication of a steady-state value for the mean height and spacing of ripples (Landry and Werner 1994).

In a different approach, Pelletier (2009) applied an improved version of Werner's dune model (Section 3.2.2) to ripple dynamics. Pelletier (2009) interpreted the (smaller) bedforms in the simulations as ripples, which then provided the roughness elements for simulations of dunes. Both ripples and dunes achieved a constant size and spacing, both increasing functions of grain size and excess shear stress (Andreotti *et al.* 2006). As explained by the author, this steady-state occurred because of the correction of the migration velocity of taller bedforms: taller dunes, subjected to larger values of bed shear stress (Figure 3.5), migrated a little faster than in the original model by Werner (1995). In this manner, the model produced a smaller rate with which small and tall dunes merge, compared to the original Werner's model, eventually achieving a nearly stationary state where ripple spacing remained constant (Pelletier 2009). The author's findings suggested an intriguing genetic linkage between ripples and dunes via the scaling between bedform size and aerodynamic roughness. Such a model however differs from Anderson's picture, since it assumes that saltation (rather than reptation) is the transport mode for ripple migration (Werner 1995).

*3.2.4.2   Continuum models of aeolian ripples*
Prigozhin (1999), following Anderson (1990), improved the continuum model of Anderson (1987a) by allowing reptating grains to roll a certain distance upon landing, due to the action of gravity. The flux of rolling particles was described using the so-called BCRE theory of surface flow of granular materials over a static sand bed, which was introduced by Bouchaud *et al.* (1994) for avalanche flows in sandpiles and adapted by Terzidis *et al.* (1998) to the problem of ripple instability. An alternative model for rolling was also proposed by Hoyle and Woods (1997) and Hoyle and Mehta (1999). These authors and Prigozhin (1999) studied numerically the nonlinear regime of ripples growth (see also the theoretical works by Valance and Rioual (1999) and Misbah and Valance (2003)).

*Towards a fully-developed state* — Following Prigozhin (1999), Yizhaq *et al.* (2004) introduced a phenomenological correction that made the reptation flux slightly larger on the lee side. This correction artificially shifted the largest flux value closer to the ripple's crest, thus acting to limit the growth of the ripples. After a sufficiently long simulation time, the system apparently approached a saturated (fully-developed) state with nearly constant ripple size (Andreotti *et al.* 2006). In an improved version of Prigozhin's model, rolling occurred not only due to gravity but also due to aeolian drag (Manukyan and Prigozhin 2009). This model also accounted for mass exchange between the bed and the saltation cloud (Sauermann *et al.* 2001). Erosion of ripple crests and deposition in throughs produced much more flat ripples than those obtained previously (Manukyan and Prigozhin 1999). These numerical simulations provided convincing evidence that the saturation of ripple growth (Andreotti *et al.* 2006) occurs due to removal of sand from crests into throughs due to the action of the wind, as suggested by Bagnold (1941). Manukyan and Prigozhin (1999) also suggested that ripples produced in a wind tunnel longer than the saturation length should display both size and spacing independent of wind strength. This prediction remains to be confirmed experimentally (Andreotti *et al.* 2006).

3.2.4.3   *Models for (granule) ripples in a bed with mixed sand sizes*
In sand beds with two grain sizes, ripples in general display coarse-grained crests, while in their inner parts and in the troughs smaller grains dominate (Sharp 1963, Hunter 1977). Ripples displaying this so-called 'inverse grading' have been named *granule ripples* or *megaripples* as they may achieve much larger wavelengths than ripples of uniform sand (Sharp 1963, Ellwood *et al.* 1975, Tsoar 1990, Fryberger *et al.* 1992, Sullivan *et al.* 2005, Jerolmack *et al.* 2006, Isenberg *et al.* 2011).

Anderson and Bunas (1993) introduced the first model for aeolian ripples incorporating two grain sizes. Their model included a phenomenological description of the wind flow around the evolving topography, such as to produce a horizontally uniform wind speed above a 'ceiling height', roughly one ripple wavelength above the mean surface altitude. The trajectory of the reptating particles was different for coarse and fine grains: since smaller grains can jump higher, they are subjected to larger wind velocities, thus experiencing larger acceleration and hopping further. Small particles, once ejected from the ripples' crest, can jump trajectories long enough to land in the throughs. In contrast, coarse grains do not hop far enough to escape the lee, thus accumulating around the crest; after being buried during migration of the ripple, they reappear on the stoss surface, thereafter skittering up the crest to repeat the process. The difference in jump lengths for coarse and fine grains thus provides an explanation for sorting and "inverse grading" of granule ripples (Werner *et al.* 1986).

Following these findings, Makse (2000) extended previous continuum modeling (Terzidis *et al.* 1998, Prigozhin 1999) to include a two-species model where coarse grains jumped shorter trajectories than smaller ones; and Manukyan and Prigozhin (2009) also adapted their model (Section 3.2.4.2) to account for two grain sizes. The latter study showed that megaripples with larger grain-size ratios (of the order of 5) than those considered in previous models (Anderson and Bunas 1993, Makse 2000) can be obtained by accounting for wind-induced motion of the coarse particles armoring the ripple crests (Manukyan and

Table 3.1: Summary of the most important models for aeolian dunes, ripples and granule ripples or megaripples. Model type may be morphodynamic ("M"), cellular automaton ("CA") or bedforms as interacting particles ("IP").

| Model | Type | Description | Comments; summary of further improvements |
|---|---|---|---|
| − *Dunes* − | | | |
| Sauermann *et al*. (2001); Kroy *et al*. (2002) | M | Includes a physically-based model for the shear stress and transient of flux saturation. | Inclusion of vegetation by Durán and Herrmann (2006b); latest improvements of the model are summarized by Durán *et al*. (2010). A linearized version of the model was also proposed (Hersen *et al*. 2002). |
| Werner (1995) | CA | Transport of sand slabs follows probabilistic rules depending on the local height. | Improved by Momishi *et al*. (2000) and Bishop *et al*. (2002) to include a phenomenological wind speed-up at the windward side and a wake at the lee; and by Baas (2002) to include vegetation. |
| Nishimori *et al*. (1998) | CA | Inspired by Werner, includes lateral diffusion and uses complex deterministic rules. | Extended by Nishimori and Tanaka (2001) to include vegetation. |
| Katsuki *et al*. (2005) | CA | Simpler than previous CA models; accounts explicitly for avalanche flux. | The model was used by the authors to simulate the collision of barchans and the emergence of barchans from a source. |
| Pelletier (2009); Narteau *et al*. (2009); Zheng *et al*. (2009) | CA | Extensions of Werner's model to include a more physical description of sand flux or turbulent shear stress. | Include first attempts to quantitatively compare CA model outputs with real field data. |
| Lima *et al*. (2002); Parteli and Herrmann (2003); Lee *et al*. (2005); Durán *et al*. (2011b) | IP | Dunes are treated as points which may exchange mass during their migration. | Diniega *et al*. (2010) presented an extensive discussion of such types of models. |
| − *Ripples* − | | | |
| Anderson (1987a) | M | First model for impact ripples. | Explained the initial growth of ripples due to a linear (screening) instability. |
| Forrest and Haff (1992) | CA | Used the idea of Anderson (1987a) that particle impact is the relevant factor for ripples formation. | The authors further simulated the non-linear regime of ripple growth and proposed that merging should lead a chain of ripples to a saturated state of constant ripple height. |
| Manakuyan and Prigozhin (2009) | M | Based on previous improvements of Anderson's model (Prigozhin 1999), includes transport through rolling due to gravity and aeolian drag. | The first model that apparently succeeded to reproduce the steady-state of ripples; supports Bagnold's hypothesis that removal of sand from crests into troughs due to the wind (missing effect in previous models; see Section 3.2.4.2) is essential for ripple growth saturation. |
| − *Granule ripples* (*megaripples*) − | | | |
| Anderson and Bunas (1993) | CA | Explained the inverse grading of granule ripples on the basis of particle size dependent hop length. | The model is the basis for all existing models for ripples in a bed of mixed sand sizes (see Section 3.2.4.3). |

Prigozhin 2009). Yizhaq (2004) also extended the model by Yizhaq et al. (2004) to describe two types of sand flux, namely reptation of coarse grains and saltation of fines. This model links the wavelength of megaripples not to reptation, as in Anderson's model, but to saltation, as suggested by Bagnold (1941).

Other recent attempts to model granule ripples considered directly solving the Navier-Stokes equations for the wind flow over the topography (Wu et al. 2008, Zheng et al. 2008). One particular model (Zheng et al. 2008), which considered three grain sizes, produced coarse-crested ripples as observed in nature and simulations. The model supported the conclusion that inverse grading of granule ripples results from the dependence of reptation length on grain size (Anderson and Bunas 1993).

*3.3   Dunes and ripples on Mars and other planetary bodies*
The study of bedform dynamics on extraterrestrial planets has become a field of broad interest for Earth and planetary scientists (Bourke et al. 2010). The occurrence and shape of ripples and dunes in an extraterrestrial field can potentially yield indirect information about local wind regimes as well as about physical attributes of the constituting sediments. Our knowledge about dune morphodynamics on Mars, Venus and Titan is moving forward every new mission with the acquisition of images from satellites and rovers with progressively higher resolution. Furthermore, numerical modeling and comparison of the observed bedforms with terrestrial analogs have provided valuable insights into the transport processes leading to dunes in diverse extraterrestrial deserts.

In Section 3.3.1 we discuss important advances achieved through missions observations in the research of Martian aeolian bedforms. Also, much of our current understanding of dune formation on Mars has been gained through numerical modeling as described in Section 3.3.2. Some highlights of the research on Venus and Titan dunes are presented in Sections 3.3.3 and 3.3.4, respectively.

*3.3.1   Aeolian bedforms on Mars: observations from orbiters and landers*
Sand dunes and ripples are amongst the most ubiquitous geological features on the surface of Mars (Bourke et al. 2010). Ripple formation or migration has been detected at practically all locations where rovers successfully landed (Arvidson et al. 1983, Moore 1985, Sullivan et al. 2005, 2008, Jerolmack et al. 2006). In some areas, changes on dune surface or net migration of dunes could be detected from satellite coverage (Fenton 2006, Bourke et al. 2008, Silvestro et al. 2010, 2011, Hansen et al. 2011, Chojnacki et al. 2011, Bridges et al. 2012a).

Most martian dune fields occur in the northern polar region and also on the floors of craters, which may act as traps for wind-blown sediments (Bridges et al. 2012). Craters may also contain sublayers of dark sediments — in general of volcanic origin and formed prior to the epoch of meteorite bombardments (Tirsch et al. 2011) — which are susceptible to wind erosion thus constituting sources for dunes sand. Since the first images of dunes by Mariner 9 (Sagan et al. 1972, Cutts and Smith 1973, McCauley 1973) significant progress in the understanding of dune processes on Mars has been made. Because barchans and transverse dunes were the prominent dune morphologies imaged by early missions, it has been suggested that sand-moving winds of Mars are predominantly unidirectional (Lee and Thomas, 1995). However, more recent orbiters' images showed a diversity of unusual dune shapes which could not form under constant wind direction (Schatz et al. 2006, Parteli and Herrmann 2007a, Hayward et al. 2007). The possibility that their constituent sand is indurated (Section 3.3.2.3) and the relative scarceness of dune activity observed on the red planet (compared to the Earth) raised the question of whether dune formation occurs on present day Mars or whether the martian dunes were formed during past climates with a denser atmosphere (Breed et al. 1979). Recent modeling and experimental works (Section 3.3.2) shed some light on formative processes of martian dunes.

A type of aeolian bedform called *Transverse Aeolian Ridges* (TARs) could not be classified as ripples or dunes (Thomas et al. 1999, Malin and Edgett 2001, Bourke et al. 2003, Wilson and Zimbelman 2004, Balme et al. 2008, Zimbelman 2010, Berman et al. 2011). TARs, found in almost all martian geologic settings, are of intermediate size between ripples and dunes, seem to align normal to local winds, and commonly share albedo with the surrounding terrain (Malin and Edgett 2001). TARs that are brighter than the surrounding terrain appear mostly underneath large dark dunes of various morphologies where these two bedform types occur together. These bright TARs were possibly formed by winds of a past climate (Thomas *et al.* 1999). A possible explanation for their low albedo is that the TARs are covered in bright dust because they are less active than the surrounding, large dunes. In some areas, dark TARs occur superimposed on the lower parts of dunes, suggesting a linkage between martian dark TARs and Earth's coarse-crested granule ripples (Balme *et al.* 2008; see Section 3.2.4.3). The coexistence of these TARs together with larger dunes of complex morphologies can be understood by recalling the concept of turnover time (Section 3.1.6.1): being smaller, TARs have a smaller value of $T_\mathrm{m}$ thus aligning transversely to the prevailing wind while the larger dunes provide a proxy for local wind directions on a longer time scale.

### 3.3.2 *Dune formation on Mars: insights from modeling and experiments*
The quantitative study of martian dune formation requires modeling studies that account for the attributes of sediment and atmosphere in an arbitrary planetary environment.

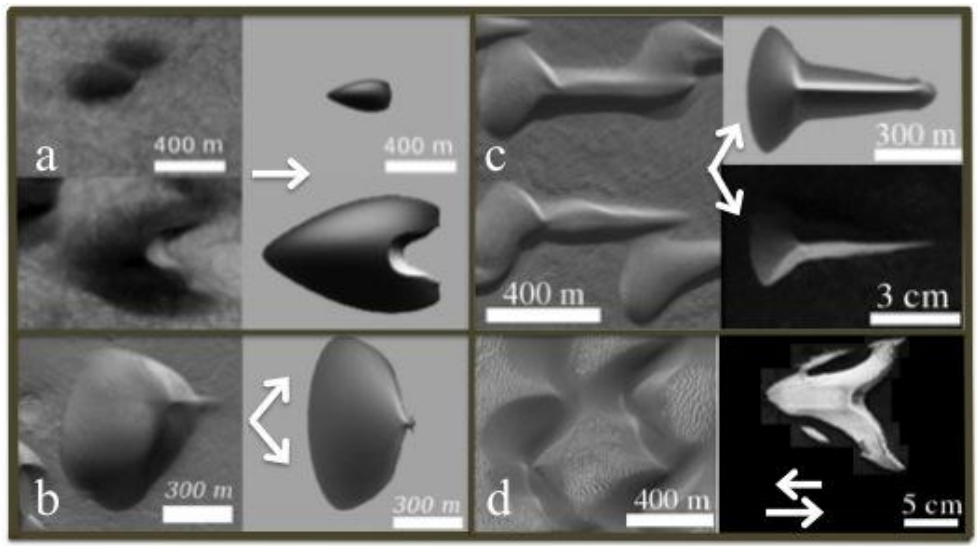

Figure 3.11 Numerical simulations and water tank experiments of dune formation under various flow conditions provided insights into the formative wind regimes of diverse martian dune shapes: (a) Barchan dunes of different sizes at Arkhangelsky crater (left) together with the corresponding shapes obtained in numerical simulations (right) accounting for the attributes of sand and atmosphere of Mars (after Parteli and Herrmann 2007a). The arrow indicates the wind direction; (b) "wedge" dune at Wirtz crater (left) and martian dune simulated by Parteli and Herrmann (2007a) (right) using a bimodal wind regime with divergence angle $\theta_\mathrm{w} = 100°$ (arrows indicate the alternating wind directions); (c) straight martian longitudinal dunes (left) could be reproduced in numerical simulations (top-right; after Parteli and Herrmann 2007a) and water tank experiments (bottom-right; after Reffet *et al.* 2010) under a bimodal wind regime with $\theta_\mathrm{w} = 120°$; (d) barchans with deformed limbs resembling those at Proctor crater (left) were obtained in water tank experiments by Taniguchi and Endo (2007) (right) using a reversing flow condition where the velocity of the secondary wind was about 75% the primary wind's speed. Images courtesy of NASA/JPL/MSSS.

#### 3.3.2.1 *The shape of intra-crater barchans*
Parteli *et al.* (2005, 2007a) presented the first calculation of the shape of barchan dunes at Arkhangelsky crater on Mars (Figure 3.11a). The simulations used the model by Kroy *et al.* (2002) considering a grain diameter $d = 500$ $\mu$m (Edgett and Christensen 1991, see also Section 2.4) and wind speeds larger than the corresponding threshold for direct entrainment under martian conditions, i.e. $u_{*\mathrm{ft}} = 2.0$–$2.5$ m/s (Parteli *et al.* 2007a). A value of impact threshold $u_{*\mathrm{t}} = 0.8\ u_{*\mathrm{ft}}$ ($\approx 2.0$ m/s) was used in analogy with saltation on Earth (Bagnold 1941) (although more recent simulations and theoretical models have suggested a substantially lower ratio $u_{*\mathrm{it}}/u_{*\mathrm{ft}}$ for Mars; Almeida *et al.* 2008, Kok 2010a, 2010b, Pähtz *et al.* 2012; see

Section 2.4). From the minimal dune width ($W_{min}$), which scales with $1/[\tau/\tau_t -1]$ (Section 3.2.1.2), the authors obtained an estimate for the average shear velocity of sand-moving winds at Arkhangelsky crater, namely $u_* = 3.0$ m/s. Excellent quantitative agreement was found between the morphological relations of the barchans obtained in the simulation and those of the Arkhangelsky barchans (Figure 3.11a).

The calculations were then adapted to reproduce the shape of barchans at other locations on Mars (Parteli and Herrmann 2007b). In spite of the different minimal threshold velocity values predicted for each field owing to variations in atmospheric pressure, the values of $u_*$ estimated from the simulations in all cases were around 3.0 m/s (Parteli and Herrmann 2007b). This value is well within the maximum range of values of $u_*$ on Mars (between 2.0 m/s and 4.0 m/s) as it has been estimated indirectly from observations of surface erosion and ripple formation by the rovers (where estimations also assumed $u_{*t} \approx$ 2.0 m/s; Arvidson *et al.* 1983, Moore 1985, Sullivan *et al.* 2005). Observations from orbiters and landers suggested that such extreme values of wind shear velocity occur occasionally on Mars, during global dust storms (Arvidson *et al.* 1983, Moore 1985, Sullivan *et al.* 2005).

As mentioned above, there is strong theoretical evidence provided by recent works (Almeida *et al.* 2006, Kok 2010a, 2010b, Pähtz *et al.* 2012) for much lower values of impact threshold on Mars than the one (2.0 m/s) previously estimated (Figure 2.18b). Future work should thus focus on reproducing the observed morphology of martian dunes using these revised estimates of the martian impact thresholds.

### 3.3.2.2 *Exotic dune shapes of Mars*

Experiments in a water tank and computer simulations have shown that some exotic dune shapes occurring on Mars could form under directionally varying flows (Figure 3.11b-d). Taniguchi and Endo (2007) performed experiments of barchan ripples under reversing flow conditions, and found barchans with deformed limbs that resembled dunes at Proctor crater (Figure 3.11d). Further, Parteli *et al.* (2009) extended the simulations of dune formation under bimodal winds (c.f. Fig. 3.7) to low sand availability condition. The authors found that the resulting dune shape depends on $\theta_w$ (which determines dune alignment, c.f. Section 3.1.5.1) as well as on the duration of the bimodal wind relative to the dune migration time, i.e. $t_w = T_w/T_m$: a value of $t_w$ smaller than $10^{-3}$ forms rounded barchans with reduced slip face ("occluded barchans"), if $\theta_w = 40°- 80°$; non-elongating "wedge" dunes (Figure 3.11b), if $\theta_w = 100°$ (also produced in water tank experiments by Hersen 2005), or elongating longitudinal bedforms (Figure 3.11c), if $\theta_w > 110°$ (see Parteli *et al.* 2009 for a phase diagram of the morphology).

### 3.3.2.3 *Sand induration*

Increased soil moisture is an important contributor to the immobilization of dunes. Crusting or induration of sand has been observed on Mars by Viking 1, Viking 2, Mars Pathfinder, Spirit, and Opportunity landers (Arvidson *et al.*, 2004, Moore *et al.*, 1999, Thomas *et al.*, 2005, Sullivan et al. 2008, 2011). The classical experiments by Kerr and Nigra (1952) illustrated the dramatic consequences sand induration can have for dune morphodynamics. The experiments consisted of spraying crude oil onto advancing barchans, in order to halt their movement. The sand arriving at the fixed dunes from the upwind was deposited at the lee, whereupon the dune was soaked with oil again. As this process continued, the slip face of the dunes became progressively smaller and each dune more elongated. The oil was sprayed on the dunes in successive stages until all motion was stopped, and the dunes achieved a rounded, elliptical dome-shaped form.

Rounded barchans occur in the north polar region of Mars known as Chasma Boreale (Schatz *et al.* 2006). Their discovery raised the question whether sand induration due to frozen carbon dioxide or the presence of salts as an intergranular cement (Clarck *et al.* 1982) could be a relevant agent for dune processes on Mars. Through using numerical modeling, Schatz *et al.* (2006) succeeded in reproducing the transition from a crescent to a rounded dune shape (Kerr and Nigra 1952). Induration (non-mobility) was simulated for each successive stage of dune development, while at the same time new sand was added from upwind (Figure 3.12). A further argument in favor of induration of the Chasma Boreale dunes is the surprising co-existence of the rounded barchans with straight-crested linear dunes aligned parallel to the orientation of the barchans — the linear dunes should simply decay in barchans if the wind was unidirectional (Section 3.2.1.4). Possibly, sand saltating downwind along both sides of the indurated linear dune could deposit at the lee due to secondary flow effects (not included in the model by Schatz *et al.* 2006) at the dune's downwind extremity (Tsoar, 2001), thus yielding dune elongation.

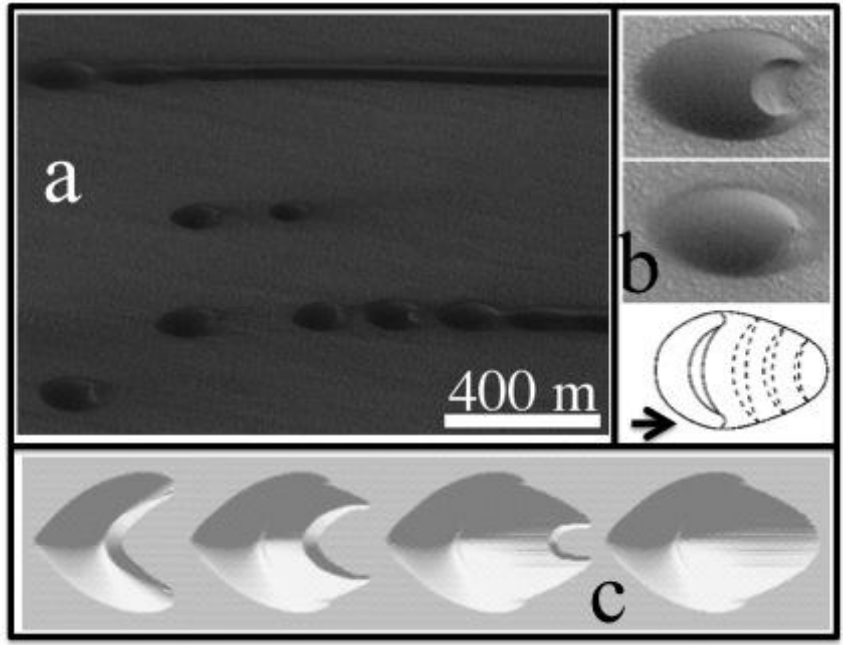

Figure 3.12 Evidence for indurated sand dunes on Mars? (a) rounded barchans occurring side-by-side with straight linear dunes at Chasma Boreale, north polar region of Mars; (b) barchans at Chasma have typically a reduced slip-face (top) or rather a dome-like shape (center); bottom: schematic diagram illustrating the conceptual model by Schatz *et al.* (2006) for the formation of the Chasma barchans: the dunes are indurated; sand incoming from the upwind is deposited at the lee, thus progressively reducing the size of the slip-face. The arrow indicates the wind direction; (d) numerical simulations by Schatz *et al.* (2006) of sand transport over a indurated barchan (time increases from left to right) produced a dome-like barchan (last snapshot on the right), thus supporting the conceptual model of the authors. After Schatz *et al.* (2006).

*3.3.3 Dunes on Venus*

Owing to the large atmospheric density of Venus, a minimal dune size of the order of 20cm is predicted from the scaling with $\ell_{drag}$ (Claudin and Andreotti 2006). Indeed, wind tunnel experiments under constant wind and atmospheric pressure condition of Venus produced small transverse bedforms, named *microdunes*, with wavelengths within the range 10–30 cm (Greeley *et al.* 1984, Marshall and Greeley, 1992). These microdunes displayed several features reminiscent of Earth aeolian dunes, such as a slip face and a separation bubble. Tests using mixed sand sizes did not lead to crest-coarsened bedforms as expected in the case of granule ripples (Sections 3.2.4.3), thus corroborating the conclusion that the microdunes were indeed small-scale prototypes of terrestrial "full-scale transverse dunes" (Marshall and Greeley 1992).

Dune fields identified in images of the Venus orbiter Magellan contain long-crested dunes much larger than the microdunes of the wind tunnel (Greeley *et al.* 1992b, 1995, Weitz *et al.* 1994). The dunes have length ranging from 500 m to 10 km, width of about 200 m, and values of interdune spacing of the order of 500 m (Greeley *et al.* 1992b) — these are dimensions akin to those of terrestrial dunes (Lancaster 1995). The orientation of nearby surficial features indicates transverse dune alignment to the prevailing wind direction. Although microdunes are not visible as surface features in the Magellan images (images' resolution is 75 m; Weitz *et al.* 1994), their occasional occurrence has been indicated by Magellan's radar data (Weitz *et al.* 1994). Because radar detection of the microdunes required their slip face be nearly perpendicular to the radar illumination (Blom and Elachi 1987, Weitz *et al.* 1994), there might be many more (small) dunes on Venus not detected by Magellan due to inadequate angle of the radar beam (Weitz *et al.* 1994, Greeley *et al.* 1995). Alternatively, the relative scarceness of dunes or microdunes on Venus might be a consequence of insufficient wind speeds, meager sand cover or a scarcity of sand-size particles on Venus, where the effect of weathering caused by water is negligible compared to Mars or Earth (Weitz *et al.* 1994).

*3.3.4    Dunes on Titan*

Titan dunes, first detected by radar images of the Cassini mission (Lorenz *et al.* 2006, Elachi *et al.* 2006), concentrate along the equatorial belt between 30°S and 30°N. They appear as hundred of kilometers long streams (i.e. linear dunes, c.f. Rubin and Hesp 2009) oriented nearly parallel to the equatorial axis. Although the minimal size predicted for dunes on Titan is about 1.5 m (Claudin and Andreotti 2006), dunes imaged by Cassini have heights of about 100 m, closely resembling longitudinal dunes in Namibia, both in size and spacing (Lorenz *et al.* 2010). Their dark appearance is consistent with an organic composition of the constituent particles (Section 2.4). Recent debate on the formative processes of dunes on Titan has advanced the knowledge on the atmosphere and wind systems of this moon of Saturn (Lunine and Lorenz 2009).

Dune crest orientation can provide valuable information on prevailing transport directions (Fenton *et al.* 2003). However, resolution of Cassini images of Titan dune fields (~ 350 m; LeGall *et al.* 2011) is too low to resolve the profiles of dunes. The elongation direction of Titan dunes could be inferred indirectly from their behavior around topographic obstacles (Radebaugh *et al.* 2008, 2010). Radebaugh *et al.* (2010) explain that the "teardrop" streamlined dune patterns in the image of Figure 3.13 can form if the dunes are elongating to the East — on Earth, linear dunes change orientation upwind of the obstacle, then resuming their original trend on the downwind side (Radebaugh *et al.* 2010).

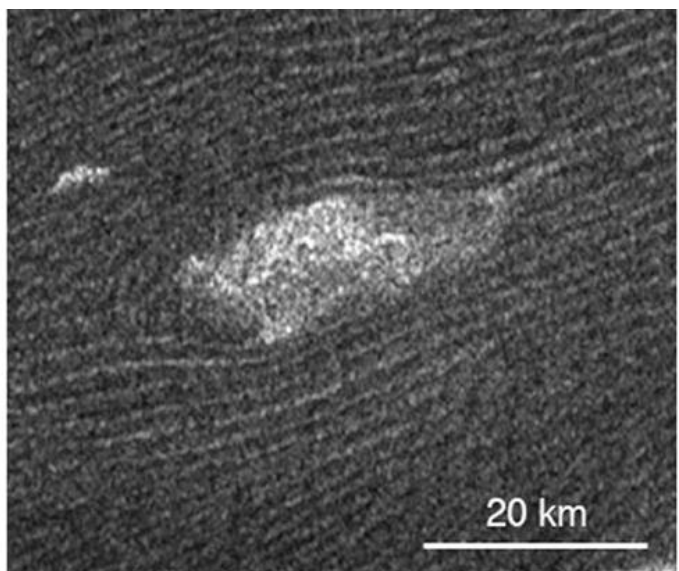

Figure 3.13 — Cassini Radar image centered on 6.5 S, 251W, acquired on Oct. 2005. North is to the top. The dunes divert around the topographic obstacle (white area in the image) and resume on the downwind side, thus implying that the direction of dune elongation is from SW-NE (L-R) in the image. After Radebaugh et al. (2010).

The global circulation model by Tokano (2010) shed light on the eastward elongation of Titan linear dunes. The model showed that exceptionally strong eastward wind gusts with velocities larger than 1.0 m/s, i.e. well above the average speed of westward winds ($\approx$ 0.6 m/s, see e.g. Lorenz 2010), occur twice a year correlated with the convergence of northward and southward winds at the equator (constituting the so-called intertropical convergence zone, ITCZ). So, if a threshold velocity of 1.0 m/s is assumed, then sand transport and dune elongation can occur only during those strong eastward wind gusts. Interestingly, this threshold wind velocity, which refers to a height of about 300 m above the surface (Tokano 2010), corresponds to a threshold shear velocity of about 0.036 m/s, which is in close agreement with the theoretical value predicted in chapter 2.4. Besides, seasonally reversing winds nearly perpendicular to the crest of the dunes lead to a bimodal wind regime with divergence angle close to 180° — which is formative of dunes of nearly "reversing" type (Section 3.2.1.4). Based on the simulations by Parteli *et al.* (2009), Tokano (2010) concluded that the lack of meandering of Titan dunes implies a large turnover time of the dunes relative to the time scale of this bimodal wind (Section 3.2.1.4).

Rubin and Hesp (2009) proposed a different scenario in which the sand of Titan dunes is indurated, such that dunes grow under unidirectional wind through a mechanism similar to the one suggested by Schatz *et al.* (2006) for the linear dunes at Chasma Boreale, Mars (Section 3.3.2.3). In this manner, dunes oriented transversely or longitudinally to the equatorial axis would form in areas with loose or indurated sediments, respectively. Such an origin of Titan dunes remains to be tested by future modeling studies.

Whatever the mechanism of formation and growth of linear dunes on Titan, their average spacing of 3 km is of the order of the height of Titan's atmospheric boundary layer, which has been estimated from *in situ* data obtained by Huygens probe as well as indirectly from a global circulation model (Lorenz *et al.* 2010). Titan dunes are giant dunes (Andreotti *et al.* 2009), in contrast to martian barchans that have sizes close to the minimal size (Claudin and Andreotti 2006, Parteli *et al.* 2007a,b, Andreotti *et al.* 2009). Possibly, the crest of Titan dunes, unresolved in current images, is armored with smaller superimposed dunes indicating dune growth through amalgamation of bedforms (Lancaster 1988, Andreotti *et al.* 2009).

## 4. The physics of dust emission

As discussed in Section 1.2.1, mineral dust aerosols affect the Earth and Mars systems through a large variety of interactions, including by scattering and absorbing radiation, and by serving as cloud nuclei (e.g., Leovy 2001, Jickells *et al.* 2005, Goudie and Middleton 2006). Understanding and simulating these interactions and their effect on, for instance, weather and climate requires a comprehensive understanding of the physics of the emission of mineral dust aerosols, which we discuss in this chapter. Because the physical conditions for dust lifting differ between Earth and Mars, we first discuss the physics of dust emission on Earth in Section 4.1, after which we discuss how the physics of dust emission differs on Mars in Section 4.2.

### *4.1 The physics of dust emission on Earth*

Soils that are most sensitive to wind erosion and dust emission usually lack protection from vegetation, have low soil moisture content (e.g., Marticorena and Bergametti 1995), and contain readily erodible sediments of fine particles (Prospero *et al.* 2002). As such, most substantial sources of dust aerosols are deserts and dry lake beds, although dust emissions from vegetated landscapes and dune fields are also commonly observed and are important especially to regional weather and climate (Bullard *et al.* 2005, 2008, Okin 2008, Rivera Rivera *et al.* 2009, Okin *et al.* 2011).

Dust particles that can be transported thousands of kilometers from their source regions, and thereby produce a substantial effect on weather and climate, predominantly have diameters smaller than 20 μm (Gillette and Walker 1977, Tegen and Lacis 1996). Particles in this size range are referred to as PM20 and experience strong interparticle forces (Section 2.1.1.2), which causes them to rarely occur loosely in soils. Rather, PM20 dust in soils occurs mostly as coatings on larger sand particles (Bullard *et al.* 2004) or as part of soil aggregates with a typical size of ~20 – 300 μm (Alfaro *et al.* 1997, Shao 2001, 2008). Dust aerosols are emitted naturally through three distinct processes (Shao 2001, 2008): (i) direct aerodynamic lifting, (ii) ejection of dust aerosols from soil aggregates by impacting saltating particles, and (iii) ejection of dust aerosols from soil aggregates that are participating in saltation (Figure 4.1). These latter two processes occur as a result of *saltation bombardment*, the impacts of saltating particles on the soil (Gillette *et al.* 1974, Shao *et al.* 1993a), which causes *sandblasting*, the release of dust aerosols from dust aggregates that are either saltating or are impacted by saltators (Alfaro *et al.* 1997, Shao 2008). The emission of dust aerosols from either soil aggregates or saltating dust aggregates is thus initiated by wind speeds exceeding the fluid threshold for saltation. In contrast, direct aerodynamic lifting of PM20 dust requires wind speeds much larger than the saltation fluid threshold (Figure 2.2), except for a small fraction of PM20 soil particles that experience very weak interparticle forces (Klose and Shao 2012). Consequently, direct aerodynamic lifting is a substantially less important source of dust aerosols than impact-induced emission from dust aggregates in the soil or in saltation (Gillette *et al.* 1974, Shao *et al.* 1993a, Loosmore and Hunt 2000).

In addition to these three natural dust emission processes, dust can be emitted through human actions such as off-road vehicles. Such emissions could be important locally but are likely insignificant on a global scale (Gillies *et al.* 2005, Goossens *et al.* 2011). Dust emissions can also be enhanced by human soil disturbances that lower the threshold shear velocity, such as grazing and other land use changes (Marticorena *et al.* 1997, Reynolds *et al.* 2001, Gillies *et al.* 2005, Neff *et al.* 2008). The resulting anthropogenic fraction of global dust emissions is highly uncertain, but possibly substantial (Tegen *et al.* 2004, Mahowald *et al.* 2004).

Dust aerosols are thus predominantly emitted through the transfer of kinetic energy onto aggregates of dust, either by the impacts of saltators onto soil dust aggregates or by the impact of saltating dust aggregates on the soil (Figure 4.1b, c). This transfer of kinetic energy onto an aggregate of bonded dust particles creates elastic waves within the aggregate that can rupture the energetic bonds between individual constituents of the aggregate (Kun and Herrmann 1999). This process creates dust aerosols either by *damaging* the aggregate, breaking it into two or more fragments with the size of the largest fragment comparable to the original size of the aggregate, or by *fragmenting* the aggregate, breaking it into a large number of fragments for which the size of the largest fragment is small compared to the original size of the aggregate (Kun and Herrmann 1999). Kok (2011a) argued that dust emission through the fragmentation of dust aggregates is analogous to the fragmentation of brittle materials such as glass, and used this analogy to derive an analytical expression for the size distribution of emitted dust aerosols (Section 4.1.3.3). The good agreement of the resulting theory with available field measurements (Figure 4.3b) suggests that dust emission commonly occurs through fragmentation of dust aerosols, although more research is required to further investigate this.

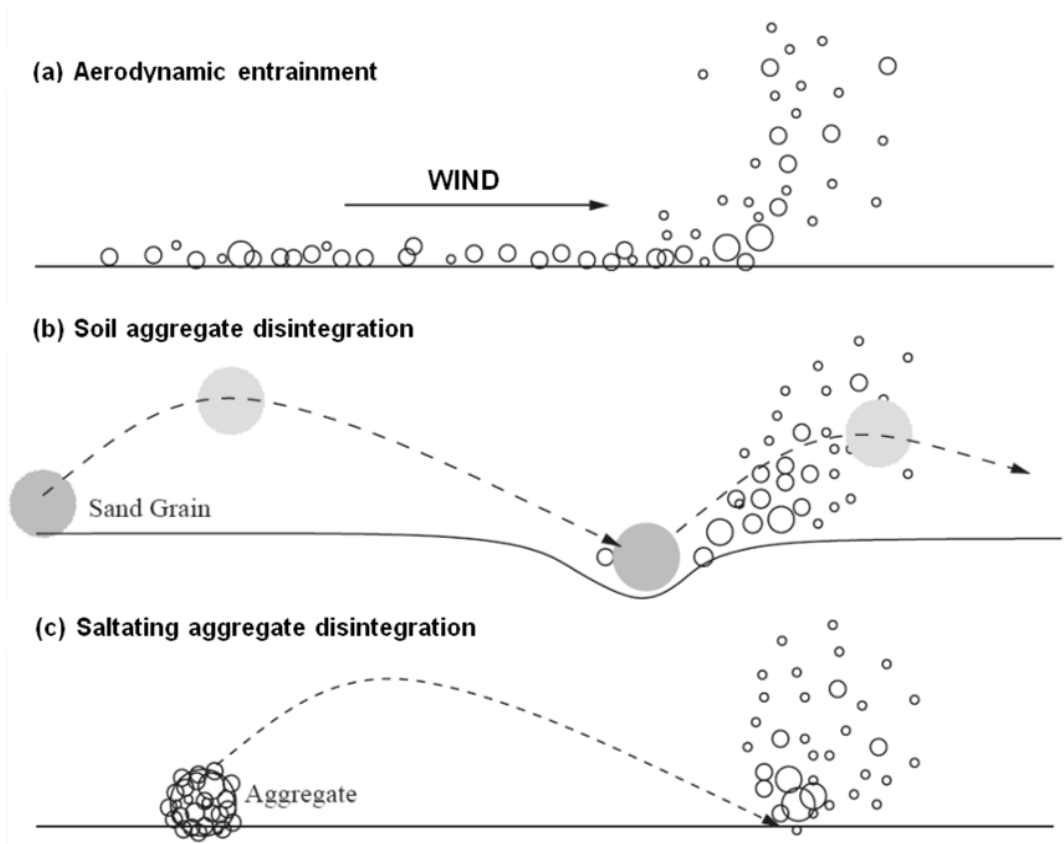

Figure 4.1. Illustration of the three dust emission mechanisms: (**a**) aerodynamic entrainment, (**b**) ejection of dust aerosols from soil aggregates by impacting saltating particles, and (**c**) ejection of dust aerosols from soil aggregates that are participating in saltation. Adapted from figure 7.5 in Shao (2008).

Since the kinetic energy supplied by saltator impacts drives the emission of dust aerosols, some theoretical models of dust emission have assumed that the vertical flux of dust emitted by an eroding soil is proportional to the total kinetic energy of saltator impacts (Shao *et al.* 1993a, 1996, Shao 2001, 2004), which is supported by wind tunnel measurements (Zobeck 1991a, Shao *et al.* 1993a, Rice *et al.* 1996). However, Lu and Shao (1999) took a slightly different approach by considering that an impacting saltator "ploughs" through the soil (Rice *et al.* 1996), thereby ejecting soil particles in its path. The volume of soil traversed by the impacting saltator in this manner was then taken to be representative of the resulting vertical dust flux. Assuming the soil to be plastic, Lu and Shao (1999) predicted that dust emission due to saltation bombardment scales with $u_*^4$ for soft, easily erodible soils, and with $u_*^3$ for harder, less erodible, soils. Shao (2001, 2004) expanded this theory to include the contribution from the disintegration of saltating dust aggregates and also proposed a theory for the size distribution of the vertical dust flux as a function of wind speed and the parent soil size distribution (Section 4.1.3.2).

The qualitative understanding of the physics of dust emission discussed above can be combined with the detailed understanding of saltation reviewed in chapter 2 to develop quantitative expressions for (i) the wind speed threshold above which dust emission occurs, (ii) the vertical dust flux emitted by the soil when this threshold is exceeded, and (iii) the size distribution of emitted dust aerosols. We discuss these three subjects in the following sections.

*4.1.1 The threshold for dust emission*

Since dust aerosol emission is primarily due to saltation bombardment and subsequent sandblasting, the threshold wind speed above which dust emission occurs is the saltation fluid threshold. However, the fluid threshold discussed in Section 2.1.1 applies to ideal soils of loose, dry sand, which are not necessarily characteristic of soils susceptible to wind erosion. These soils have several components that can enhance their resistance to wind erosion, including crusts (Cahill *et al.* 1996, Belnap *et al.* 1998, Gomes *et al.* 2003, Ravi *et al.* 2011), soil aggregates (Chepil 1950), soil moisture (Belly 1964), and non-erodible roughness elements such as pebbles, rocks, or vegetation (Lancaster and Baas 1998). In the following sections, we discuss the corrections to the saltation fluid threshold that are required to account for the presence of soil moisture and non-erodible roughness elements.

*4.1.1.1 Effect of soil moisture on the saltation fluid threshold*

The presence of soil moisture can create substantial interparticle forces that inhibit the initiation of saltation, especially for sandy soils (Chepil 1956, Belly 1964, McKenna Neuman and Nickling 1989, Sherman *et al.* 1998). For low relative humidities (below ~65 %), these interparticle forces are produced primatirly by bonding of adjacent adsorbed water layers (hygroscopic forces), whereas for high relative humidities (above ~65 %) this occurs primarily through the formation of water wedges around points of contact (capillary forces) (Hillel 1980, Ravi *et al.* 2006, Nickling and McKenna Neuman 2009).

Water adsorption is governed by electrostatic interactions of the mineral surface with the polar water molecules. Since sandy soils generally contain a lower density of net electric charges, substantially less water can be adsorbed onto sandy soils than onto clayey soils (Hillel 1980). Consequently, water bridges form in sandy soils at a relatively low soil moisture content, thereby producing substantial capillary forces (e.g., Belly 1964). McKenna Neuman and Nickling (1989) used this observation to derive an expression for the increase of the saltation fluid threshold for sand due to the presence of soil moisture. Fécan *et al.* (1999) then generalized the McKenna Neuman and Nickling expression to all soil types, obtaining the following empirical expressions

$$\frac{u_{*\mathrm{wt}}}{u_{*\mathrm{ft}}} = 1, \qquad (w < w') $$
$$\frac{u_{*\mathrm{wt}}}{u_{*\mathrm{ft}}} = \sqrt{1 + 1.21(w - w')^{0.68}} \qquad (w \geq w') \tag{4.1}$$

where $u_{*\mathrm{wt}}$ and $u_{*\mathrm{ft}}$ are respectively the fluid saltation thresholds in the presence and absence of soil moisture, and $w$ is the soil's volumetric water content in percent. Since water adsorbs more readily onto clay soils (Hillel 1980), the volumetric fraction of water that the soil can absorb before capillary forces become substantial, $w'$, increases with the soil's clay content. Using curve fitting of Eq. (4.1) to experimental results, Fécan *et al.* (1999) obtained

$$w' = 0.17 c_\mathrm{s} + 0.0014 c_\mathrm{s}^2, \tag{4.2}$$

where $c_\mathrm{s}$ is the soil's clay content in percent.

Although the empirical parameterization of Eqs. (4.1) and (4.2) ignores interparticle forces due to bonding of adsorbed water layers, which is likely substantial for soils with either a high clay content or a low soil moisture content (Hillel 1980, Ravi *et al.* 2004, 2006), this fairly straightforward parameterization seems to reasonably match measurements and is widely used to parameterize the effect of soil moisture on the dust emission threshold in atmospheric circulation models (e.g., Zender *et al.* 2003a).

*4.1.1.2 Effect of non-erodible roughness elements on the saltation fluid threshold*

In addition to soil moisture, the saltation fluid threshold is also affected by the presence of non-erodible roughness elements such as pebbles, rocks, and vegetation. Since the flow resistance of these objects is substantially larger than that of bare soil, they extract momentum from the wind. The presence of roughness elements therefore reduces the wind shear stress on the intervening bare soil and increases the total threshold wind stress required to initiate saltation and dust emission (Raupach *et al.* 1993).

The effect of roughness elements on the saltation fluid threshold depends on the partitioning of the total wind stress between the fraction absorbed by the bare soil ($\tau_S$) and the fraction absorbed by the roughness elements ($\tau_R$) (Schlichting 1936),

$$\tau = \tau_S + \tau_R . \tag{4.3}$$

The difficulty now lies in relating the partitioning of the fluid drag to the properties of the roughness elements. The absorption of fluid drag by roughness elements is largely determined by the frontal area presented to the flow (Marshall 1971), which is captured in the roughness density $\lambda$ (e.g., Raupach 1992),

$$\lambda = nbh, \tag{4.4}$$

where $n$ is the number of roughness elements per unit area, and $b$ and $h$ are the average width and height of the roughness elements. By first relating the fluid drag absorbed by roughness elements to the roughness density $\lambda$ and then assuming that the area and volume of the wakes of different roughness elements are randomly superimposed, Raupach (1992) derived an approximate expression for the fraction of the fluid drag absorbed by the bare soil,

$$\frac{\tau_S}{\tau} = \frac{1}{1+\beta\lambda}. \tag{4.5}$$

where $\beta$ is the ratio of the drag coefficients of a typical roughness element and the bare soil, which is on the order of ~100 for typical conditions. Raupach *et al.* (1993) then expanded the Raupach (1992) drag partitioning theory to estimate the effect of roughness elements on the saltation fluid threshold and obtained an expression for the ratio $R$ of the fluid threshold without and with roughness elements (Gillette and Stockton 1989):

$$R = \frac{u_{*ft,bare}}{u_{*ft,rough}} = \left[\frac{1}{(1-m\sigma\lambda)(1+m\beta\lambda)}\right]^{1/2}, \tag{4.6}$$

where $\sigma$ is the ratio of the basal area to the frontal area for the roughness elements, and $m$ ($0 < m \leq 1$) is a factor meant to account for the fact that the fluid saltation threshold of a homogeneous soil is not determined by the average shear stress required for particle mobilization, but rather by the maximum shear acting on any erodible point on the surface. Raupach *et al.* (1993) recommended $m \approx 0.5$ for flat, erodible surfaces based on comparisons with the experimental data of Gillette and Stockton (1989). Subsequent wind tunnel and field measurements of the drag partition and the ratio $R$ have generally confirmed the validity of Eqs. (4.5) and (4.6), although most experiments had to adapt the values of $\sigma$, $m$, and $\beta$ from the values proposed by Raupach (Wyatt and Nickling 1997, Crawley and Nickling 2003, King *et al.* 2005, Gillies *et al.* 2007).

Despite the reasonable agreement of Eqs. (4.5) and (4.6) with measurements, the roughness density $\lambda$ does not fully capture the dependence of the drag partition on the physical properties, dimensions, spatial distribution, and orientation of the roughness elements (Gillies *et al.* 2006, Nickling and McKenna Neuman 2009). For instance, the drag partition is affected by the porosity (Shao 2008) and Reynolds number dependency of the roughness element drag coefficients (Gillies et al. 2002), which is highly relevant for determining the effects of vegetation on the saltation fluid threshold (Okin 2005, 2008). Moreover, the spatial distribution of the roughness elements can play an important role in determining their effect on the fluid threshold. For example, Okin and Gillette (2001) found that the distribution of vegetation in a semi-arid mesquite landscape adapts in such a manner as to form "streets" of readily erodible bare soil, for which the saltation fluid threshold is substantially lower than would be predicted with Eq. (4.6). The non-uniformity of the spatial distribution of vegetation thus likely results in higher saltation and dust emission fluxes than would be predicted using the Raupach model (Okin 2005, 2008), which assumes a random distribution (Raupach 1992).

*4.1.2 The vertical dust flux*

When the saltation fluid threshold discussed in the previous sections is exceeded, saltation bombardment of the soil is initiated. As discussed above, dust aerosols are mainly emitted through the transfer of kinetic energy to dust aggregates through saltator impacts. The vertical flux of dust aerosols $F_d$ emitted by an eroding soil thus depends on (i) the flux $n_s$ of saltators impacting onto the soil, (ii) the average kinetic energy $\overline{E_s}$ delivered by these saltators, and (iii) the efficiency $\varepsilon$ with which this kinetic energy is converted to emitted dust aerosols. That is, the vertical dust flux can be expressed as

$$F_d = n_s \overline{E_s} \varepsilon, \tag{4.7}$$

where the dust emission efficiency $\varepsilon$ has units of kg/J and is defined as the average mass of dust aerosols produced by a unit of impacting energy. The number of saltation impacts on the soil surface per unit time and area $n_s$ can be derived from the balance of horizontal momentum in the saltation layer (Eqs. 2.20 and 2.21), which yields (Shao *et al.* 1993a)

$$n_s = \frac{\rho_a (u_*^2 - u_{*\text{sfc}}^2)}{m_s (\overline{v_{\text{imp},x}} - \overline{v_{\text{lo},x}})}, \tag{4.8}$$

where $m_s$ is the typical mass of saltators, $u_{*\text{sfc}}$ is the shear velocity at the surface as defined by Eq. (2.28), and $\overline{v_{\text{imp},x}}$ and $\overline{v_{\text{lo},x}}$ are respectively the mean horizontal components of the speeds with which saltators impact and lift off from the surface. Using that the lift-off speed of saltating particles is $\sim v_{\text{imp}}/2$ and that the impact and rebounding angles are $\sim 12°$ and $\sim 35°$ (see Section 2.1.3), Eq. (4.8) reduces to

$$n_s \cong \frac{2\rho_a (u_*^2 - u_{*\text{sfc}}^2)}{m_s \overline{v_{\text{imp}}}}, \tag{4.9}$$

where $\overline{v_{\text{imp}}}$ is the mean saltator impact speed. The final term in Eq. (4.7) is the mean impact energy of saltating particles $\overline{E_s}$. Since the saltating particle impact speed $v_{\text{imp}}$ approximately follows an exponential distribution (Kok 2010a, Durán *et al.* 2011a),

$$\overline{E_s} \approx m_s \overline{v_{\text{imp}}}^2, \tag{4.10}$$

where $\overline{v_{\text{imp}}}$ is the mean impact speed of saltating particles. Substituting Eqs. (4.8) – (4.10) into Eq. (4.7) and approximating $u_{*\text{sfc}}$ with $u_{*\text{it}}$ (Shao *et al.* 1993a) (see Section 2.3.2.3 for a discussion of this approximation) then yields

$$F_d = 2\rho_a (u_*^2 - u_{*\text{it}}^2) \varepsilon \overline{v_{\text{imp}}}. \tag{4.11}$$

As was the case for parameterizations of the saltation mass flux (Section 2.3.2.3), different assumptions for the dependence of $\varepsilon$ and $\overline{v_{\text{imp}}}$ on $u_*$ result in different relationships for the vertical dust flux $F_d$ as a function of $u_*$. Shao *et al.* (1993) inferred from their wind tunnel measurements that the dust emission efficiency $\varepsilon$ is constant with $u_*$ (at least for the loose dry soil used in their experiment), which was later confirmed by Rice *et al.* (1996). Shao *et al.* further assumed that $\overline{v_{\text{imp}}}$ is proportional to $u_*$, resulting in

$$F_d = C_S \rho_a u_* (u_*^2 - u_{*\text{it}}^2). \tag{4.12}$$

However, we now know from wind tunnel and field measurements that $\overline{v_{\mathrm{imp}}}$ is independent of $u_*$ for transport limited saltation (see Figure 2.12) and thus that the assumption that $\overline{v_{\mathrm{imp}}}$ is proportional to $u_*$ is incorrect. A more physical approximation of Eq. (4.11) is thus

$$F_{\mathrm{d}} = C_{\mathrm{F}} \rho_{\mathrm{a}} u_{*\mathrm{it}} \left( u_*^2 - u_{*\mathrm{it}}^2 \right), \tag{4.13}$$

where the proportionality constants $C_S$ and $C_F$ have units of kg / J, and where we used that $\overline{v_{\mathrm{imp}}} \propto u_{*\mathrm{it}}$ (see Section 2.3.2.3).

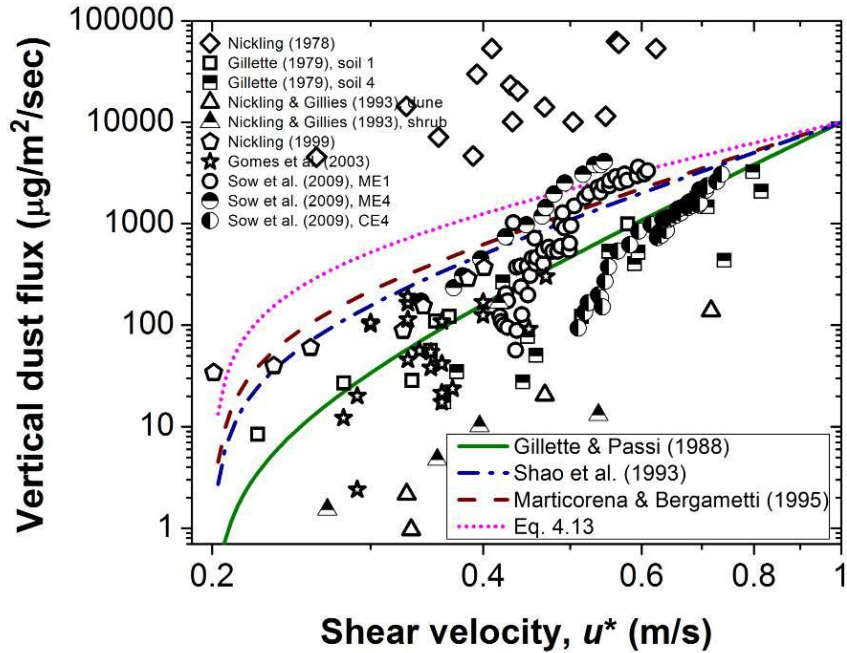

Figure 4.2. Compilation of field measurements of the vertical dust flux emitted by eroding soils. The large differences between the data sets are mostly due to variations in soil erodibility and saltation fluid threshold. Included for illustration are several vertical dust flux relations (lines), which use $u_{*\mathrm{it}} = 0.20$ m/s and are normalized to yield 10 mg/m$^2$/sec at $u_* = 1$ m/s.

An alternative approach to deriving an expression for $F_d$ is to assume that the vertical dust flux is proportional to the horizontal saltation flux (Marticorena and Bergametti 1995). This appears to be a reasonable assumption since the saltation flux and the amount of kinetic energy impacting on the soil surface have a similar scaling with $u_*$ (see Table 2.1 and Eq. (4.11)). This assumption results in (Gillette 1979, Marticorena and Bergametti, 1995)

$$F_{\mathrm{d}} = \alpha Q, \tag{4.14}$$

where $\alpha$ is the *sandblasting efficiency*, which is on the order of $10^{-5} - 10^{-2}$ m$^{-1}$ (Gillette 1979, Marticorena and Bergametti 1995, Gomes *et al.* 2003, Rajot *et al.* 2003). The equation for $Q$ used with Eq. (4.14) is most commonly that of Kawamura (1951) (see Table 2.1 and Marticorena and Bergametti 1995, Zender *et al.* 2003a), resulting in

$$F_{\mathrm{d}} = C_{\mathrm{K}} \frac{\rho_{\mathrm{a}}}{g} u_*^3 \left( 1 - \frac{u_{*\mathrm{it}}^2}{u_*^2} \right) \left( 1 + \frac{u_{*\mathrm{it}}}{u_*} \right), \tag{4.15}$$

where the proportionality constant $C_K$ has units of m$^{-1}$. Note that although a range of studies have indeed found that the vertical dust flux scales with the saltation mass flux (Gillette 1979, Shao et al. 1993a, Gomes et al. 2003), results from other investigators indicate that the vertical dust flux increases more rapidly with wind speed than the horizontal saltation flux (e.g., Nickling et al. 1999). A possible cause of these divergent results is that the dust emission efficiency $\varepsilon$ increases with $u_*$ for certain soils. Further research is required to resolve this issue.

In addition to Eqs. (4.12), (4.13), and
(4.15), an expression for $F_d$ based on an unpublished analysis by P.R. Owen was provided by Gillette and Passi (1988),

$$F_d = C_{GP} u_*^4 \left(1 - u_{*it}/u_*\right), \tag{4.16}$$

where the proportionality constant $C_{GP}$ has units of kg m$^{-6}$ s$^3$.

The different vertical dust flux equations are plotted in Figure 4.2, which also includes a compilation of field measurements. The large spread in the vertical dust flux measured in different field studies is indicative of both the experimental difficulties in measuring the vertical dust flux (Zobeck et al. 2006), and the large sensitivity of dust emission to differences in saltation fluid threshold and soil properties (Marticorena and Bergametti 1995, Shao 2008). We discuss the dependence of dust emissions on soil properties in the next section.

*4.1.2.1 Dependence of the vertical dust flux on soil properties*

Although the proportionality constants $C_S$, $C_F$, $C_K$, and $C_{GP}$ in Eqs. (4.12) – (4.16) are independent of $u_*$, these parameters depend on how efficiently impacting kinetic energy is converted into emitted dust aerosols. Specifically, soil properties such as the clay content (Marticorena and Bergametti 1995), the bonding strength of soil dust aggregates (Shao et al. 1993a, Rice et al. 1996), the dry aggregate size distribution (Chepil 1950, Zobeck 1991b), the presence of soil crusts (Belnap et al. 1998, Gomes et al. 2003), and the soil plastic pressure (Lu and Shao 1999) are thought to determine these proportionality constants. Unfortunately, the exact dependence of the vertical dust flux on these physical parameters is both poorly understood and requires detailed knowledge of soil properties that are not normally available on regional or global scales (e.g., Shao 2001, Laurent et al. 2008). Extensive further study is thus required to better relate the vertical dust flux to soil properties.

In order to parameterize observed regions of high dust emission potential despite these problems, atmospheric circulation models generally use empirical parameterizations (e.g., Marticorena and Bergametti 1995, Ginoux et al. 2001, Tegen et al. 2002, Zender et al. 2003a). In particular, the soil's ability to produce dust aerosols is often captured in an empirical *soil erodibility* function. Prospero et al. (2002) noted from multi-decadal observations by the Total Ozone Mapping Spectrometer (TOMS) onboard the Nimbus 7 satellite that areas with high dust loading coincided with topographic lows, which they hypothesized to be due to the fluvial deposition of fine-grained material in topographic lows. This finding was subsequently used to propose empirical soil erodibility parameterizations (Ginoux et al. 2001, Tegen et al. 2002, Zender et al. 2003b); for instance, Ginoux et al. (2001) proposed an empirical formulation of the soil erodibility based on the relative height of a model grid box compared to its surroundings. However, advection of dust plumes after emission causes the high dust loading observed by TOMS to be shifted downwind from dust source regions (Schepanski et al. 2007), which later studies indicate are actually concentrated in mountain foothill regions and dry lake beds (Schepanski et al. 2009). Although the use of soil erodibility parameterizations improves model agreement with measurements (Zender et al. 2003b, Cakmur et al. 2006), the empirical nature of these parameterizations suggests that improvements in understanding the dependence of the vertical dust flux on soil properties could bring corresponding improvements in simulations of the dust cycle. Indeed, current model simulations of the global dust cycle still show substantial discrepancies with measurements (e.g., Cakmur et al. 2006).

*4.1.3 The size distribution of emitted dust aerosols*
Accurately assessing the effects of mineral dust aerosols on the Earth and Mars systems requires not only a detailed understanding of what processes determine the vertical dust flux, but also an understanding of the size distribution of the emitted dust. Indeed, the size distribution of mineral dust governs the interactions with radiation, clouds, ecosystems, the cryosphere, and public health (Tegen et al. 1996, Goudie and Middleton 2006, Mahowald et al. 2011, Ito et al. 2012).

As discussed in Section 4.1, the dominant source of dust aerosols is their ejection from dust aggregates through the rupture of interparticle bonds by elastic waves created by mechanical saltator impacts. The exact pattern in which these interparticle bonds are ruptured then determines the size distribution of emitted dust aerosols. Unfortunately, the interparticle forces bonding dust particles to each other and to other soil components are due to a large variety of complex and poorly understood interactions (see Section 2.1.1.2) (Zimon 1982, Shao 2001, Castellanos 2005). Nonetheless, several theories for dust emission have been proposed that relate the size distribution of emitted dust aerosols to the wind speed and the soil size distribution.

*4.1.3.1 The Dust Production Model (DPM)*
The Dust Production Model (DPM) assumes that sandblasting results in the emission of three separate lognormal modes of dust aerosols, with the relative contribution of each mode determined by its bonding energy and the kinetic energy of impacting saltators (Alfaro *et al.* 1997, Alfaro and Gomes 2001). Using wind tunnel experiments with two separate soils, Alfaro *et al.* (1998) determined the median diameters of the three modes to be 1.5, 6.7, and 14.2 μm. However, recent field measurements suggest that the bonding energies and median diameter of these three lognormal modes might need to be adapted for specific soils (Sow *et al.* 2011).

Since the rupturing of interparticle bonds requires energy, the DPM predicts that larger saltating particle impact energies produce more disaggregated and thus smaller dust aerosols, as also suggested by experiments showing that glass broken at larger input energies produces a larger fraction of small fragments (Scheibel *et al.* 1990, Weichert 1991). Because the DPM further assumes that the saltator impact speed is proportional to wind speed, it predicts a shift to smaller aerosol sizes with increasing wind speed. Although the assumption that saltator impact speeds scale with $u_*$ conflicts with measurements for transport limited saltation (Figure 2.12), several wind tunnel studies indeed observed a shift to smaller aerosol sizes with increasing wind speed (Alfaro *et al.* 1997, 1998, Alfaro 2008). Consequently, the theory of Alfaro and Gomes (2001) is in agreement with these wind tunnel measurements (Figure 4.3a).

The possible dependence of the dust size distribution on the wind speed is discussed in more detail in Section 4.1.3.4.

*4.1.3.2 The Shao theory*
A second theory for the size distribution of emitted dust aerosols was formulated by Shao (2001, 2004) and is based on the insight that the emitted dust size distribution must be bound by the two extreme states of the soil: the undisturbed, minimally disaggregated state, and the disturbed, fully disaggregated state. That is,

$$p_d(D_p) = \gamma p_m(D_p) + (1-\gamma) p_f(D_p), \qquad (4.17)$$

where $p_d$, $p_m$, and $p_f$ are respectively the particle size distributions of the emitted dust aerosols, the minimally disturbed soil, and the fully disaggregated soil. Using the assumption that the saltator impact speed scales with $u_*$, Shao postulated a weighting factor $\gamma$ that increases monotonically with the energy of impacting saltators:

$$\gamma = \exp\left[-k(u_* - u_{*t})^n\right], \qquad (4.18)$$

where $k$ and $n$ are empirical coefficients that are determined from fitting of Eq. (4.17) and (4.18) to experimental data. An important similarity of the Shao model to the DPM is that it also predicts that an increase in $u_*$ results in more disaggregated, and hence smaller, dust aerosols. Like the DPM, the Shao (2001, 2004) model is in agreement with the Alfaro wind tunnel measurements (Figure 4.3a), although it requires the *n* and *k* parameters to be tuned accordingly. However, Shao *et al.* (2011b) recently found that a constant value of *γ* (i.e., independent of $u_*$) produced the best agreement with field measurements.

*4.1.3.3 Brittle fragmentation theory*
The theories of Alfaro and Gomes (2001) and Shao (2001, 2004) provide a framework for predicting the dust size distribution of emitted dust aerosols. However, the use of these theories for regional and global circulation model parameterizations of dust emission is limited by the poor understanding of the interparticle forces that determine both the energy required to rupture the cohesive bonds between soil particles (in the case of the DPM; see Sow *et al.* 2011) and the soil's aggregation state (in the case of Shao's theory). Moreover, the DPM and especially Shao's theory require knowledge of the soil size distribution, which is not normally available on regional and global scales.

As a way to circumvent the problem of the poorly understood interparticle forces and bonding energies and the resulting use of empirical coefficients, Kok (2011a) developed an alternative theory for the size distribution of emitted dust aerosols, assuming that most dust emission is due to brittle material fragmentation. This idea was previously noted by Gill *et al.* (2006) based on the observation from soil science that stressed dry soil aggregates fail as brittle materials (Lee and Ingles 1968, Braunack *et al.* 1979, Perfect and Kay 1995, Zobeck *et al.* 1999). Since the size distribution of fragments resulting from brittle material fragmentation is determined by the patterns in which cracks nucleate, propagate, and merge in the material (Astrom 2006), the analogy of dust emission with brittle material fragmentation mostly eliminates the need for a detailed understanding of interparticle forces to predict the emitted dust size distribution. Moreover, since the pattern with which cracks propagate and merge is largely *scale-invariant* (Astrom 2006), the size distribution of fragments produced by brittle material fragmentation follows a simple power law, with the exponent of the power law only weakly dependent on the brittle object's shape (Oddershede *et al.* 1993) and seemingly invariant to the material type (see Figure 1 in Kok 2011a). (Note that scale invariance occurs for many natural phenomena, including avalanches, the fragmentation of rocks and atomic nuclei, and earthquakes (Gutenberg and Richter 1954, Turcotte 1986, Bak *et al.* 1987, Bondorf *et al.* 1995, Astrom *et al.* 2004).)

Evidence that the fragmentation of dust aggregates is indeed a form of brittle material fragmentation is that the power law that describes the size distribution of fragments produced by brittle fragmentation also describes the size distribution of emitted dust aerosols in a limited size range (~2 – 10 μm; see figure 2 in Kok, 2011a). The emission of aerosols > 10 μm is depleted relative to this power law because of the finite propagation distance *λ* of the side branches of cracks created in the dust aggregate by a fragmenting impact (Astrom 2006). Conversely, the emission of aerosols < 2 μm is depleted relative to the power law because the discrete particles composing the dust aggregates have a typical size of ~1 μm, which inhibits the creation of smaller fragments. Since many dust aerosols are aggregates themselves (Okada *et al.* 2001, E. A. Reid *et al.* 2003, Chou *et al.* 2008), Kok assumed that the production of dust aerosols with size $D_d$ is proportional to the volume fraction of soil particles with size $D_s \leq D_d$ that can contribute to the formation of the aerosol. By combining this assumption with the size distribution resulting from brittle material fragmentation (reviewed in Astrom 2006), Kok (2011a) derived relatively straightforward analytical expressions for the number and volume size distribution of emitted dust aerosols:

$$\frac{dN_d}{d\ln D_d} = \frac{1}{c_N D_d^2}\left[1+\mathrm{erf}\left(\frac{\ln(D_d/\overline{D_s})}{\sqrt{2}\ln\sigma_s}\right)\right]\exp\left[-\left(\frac{D_d}{\lambda}\right)^3\right], \text{ and} \tag{4.19}$$

$$\frac{dV_d}{d\ln D_d} = \frac{D_d}{c_V}\left[1+\mathrm{erf}\left(\frac{\ln(D_d/\overline{D_s})}{\sqrt{2}\ln\sigma_s}\right)\right]\exp\left[-\left(\frac{D_d}{\lambda}\right)^3\right], \tag{4.20}$$

where $N_d$ and $V_d$ are respectively the normalized number and volume of dust aerosols with size $D_d$, $c_N = 0.9539$ μm$^{-2}$ and $c_V = 12.62$ μm are normalization constants, $\overline{D_s} \approx 3.4$ μm and $\sigma_s \approx 3.0$ are the median diameter by volume and the geometric standard deviation of the log-normal distribution that describes a "typical" arid soil size distribution in the PM20 size range, erf is the error function, and $\lambda \approx 12$ μm denotes the propagation distance of the side branches of cracks created in the dust aggregate by a fragmenting impact.

The brittle fragmentation theory applies only to dust emission events that are predominantly due to the fragmentation of soil aggregates. Therefore, Eqs. (4.19) and (4.20) are for instance not valid for (i) aerodynamically lifted dust, (ii) dust emitted mainly by impacts in the damage regime (Kun and Herrmann 1999), which could occur either for very cohesive soils or for sandy soils where most of the PM20 dust exists as coatings on larger sand grains (Bullard *et al.* 2004), and (iii) dust with diameters larger than ~20 μm, which are more likely to occur as loose particles in the soil (Alfaro *et al.* 1997, Shao 2001), such that their emission is not predominantly due to fragmenting impacts. Despite these limitations, the good agreement of Eqs. (4.19) and (4.20) with 7 different data sets of field measurements (Figure 4.3b) suggests that many natural dust emission events are due to aggregate fragmentation and can thus be reasonably described by brittle fragmentation theory. Further research is required to investigate this issue.

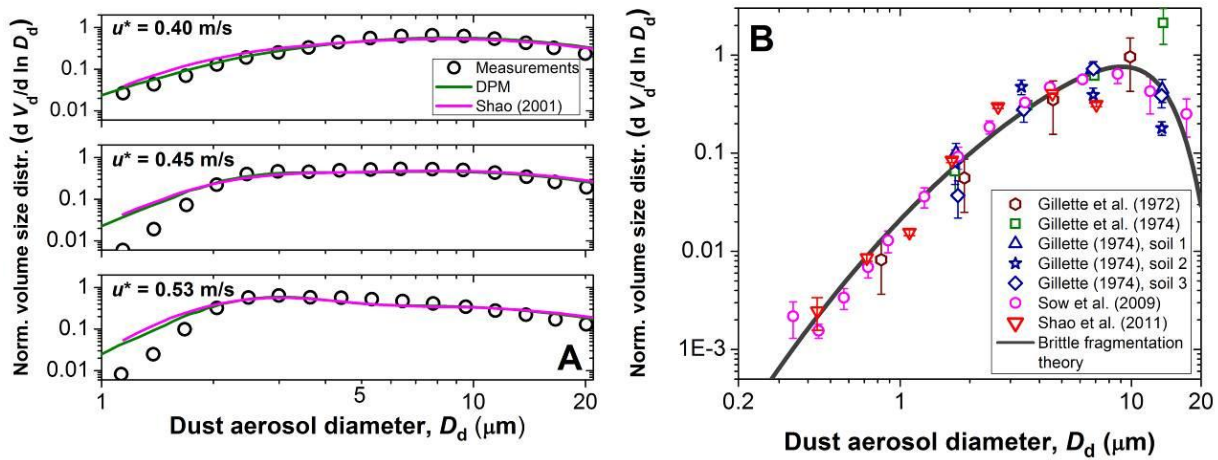

Figure 4.3. (**a**) Wind tunnel measurements of the size distribution of dust aerosols produced by saltation bombardment of a clay substrate (symbols; Alfaro *et al.* 1997). Both the DPM (green line) and the theoretical model of Shao (magenta line) can reproduce these measurements after tuning of the model parameters. (**b**) Compilation of field measurements of the size distribution of emitted dust aerosols (assorted symbols). The brittle fragmentation theory of dust emission (solid line) is in good agreement with these measurements, including the data set of Shao *et al.* (2011b), which was published subsequent to the brittle fragmentation theory (Kok 2011a). Note that the emitted dust size distribution predicted by the theories of Alfaro and Gomes (2001) and Shao (2001, 2004) cannot be included on this plot because they depend on both the soil size distribution and the wind speed. The processing of the field measurements is described in Kok (2011a).

*4.1.3.4 Dependence of emitted dust size distribution on wind and soil conditions*
Measurements and theories have yielded conflicting results on whether the dust size distribution depends on the wind speed at emission. Whereas the Alfaro and Gomes (2001) and Shao (2001, 2004) theories predict a decrease in the mean dust aerosol size with $u_*$, the brittle fragmentation theory of dust emission (Eqs. (4.19) and (4.20)) predicts that the size distribution is independent of $u_*$. Similarly, the wind tunnel measurements of Alfaro and colleagues (Alfaro *et al.* 1997, 1998, Alfaro 2008) reported a strong dependence of the emitted dust size distribution on $u_*$, whereas some field measurements have not found such a dependence (Gillette *et al.* 1974, Shao *et al.* 2011b). The study of Kok (2011b) attempted to settle this issue using a statistical analysis of the mean aerosol diameter of published field measurements (Gillette 1974, Gillette *et al.* 1974, Sow *et al.* 2009, Shao *et al.* 2011b). This study showed that individual data sets had opposing and statistically insignificant trends with $u_*$. Moreover, a compilation of all published field measurements showed no statistically significant trend with $u_*$, indicating that the size distribution of emitted dust aerosols does not depend on the wind speed at emission. A similar conclusion was reached by Reid *et al.* (2008), based on the similarity of measured size distributions of dust advected with different wind speeds at emission.

Kok (2011b) also offered a possible explanation for the discrepancy between field measurements on the one hand and wind tunnel measurements and the theories of Alfaro and Gomes (2001) and Shao (2001, 2004) derived from them on the other hand. The measurements of Alfaro *et al.* (1997, 1998) were probably not in steady state since the wind tunnel was only 3.1 meters in length (Alfaro *et al.* 1997, p. 11243). Consequently, the saltator impact speed could have increased with $u_*$, which in turn could have produced a shift to smaller aerosols as proposed by Alfaro and colleagues. Similarly, the dependence of the emitted dust size distribution on $u_*$ in the DPM and Shao theories is due to the assumption in these models that the saltator impact speed scales with $u_*$, which measurements indicate is incorrect for transport limited steady state saltation (Figure 2.12).

All three theories reviewed above also include a dependence on the soil size distribution, and many regional and global circulation models also account for a soil dependence of the emitted dust size distribution (e.g., Ginoux *et al.* 2001, Laurent et al. 2008). However, the small amount of scatter between the dust flux measurements compiled in Figure 4.3b, despite the widely varying soil types for which these measurements were made, suggests that changes in soil conditions have only a limited effect on the emitted dust size distribution. A similar conclusion is suggested by the insensitivity of dust aerosol size distributions to changes in the source region (Reid *et al.* 2008). More research is clearly needed to better quantify the influence of the soil size distribution on the emitted dust size distribution, but the apparently limited dependence of the emitted dust size distribution to the soil size distribution is highly fortuitous for regional and global dust modeling of the dust cycle.

*4.1.4 Deposition of dust aerosols*

After emission, mineral dust aerosols are removed from the atmosphere by either *dry deposition* or *wet deposition*. Dry deposition is due to the combined action of gravitational settling with turbulent diffusion in the atmospheric boundary layer and molecular diffusion in the laminar sublayer near surfaces, such as vegetation canopies (Slinn 1982). Wet deposition includes both in-cloud scavenging, in which dust aerosols serve as cloud condensation or ice nuclei and subsequently precipitate (DeMott *et al.* 2003, 2010), and below-cloud scavenging (Jung and Shao 2006), in which precipitating raindrops collect dust aerosols. Wet deposition generally dominates for aerosols smaller than ~5 μm in diameter, whereas dry deposition dominates for aerosols larger than ~5 μm (Woodward 2001, Zender *et al.* 2003a, Miller *et al.* 2006). The resulting lifetime of a dust aerosol decreases with its size and ranges from ~1-2 weeks for clay aerosols (with diameter < 2 μm), to several hours or days for silt aerosols (>2 μm) (Tegen and Lacis 1996, Zender *et al.* 2003a, Miller *et al.* 2006). Consequently, only aerosols smaller than ~20 μm in diameter remain suspended in the Earth's atmosphere for sufficient time periods to substantially affect weather and climate (e.g., Tegen and Lacis 1996).

*4.2 Dust emission physics on Mars*

Orbital and landed spacecraft observations have made it obvious that airborne and surficial dust is an important and ubiquitous aspect of the present martian environment (e.g., Landis *et al.* 2000, Tomasko *et al.* 1999). Aeolian dust emission processes on Mars are in many ways similar to those in the most arid locales on Earth. However, such processes on Mars are not modulated in a significant way by either liquids or vegetation, and the emitted dust has a unique martian character (e.g., Morris *et al.* 2000).

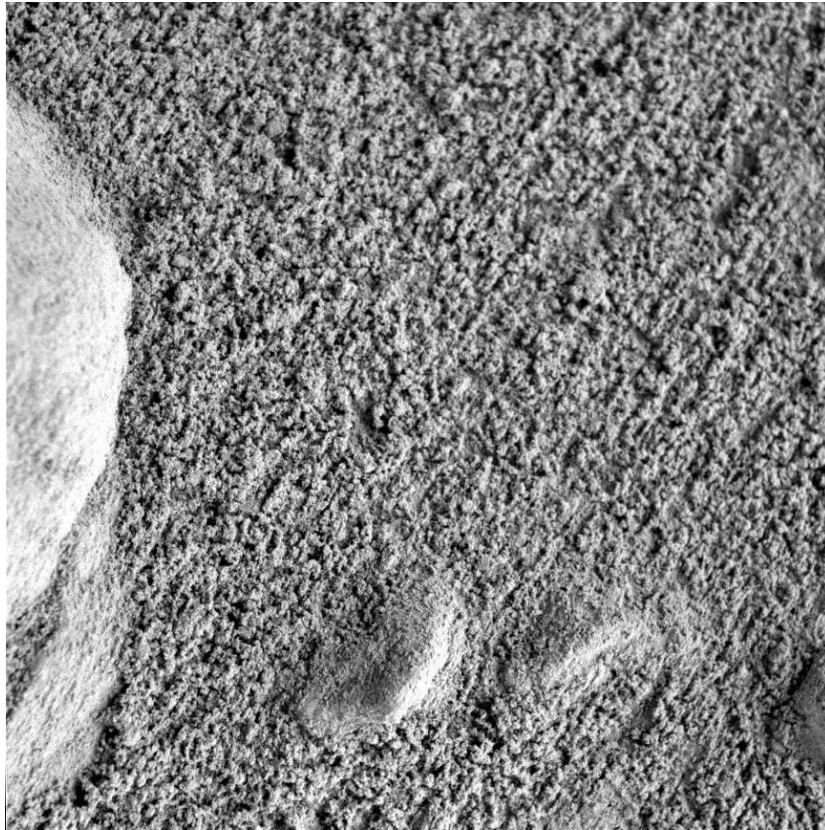

Figure 4.4. Mars Exploration Rover Spirit image (~31x31 mm) from its Microscopic Imager instrument, illustrating surface agglomerates composed of air-fall dust that likely formed via the interaction of wind and electrostatic forces (e.g., Sullivan *et al.* 2008). The oriented texture in this image is likely due to the dust agglomerates gradually growing in size as they are gently rolled through the surrounding air-fall dust by the wind. This spatial texture is finer on and near the rocks in this image, perhaps since the increased exposure to the wind in such locations reduces the time that such a growth process can act before the dust agglomerates are entrained into the flow. Acquired on Sol 70 within Gusev Crater, Mars. Image credit: NASA/JPL/Cornell/USGS.

*4.2.1 Mars surface dust emission*

Plausible surface dust emission processes on contemporary Mars are numerous, and it is probable that all may occur at some location and time on the planet (either singly or in concert). Mars' often ultra-arid surface environment and known populations of micrometer-scale particles (e.g., Wolff *et al.* 2006) suggests that electrostatic, magnetic, and even van der Waals forces may play important roles in modulating dust emission from Mars' surface. Similar to Earth (Section 4.1), atmospheric dust entrainment induced by sand-sized particle saltation is likely a major emission process on Mars (e.g., Greeley 2002). Relevant *in situ* observations have been obtained by the twin NASA Mars Exploration Rovers (MERs), specifically by their optical Microscopic Imager (MI) instruments, definitively identifying solid sand-sized particles that could participate in such a dust entrainment process (e.g., Soderblom *et al.* 2004). Such particles have been generally inferred to exist on much of Mars' surface (e.g., due to the presence of dunes, see Section 3.3). The recent result that the impact threshold for saltation is approximately an order of magnitude below the fluid threshold (Section 2.4 and Figure 2.18b) would allow saltation-induced dust emission to occur for substantially lower wind speeds than previously thought (Almeida *et al.* 2008, Kok 2010b).

In addition to traditional saltation-induced dust emission, the MER MI instruments discovered that the dust itself often formed low-strength sand-sized spheroidal agglomerates (see Figure 4.4, possibly due to electrostatic processes (e.g., Sullivan *et al.* 2008). Such dust structures may potentially be disrupted by aerodynamic stresses that are significantly less than those required to initiate saltation of a plausible solid sand-sized particle, injecting dust into the atmosphere without a need for stronger winds to initiate saltation. Conditions within dust devils (see Section 5.3.2) may also enhance the abilities of those vortices to remove dust from the surface (e.g., Balme and Hagermann 2006, Neakrase and Greeley 2010b).

Due to the large vertical thermal gradient capable of existing in the upper millimeter of the ground, thermophoresis has also been advanced as a possible mechanism (alone or in conjunction with another process) for dust injection into the atmosphere (e.g., Wurm *et al.* 2008). In this context, thermophoresis occurs when solar insolation warms the upper millimeter of the surface (dust particles and pore space gas molecules) much more than the atmospheric gas molecules directly above. The higher temperature pore space gas molecules can transfer enough momentum so that a subset of the surface dust particles are able to jump from the surface and be entrained by the atmosphere (overcoming gravity, the competing momentum transfer from the cooler atmospheric gas molecules, and any interparticle cohesive forces). Rapid sublimation of carbon dioxide ice that underlies or surrounds (e.g., as in dirty ice) dust particles may also provide the impulse required to inject dust into the atmosphere (e.g., Holstein-Rathlou *et al.* 2010).

*4.2.2   Properties of martian dust*
The average chemical composition of airborne and surficial dust on Mars is thought to be approximately constant, since: 1) comparisons between landed and orbital spacecraft measurements indicate few, if any, significant differences in gross dust composition (cf., Bell, III, *et al.* 2000, Christensen *et al.* 2004, Bandfield and Smith 2003, Aronson and Emslie 1975), and 2) the unfocused redeposition of dust on the surface after planet-encircling dust storms and other forms of long-distance atmospheric mixing and transport will tend to spatially homogenize the dust composition over time. Mechanical weathering (e.g., impact of saltating particles, thermal cycle stresses) of basaltic rocks across Mars is the presumed genesis of the majority of the dust (e.g., Pleskot and Miner 1981, Christensen 1988), although minor chemical alteration to the source rocks and/or generated dust may occur (e.g., Hamilton *et al.* 2005). One plausible bulk composition of the current Mars dust suggests composite particles made up of a large fraction of framework silicate minerals (such as plagioclase feldspar and zeolite) with lesser amounts of iron oxides such as hematite and magnetite, along with many other plausible minor components (Hamilton *et al.* 2005). The iron oxides give Mars its reddish hue and cause much of the dust to be magnetic to varying degrees (e.g., Goetz *et al.* 2005). Similar to terrestrial dust, martian dust likely exhibits a quasi-continuum of particle diameters, ranging from ten micrometers to much less than one micrometer – consistent with available *in situ* observational studies such as Goetz *et al.* (2010).

# 5. Atmospheric dust-entrainment phenomena

As discussed in the previous chapter, dust emission is generated through saltation bombardment by winds that exceed the saltation threshold. After discussing the main dust source regions on Earth in the next section, we discuss the main meteorological features forcing dust emission on Earth (Section 5.1.2) and Mars (Section 5.2). Section 5.3.2 also discusses a special case of dust lifting phenomena: small tornado-like vortices called *dust devils* that occur on both Earth and Mars.

## 5.1 Dust storms on Earth
### 5.1.1 Main source regions

Satellite observations have shown that, on a global scale, the dominant sources of natural mineral dust aerosols are located in the Northern Hemisphere and form what is called the 'Afro-Asian belt' of deserts or the 'dust belt' (Figure 5.1a). It stretches from the west coast of Africa all the way to Mongolia and China through the Middle-East, Iran, Afghanistan and Pakistan (Middleton, 1986a, b; Prospero *et al.*, 2002; Leon and Legrand, 2003). Within this dust belt, major dust activity is evident in the Sahara desert of North and West Africa, in the Arabian Peninsula, in the Middle East and in southwest and central Asia including Iran, Turkmenistan, Afghanistan, Pakistan, Northern India, the Namib and Kalahari deserts, the Gobi desert in Mongolia and the Tarim Basin in China (e.g. Herman *et al.* 1997; Torres *et al.* 1998). Some minor dust activity occurs in the Great Basins (Western United States), in Mexico, in central Australia, in southern Africa and in Bolivia (e.g. Engelstaedter *et al.*, 2006, Huneeus *et al.* 2011). Global circulation models estimate the global dust emission rate of PM20 dust to equal around 1000 – 5000 Tg/year (Huneeus *et al.*, 2011), although some models might underestimate the global dust emission rate due to an underestimation of the emission of silt aerosols (Kok 2011a).

Most active dust source areas in the Sahara are located in the foothills of mountain ranges, where fluvial abrasion provides sediments and deflatable materials (e.g. Middleton and Goudie, 2001, Mahowald *et al.*, 2003, Zender *et al.*, 2003a; Schepanski *et al.*, 2009). Also, dry lake beds, where fine sediments are abundantly present, constitute important dust source areas (Goudie and Middleton, 2001, Prospero *et al.*, 2002, Mahowald *et al.*, 2003). In particular, the area of the paleo-lake of Chad, known as the Bodélé Depression, is considered as the most important source of dust emissions in the Sahara (Prospero *et al.*, 2002; Washington *et al.*, 2003).

In Central and Eastern Asia, the major source regions are the Tarim Basin (Taklimakan Desert) in western China, the upper reach of the Yellow River (Gobi Desert) in southern Mongolia and northwestern China, the east part of Inner Mongolia, and the northern part of the Indian Subcontinent (Figure 5.1b; Sun *et al.* 2001).

### 5.1.2 Main meteorological features forcing dust emission

Based on the World Meteorological Organization protocol, dust events are classified according to visibility into one of four categories (Shao, 2008). The first of these, *dust haze*, consists of aeolian dust particles homogeneously suspended in the atmosphere. These are not actively entrained, but have been uplifted from the ground by a dust event that occurred prior to the time of observation or from a considerable distance. Visibility may sometimes be reduced to 10 km. *Blowing dust* is the state where dust is emitted locally through strong winds and at the time of observation, reducing visibility to 1 to 10 km. A *dust storm* is the result of strong turbulent winds entraining large quantities of dust particles, reducing visibility to between 200 meters and 1km. And finally, a *severe dust storm* is characterized by very strong winds that uplift large quantities of dust particles, reducing visibility to less than 200 m.

All atmospheric phenomena forcing dust emission must generate strong wind speeds that exceed the local dust emission threshold (see Sections 2.1.1 and 4.1.1). The atmospheric phenomena that produce dust events are of a variety of scales, including synoptic, regional, local, as well as turbulent scales. In ascending order of their length scales, important meteorological phenomena able to provide atmospheric conditions suitable for dust mobilization include dust devils (see Section 5.3.2), the downward turbulent mixing of momentum from nocturnal low-level jets (LLJs), density currents associated with moist-convection, cyclones, and synoptic scale cold fronts. We discuss these phenomena in the subsequent sections.

(a)
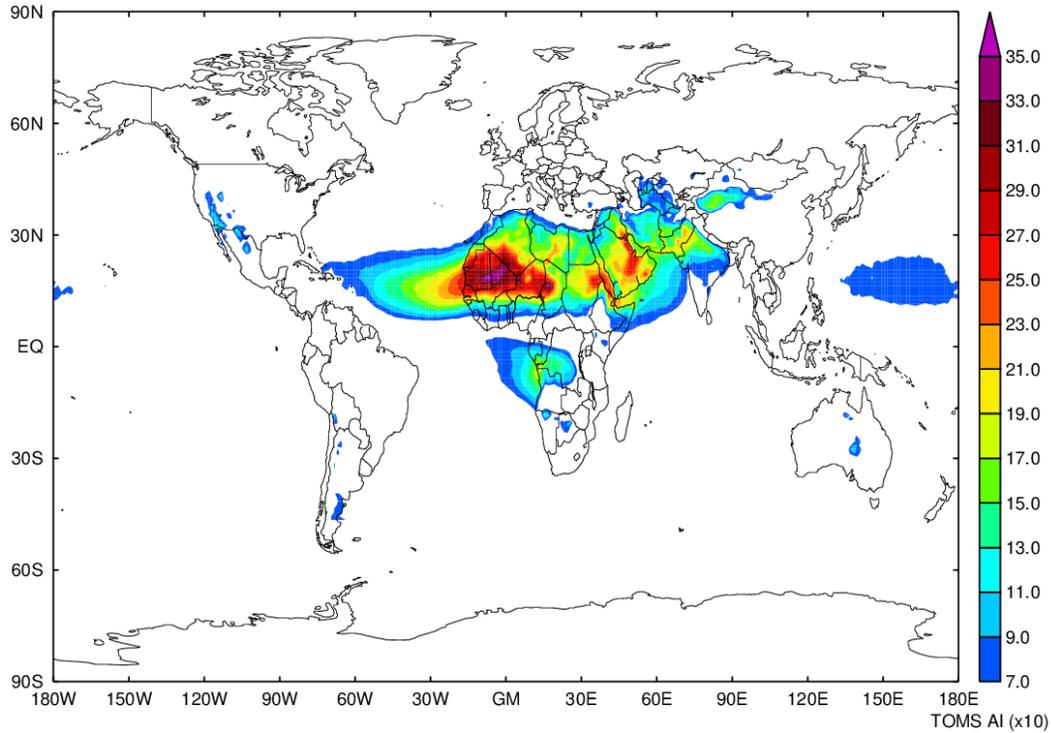

(b)
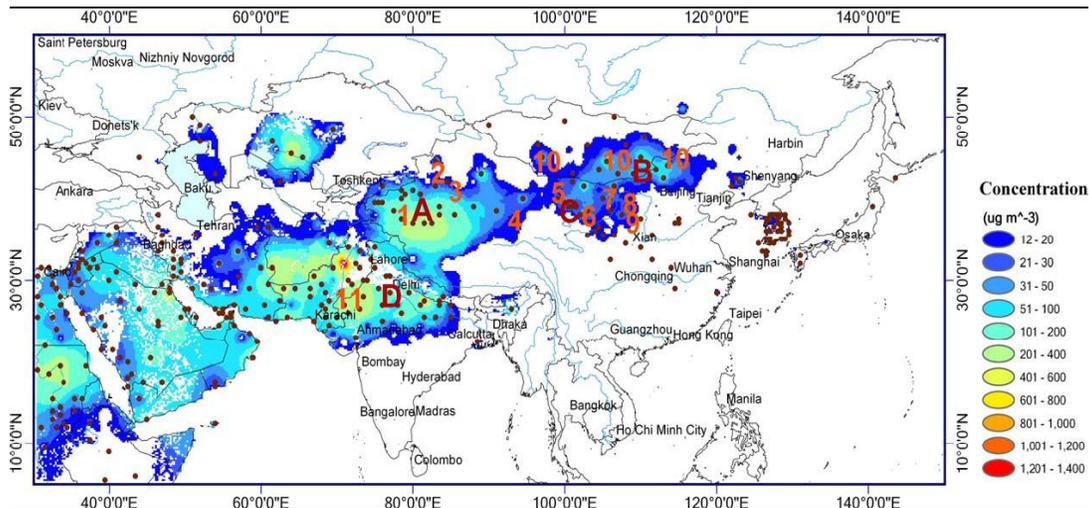

Figure 5.1: (a) May-July seasonal mean for the period 1980–1992 of Aerosol Index (AI) derived from the Total Ozone Mapping Spectrometer (TOMS) satellite observations showing the main dust sources on global scale forming the dust belt (after Engelstaedter *et al.*, 2006). (b) Mean dust concentration (averaged over time) for Asia. Data used for this graph are derived from visibility observations from 27 May 1998 to 26 May 2003. Main deserts in region are enumerated: 1 Taklamakan (Tarim Basin); 2 Gurbantunggut (Junggar Basin); 3 Kumutage; 4 Tsaidam Basin; 5 Badain Juran; 6 Tengger; 7 Ulan Buh; 8 Hobq; 9 Mu Us; 10 Gobi and 11 Thar. Four regions of frequent dust events, i.e., the Tarim Basin, Inner Mongolia, the Gobi region and the Indian Subcontinent, are denoted with A, B, C and D, respectively. Reprinted from Shao and Dong (2006), with permission from Elsevier.

*5.1.2.1 Nocturnal Low-Level Jet (LLJ)*

LLJs are horizontal winds characterized by a maximum of about 15 m s$^{-1}$ in the frictionally-decoupled layer immediately above the surface layer (e.g. Blackadar, 1957, Holton, 1967). They are most commonly observed at nighttime and can occur over all continents above both flat and complex terrain and may extend over tens to hundreds of kilometers (May, 1995; Davis, 2000; Schepanski *et al.*, 2009). During nights with low surface wind speeds, near-surface air layers are well stratified and turbulence is suppressed. In such conditions, air layers above the near-surface are frictionally decoupled from the surface and hence are associated with high wind speeds (e.g. Hoxit, 1975, Garratt, 1992, Mauritsen and Svensson, 2007, Mahrt, 1999).

After sunrise, convective turbulence arises with the onset of solar heating at the surface. The decoupled air layer aloft becomes frictionally coupled to the surface, and LLJ momentum is mixed down (Blackadar, 1957, Lenschow and Stankov, 1979). As a consequence, high surface wind speeds occur until the LLJ is eroded by the convection in the planetary boundary layer set on by the solar heating at the surface. The LLJ peaks at nighttime while surface wind peaks in mid-morning when LLJ momentum is mixed down to the surface (Figure 5.2, Washington *et al.*, 2006, Todd *et al.*, 2008). Over the Sahara, LLJs occur under clear skies and low surface wind speed conditions (e.g. Thorpe and Guymer, 1977) and have been observed within the northeasterly 'Harmattan' flow (Washington and Todd, 2005). About 65% of dust activity over the Sahara is due to the break-down of the nocturnal LLJs (e.g. Schepanski *et al.*, 2009). In particular, dust emission throughout the year in the Bodélé region in Chad is mainly related to LLJs dynamics (Washington *et al.*, 2006).

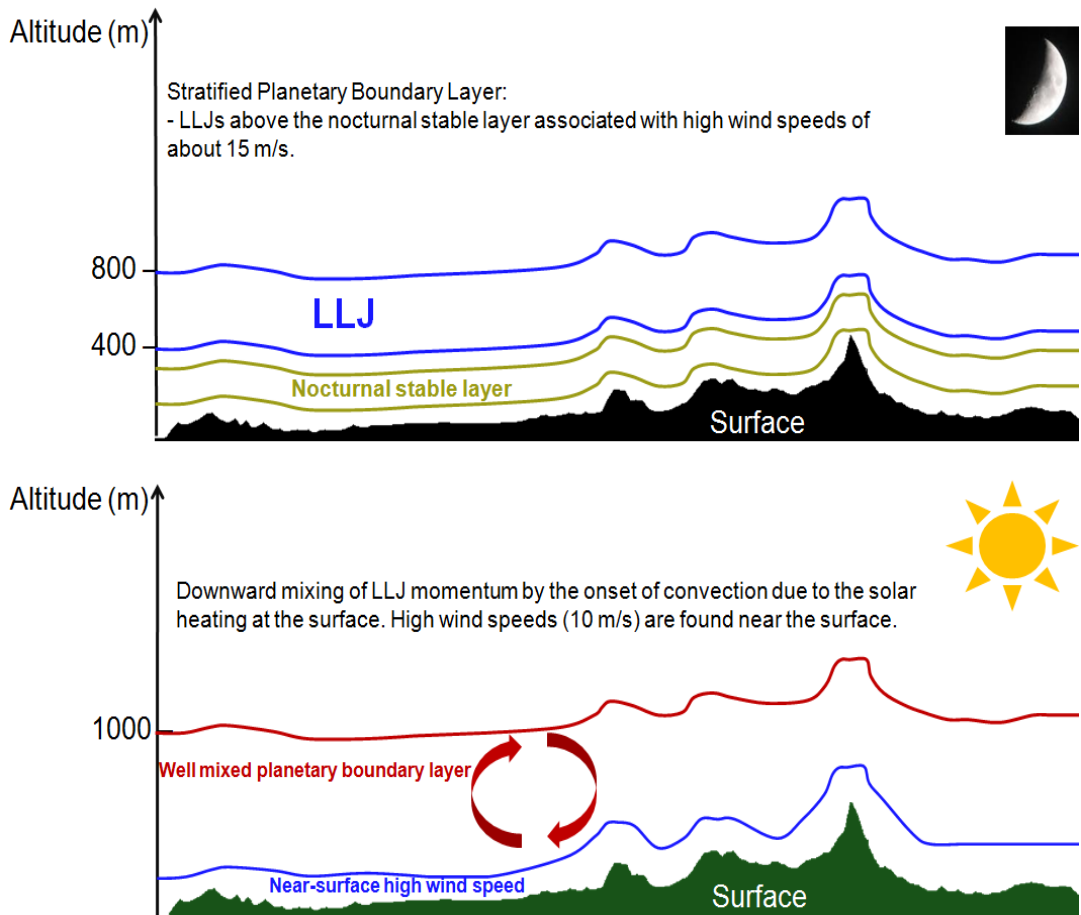

Figure 5.2: Illustration of Low Level Jets (LLJs) during the night (top) and the associated high near-surface wind speeds during the morning hours (bottom). The lines represent streamlines of the atmospheric flow in the different atmospheric layers.

*5.1.2.2 Moist convection*

Deep moist convection and the associated downdrafts of cold, humid air is another meteorological phenomenon capable of generating the high surface wind speeds required for dust emission. Such outflows take the form of density currents and can propagate many hundreds of kilometers away from the parent convective system (Simpson, 1997; Williams et al; 2009). Over arid and semiarid areas, these density currents cause strong dust emission (of the order of 1 Tg per event (Bou Karam *et al.*, 2011)). Over North and West Africa, such events are called 'haboobs' (Figure 5.3) and are observed during the summer monsoon season. Haboobs are frequent during afternoons and evenings, when moist convection is at maximum (Sutton, 1925, Idso *et al.*, 1972, Droegemeier and Wilhelmso, 1987, Flamant *et al.*, 2007, Bou Karam *et al.*, 2008, 2011).

Deep moist convection can also occur in mountain areas due to blocking of the atmospheric flow by the orography. Large scale density currents can form resulting from the subsequent evaporative cooling of cloud particles or precipitation. The high surface wind speeds associated with these density currents can lead to dust emission over arid and semiarid areas (Droegemeier and Wilhelmson, 1987). Over the Sahara, such dust events can be observed for example in the Atlas Mountains (Knippertz *et al.*, 2007) and in the Hoggar Mountains domain (Cuesta *et al.*, 2010).

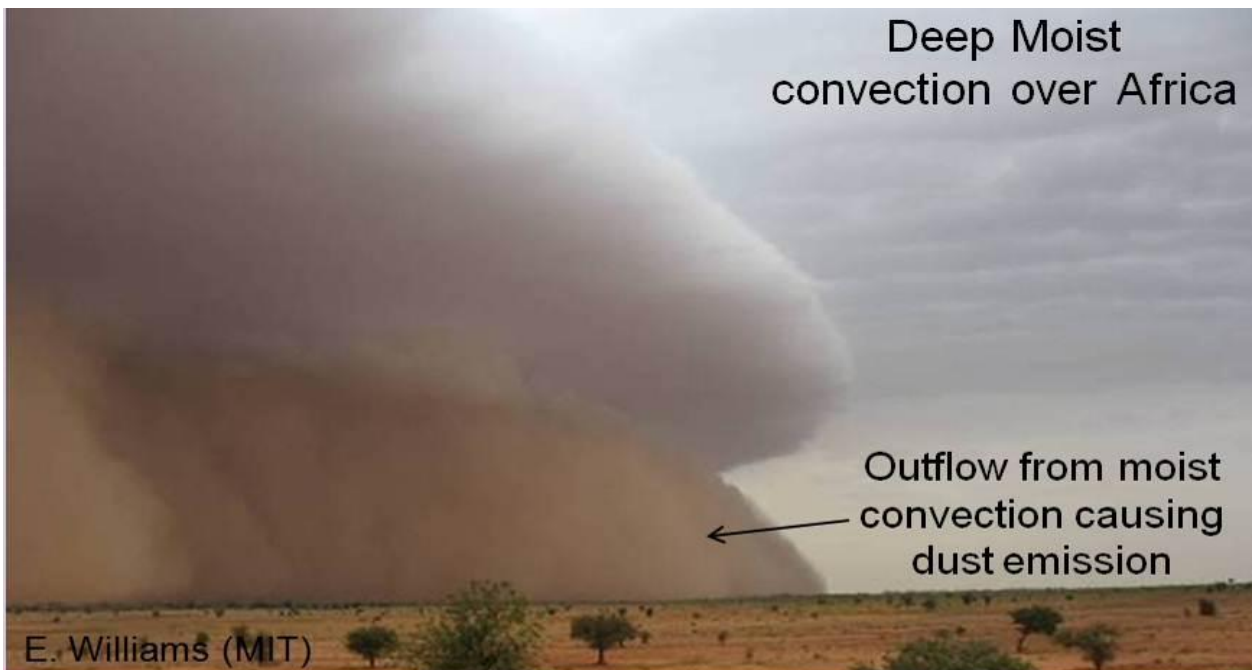

Figure 5.3: A density current emanating from a deep mesoscale convective system over Niger in summer 2006 and the associated dust emission.

*5.1.2.3 Cyclones*

Cyclones are considered as major dynamic features for the mobilization and the upward mixing of dust by virtue of the associated strong near-surface cyclonic winds and convective turbulence (e.g. Liu et 2003; Bou Karam *et al.*, 2010). This mechanism operates in two steps; first the dust is mobilized by strong momentum from the cyclonic surface wind and then it is mixed upward to high altitudes by the strong turbulence and the systemic ascending motion associated with the cyclone dynamics (Liu *et al.*, 2003; Bou Karam *et al.*, 2010). For example, the Mongolian Cyclone (Figure 5.4a), affects eastern Asia in March, April and May. This intense system is associated with the East Asian trough (a trough is an elongated region of low atmospheric pressure, often associated with fronts) and leads to strong northwesterly near-surface winds up to 18 m s$^{-1}$, thereby generating dust storms (Shao and Wang, 2003). In general, Asian dust storms are mainly associated with cyclonic cold fronts during the surges of cold continental air masses in late winter and early spring when most of the area is under the influence of the powerful Siberian–Mongolian anticyclone (e.g. Watts, 1969; Littmann 1991).

Cyclones are also observed over the Sahara desert in spring months when temperature gradients between the North African coast and the Mediterranean Sea are strongest (Pedgley, 1972, Alpert and Ziv, 1989, Trigo *et al.*, 2002). Observations indicate a frequent initiation of Saharan cyclones on the leeward side east and south of the Atlas Mountains (Barkan *et al.*, 2005, Prezerakos *et al.*, 1990, Alpert *et al.*, 1990, Alpert and Ziv, 1989). Hence, both the lee-effect of the mountains and the coastal thermal gradient effect can explain the spring cyclogenesis, initiated by the presence of an upper-level trough to the west (e.g. Horvath *et al.*, 2006, Egger *et al.*, 1995; Schepanski *et al.*, 2009; Bou Karam *et al.*, 2010). Saharan cyclones are characterized by an active warm front, heavy dust storms and low visibilities, and a sharp cold front that is well defined at the surface by changes in temperature of 10-20 K (5.4b). They move quickly eastward (faster than 10 ms-1) mostly following the North African coast (Alpert *et al.*, 1990, Alpert and Ziv, 1989; Bou Karam *et al.*, 2010). The dust load associated with Saharan cyclones is estimated to be of the order of 8 Tg per event (Bou Karam *et al.*, 2010).

(a)  (b)

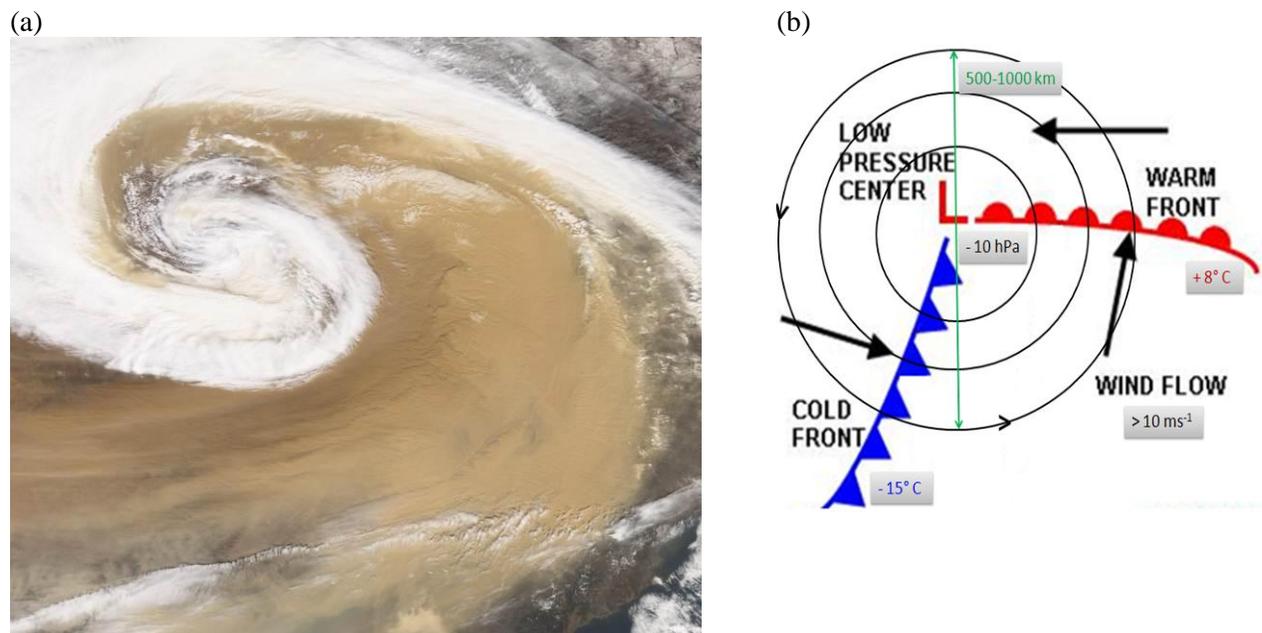

Figure 5.4: (a) Satellite image of the Mongolian dust Cyclone on 7 April 2001 taken by the MODIS (Moderate Resolution Imaging Spectroradiometer) on NASA's Aqua satellite. The cyclone measured around 3 km vertically and up to 2000 km horizontally. (b) Illustration of the main characteristics of the (Northern Hemisphere) Saharan Cyclone (Bou Karam *et al.*, 2010).

### 5.1.2.4 *Synoptic scale cold fronts*

Large and persistent outbreaks of dust over arid and semiarid regions can also be caused by cold fronts of synoptic scale (e.g., Figs. 1.1a and 5.5). These cold fronts mainly originate from the penetration of upper-level troughs from high latitudes into low latitudes (e.g. Jankowiak and Tanré, 1992; Knippertz and Fink, 2006; Tulet *et al.*, 2008; Liu *et al.*, 2003). The dust fronts generated during these events are related to density currents caused by strong evaporative cooling along the precipitating cloud-band that accompany the penetration of the upper-level cold front into arid regions (e.g. Knippertz and Fink, 2006). During such storms, the dust is pushed ahead of the cold front into the rising warm air over the desert. Over northern China, most of the dust storms during spring are caused by cold fronts that form in connection with the dynamic of the Siberian High.

Furthermore, cold-front dust emission can occur together with the surge of monsoon flows from the ocean toward the continent. Bou Karam *et al.* (2008) have identified this new mechanism for dust emission over West Africa during the summer monsoon season. Highly turbulent winds at the leading edge of the monsoon nocturnal flow in the Inter Tropical Front ((ITF), the ITF marks the interface in the lower atmosphere between the moist southwesterly monsoon flow and the hot and dry northeasterly harmattan flow over Africa) region have been identified to generate dust uplifting. Also, Marsham *et al.*, (2008) have described, via in situ measurements, the characteristics of the dusty layer at the leading edge of the monsoon flow over West Africa. Bou Karam *et al.*, (2009) have estimated dust emission associated with the nocturnal monsoon flow over Niger to be of the order of 0.7 Tg per day. This mechanism is most active during the night when the monsoonal surge usually occurs.

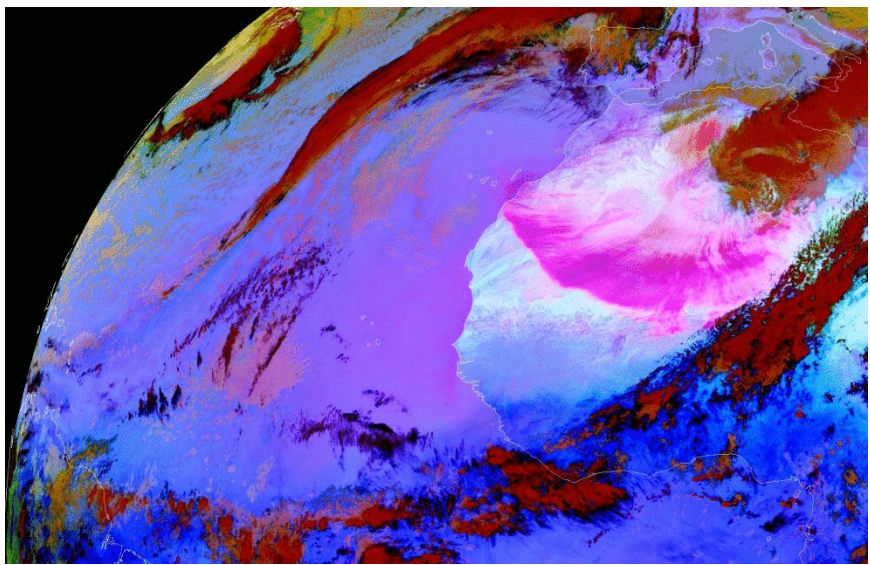

Figure 5.5: False-color image derived from the Spinning Enhanced Visible and Infrared Imager (SEVIRI) satellite observations on March 2006 showing a synoptic cold front and the associated dust storm (pink and magenta colors) over Africa.

## 5.2  Dust storms on Mars

The vast dust storms that sporadically occur on Mars play an important role in defining the planet's climate and surface expression (Kahn *et al.* 1992). Compared to aeolian dust storms on Earth, those on Mars are often larger, deeper (40 km or more), and longer-lasting (tens of Mars-days) (e.g., Kahn *et al.* 1992, Cantor *et al.* 2001). These phenomena are also an integral part of Mars' seasonal dust cycle, which strongly modulates the planet's climate, by significantly altering the absorption and/or emission of infrared and visible radiation by the atmosphere and surface (e.g., Gierasch and Goody 1968, Kahn *et al.* 1992). The following subsections explore the general processing and transport of atmospheric dust within Mars' atmosphere and the current understanding of the (more visually explicit) dust storms.

### 5.2.1  Atmospheric transport and processing of martian dust

Once airborne, dust on Mars is acted upon by three primary genres of physical processes: 1) direct gravitational sedimentation, 2) particle scavenging and alteration by water- and $CO_2$-ice cloud microphysical processes, and 3) transport and mixing by many scales of atmospheric circulations. A continuous dust loading within the lower atmosphere which varies on seasonal and faster time scales (e.g., Pollack *et al.* 1979, Smith 2004) is maintained by these processes. The overall depth of this dust loading varies with season and location, but in general may be as shallow as the planetary boundary layer depth (wide range between 0-15 km; cf. Hinson *et al.*, 2008) and perhaps as deep as 70 km or more based on Mars Express SPICAM results (Spectroscopy for Investigation of Characteristics of the Atmosphere of Mars; e.g., Montmessin *et al.* 2006). The vertical distribution of dust mixing ratio (i.e., the ratio of dust mass to atmospheric mass in a given volume) generally cannot be considered constant with height, although historically such a simplifying assumption has been widely used (e.g., Haberle *et al.* 1993, Forget *et al.* 1999). Instead, recent observations show that an elevated peak of dust mixing ratio at approximately 15-25 km often exists (e.g., McCleese *et al.* 2010), and it is likely that dust mixing ratio also peaks within the planetary boundary layer (PBL). Although robust observational confirmation of this local maximum within the PBL does not yet exist for Mars, it is assumed similar to what has been observed in Earth's atmosphere over deserts, since the PBL is where the atmospheric dust emission/injection occurs, and above the PBL vertical atmospheric transport of dust is significantly hindered by a temperature inversion (e.g., Arya 1999).

The terminal fall speed of a particle, driven by gravity and counteracted by external fluid drag, depends strongly on its shape, size, and the mass density of the ambient atmospheric medium, among other parameters (e.g., Pruppacher and Klett, 1997). The proportion of small to large dust particles in the atmosphere increases nonlinearly with altitude, due to the increasingly disparate fall speeds of the larger versus the smaller particles. This gravitational sedimentation process also results in semi-continuous dust deposition at the surface, as observed *in situ* by every successful Mars lander (e.g., Bell, III, *et al.* 2000, Goetz *et al.* 2005, Vaughan *et al.* 2010).

Dust particles also likely serve as condensation nuclei for water- and CO2-ice cloud particles on Mars (e.g., Colaprete *et al.* 1999, Colaprete and Toon 2002). The resulting ice crystals are significantly larger than the dust that they nucleated on (now entombed), and thus generally have a correspondingly higher terminal fall speed. Larger dust particles will generally nucleate ice crystals before smaller ones do, due to microphysical considerations (Pruppacher and Klett 1997). This process can both dramatically enhance the descent rate of all sizes of dust and impede the upward or lateral transport of dust through areas of the atmosphere where conditions are conducive for ice-cloud formation.

To maintain and modulate the observed atmospheric dust loading over seasonal (and shorter) timescales in the presence of gravitational sedimentation, there must be vertical and horizontal transport processes within the atmosphere which resupply the higher reaches of Mars' atmosphere with dust. Significant dust emission events (or *dust storms*; see Section 5.2.2, below) and small-scale vertical vortices (*dust devils*; see Section 5.3.2, below) are clearly part of this process. However, it is unclear that, taken alone, these two types of readily observable atmospheric phenomena can fully explain the observed details of the atmospheric dust loading for myriad reasons, which include the dust storms' stochasticity (in time) and the dust devils' confinement to the planetary boundary layer.

Numerical atmospheric models have been used to provide insight into further (nearly invisible) circulations that may play important roles in redistributing dust on Mars. Modeling studies generally agree that larger-scale horizontal circulations are both present and capable of redistributing airborne dust laterally. However, the nature of the vertical circulations needed to maintain the depth of the atmospheric dust loading is less clear. Global circulation model (GCM) studies, which typically use relatively coarse computational grids (i.e., with grid cells that are rarely less than 3x3 degrees in latitude and longitude, or ~180x180 km at the equator), have suggested that broad-area rising motions (seasonally varying, but occurring every day) along with saltation-based and dust devil dust emission processes may provide the mechanism by which the global dust loading is continuously maintained (e.g., Newman *et al.* 2002, Basu *et al.* 2004, Kahre *et al.* 2006). However, some higher-resolution (i.e., 5-40 km grid cells) atmospheric modeling studies have suggested that: 1) deep (up to ~60 km above the areoid) vertical transport is thermally-induced on a daily basis by the largest mountains on Mars (e.g., those of the Tharsis plateau), and appears capable of creating and maintaining the observed depth of the atmospheric dust loading (e.g., Rafkin *et al.* 2002, Michaels *et al.* 2006), and 2) vertical jets thermally-forced by smaller-scale severe topography (e.g., the rims of larger craters, canyon walls) effectively pierce the top of the planetary boundary layer (PBL), allowing dust that is otherwise trapped within the PBL to be entrained into larger-scale horizontal circulations between approximately 5-20 km above the surface (e.g., Rafkin and Michaels 2003, Michaels *et al.* 2006). Note that when spatially and temporally averaged, these two types of phenomena would approximate the broad-area rising motions discussed by the previously mentioned GCM studies (e.g., Rafkin 2011).

*5.2.2    A menagerie of Mars dust storms*

Mars is a desert planet (by terrestrial standards) with a substantial and dynamic atmosphere, so it is not surprising that it exhibits a broad range of dust storm activity. As on Earth (Section 5.1.2), martian dust storms often have a morphology that hint at the atmospheric phenomena (e.g., cold fronts, thermally-driven slope circulations) responsible for their genesis and evolution (e.g., Kahn *et al.* 1992). Opaque dust clouds associated with martian dust events commonly have a "cauliflower" appearance (e.g., Strausberg *et al.* 2005; see Figures 5.6 and 5.7), caused by atmospheric convection (turbulence) that is likely significantly driven by radiative transfer attributable to the dust itself. Sand-sized particles are unlikely to be a significant component of the airborne material more than several meters from the surface (e.g., Almeida *et al.* 2008) within a dust storm on Mars, unlike on Earth, due to the relatively thin martian atmosphere (global mean surface pressure of ~6 hPa, and thus greater particle fall speeds; see Figure 2.19).

The causation of many Mars dust storms is still largely unknown, in large part due to a paucity of relevant *in situ* meteorological observations. Clearly, however, one or more dust emission processes (see Section 4.2.2) must be operating, nearly all of which require a substantial near-surface horizontal wind magnitude. Although dust devils (see section 5.3.2) are common phenomena that enhance relatively calm winds at smaller scales, spacecraft observations suggest that dust storms are not triggered by such activity (e.g., Cantor *et al.* 2002, Balme *et al.* 2003, Cantor *et al.* 2006). Wind gusts due to boundary layer convective turbulence may also be capable of spontaneously initiating dust emission processes (Fenton and Michaels 2010), but since this process occurs on a daily basis over much of Mars, an infrequent triggering event (e.g., the passage of a front) appears necessary to generate a dust event this way.

It is also not well constrained when and where the surface dust emission is actually occurring during a given storm. Current observational capabilities for Mars relevant to measuring dust entrainment activity cover only the middle and late afternoon hours well. Even in the available afternoon images, the height of the dust cloud tops is uncertain, and as such poorly constrains whether the dust emission at the surface mirrors the spatial distribution of the dust clouds (or is still occurring at all). All clouds of dust are subject to transport by the atmosphere, and may potentially translate with their progenitor disturbance long after dust emission has ceased.

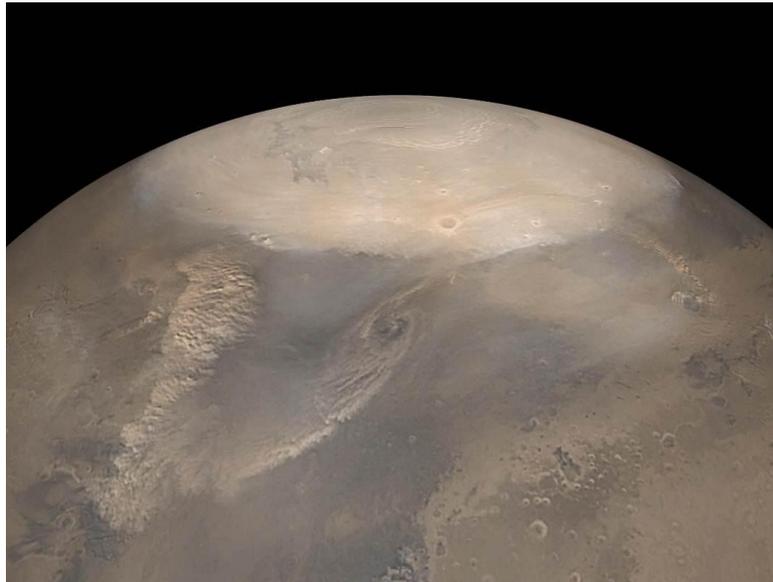

Figure 5.6. Examples of early spring dust storms associated with the retreated north polar seasonal cap of Mars (brighter circular feature). A mosaic of approximate-color wide angle images from NASA Mars Global Surveyor's Mars Orbiter Camera, taken in 2002 (note that this mosaic exhibits a significantly limited range of local times due to the spacecraft's orbit). At least three arcuate dust events are visible. Image credit: NASA/JPL/Malin Space Science Systems.

Polar cap dust storms (see example in Figure 5.6) are perhaps the best understood of Mars dust storms. These dust events typically are of the local, and rarely, the regional dust storm varieties of Martin and Zurek (1993) and Cantor *et al.* (2001). In the classification scheme of Cantor *et al.* (2001) regional dust storms (areal coverage of >= 1.6x106 km2) have a duration of >3 sols (Mars-days), and local dust storms exhibit a lesser areal extent and duration. This type of dust storm typically occurs in local spring and summer along and near the retreating seasonal carbon dioxide ice cap. Their commonly curvilinear morphology, circumpolar translation, and common occurrence strongly suggest that strong winds associated with cold fronts trigger these events (e.g., Barnes *et al.* 1993, James *et al.* 1999, Cantor *et al.* 2001). The lag deposit of dust left behind by the sublimation of dusty $CO_2$ ice in the retreating seasonal cap may also provide an additional dust source for these phenomena (James *et al.* 1999). Figure 5.6 illustrates (at the upper left of the image) that such storms can also extend significantly over the ice cap.

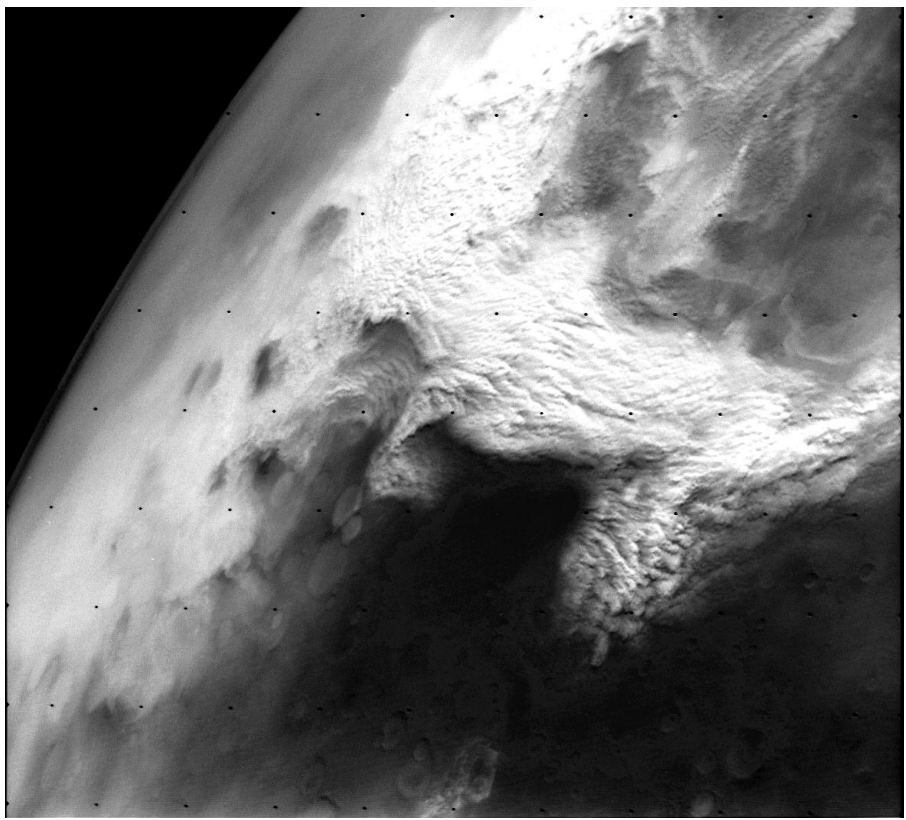

Figure 5.7. Example of a regional dust storm over the Thaumasia region of Mars (in the midlatitudes of the southern hemisphere). A Viking Orbiter 2 image (176B02) acquired in 1977. The textured cloud of dust is approximately 1000 km in length. Soon after this image was obtained, this storm became the progenitor of the first planet-encircling dust storm observed by the two Viking Orbiter spacecraft.

Local and regional (non-polar) dust storms (see example in Figure 5.7) are also relatively common on Mars, with hundreds occurring each Mars-year without a bias towards either the north or south hemisphere (Cantor *et al.* 2001). Some of these storms originate in the high latitudes and travel equatorward, with a subset even crossing the equator (e.g., Cantor *et al.* 2001, Wang *et al.* 2003). Occasional, cooperative interaction between different types of atmospheric circulations may explain the occurrence and behavior of such storms (Wang *et al.* 2003, Wang 2007). Other forms of these dust events may be ultimately due to similar interactions between slope winds, atmospheric tides, and other atmospheric phenomena. Approximately half of all regional dust storms appear to result from the merger of two or more local dust storms (Cantor *et al.* 2001). Similarly, planet-encircling dust storms seem to result from the combined effects of multiple regional events (e.g., Cantor 2007). Regional dust events can result in an enhanced surface dust cover (often temporary) over areas downwind as entrained dust falls out of the atmosphere (e.g., Thomas and Veverka 1979, Geissler 2005).

A planet-encircling dust storm on Mars (see example in Figure 1.2b) is a dust event that shrouds one or both hemispheres of the planet in dust for hundreds of sols (Martin and Zurek 1993, Cantor 2007) (up to a season in duration; the polar regions often escape much of the dust pall). These events generally occur stochastically with a multiyear incidence (e.g., 1971, 1973, 1977a, 1977b, 2001) (Martin and Zurek 1993, Cantor 2007). The genesis of such an event is not clear, but appears to involve one or more relatively discrete regional dust storms intensifying and somehow inducing additional lifting centers (areas where surface dust emission is occurring) near the edge of the overall area of dust obscuration as it expands, which leads to a great quantity of entrained dust that is spread more globally by the atmospheric circulation (e.g., Thorpe 1979, Cantor 2007). Dust can be lofted to 60 km or more in altitude during these events (Cantor 2007). It is widely believed that a primary mechanism by which the dust emission of such a storm ends is when the associated dust pall modifies the atmospheric circulation and structure in a way that inhibits further lifting centers from forming, largely via the radiative transfer effects of the atmospheric dust loading (e.g., Pollack *et al.* 1979). Widespread enhanced surface dust cover results from a planet-encircling dust storm, but appears to be sporadically removed (generally at a rate that does not produce significant clouds of dust) as the atmosphere gradually clears (e.g., Thomas and Veverka 1979, Geissler 2005, Cantor 2007).

*5.3 Dust entrainment by small-scale vertical vortices on Earth and Mars*

The dust storm discussions in Sections 5.1 and 5.2 implicitly excluded smaller-scale and more ephemeral atmospheric dust entrainment phenomena. Chief among these, especially in terms of instantaneous dust mass flux (due to sustained wind speeds much greater than ambient conditions), are the near-surface vertical vortices. A quasi-continuum of such vortices are a natural consequence of three-dimensional atmospheric thermodynamics, and they can conveniently be classified by energy source.

*5.3.1 Vertical vortices with greater sources of energy*

Tornadoes, the vertical vortex type with the greatest near-surface wind magnitude (often 100 m s-1 or more), are ultimately powered by the release of latent heat due to the phase change of water substance in the atmosphere (e.g., cloud formation). The quantity of water substance in Mars' atmosphere is far too small to support similar phenomena. These phenomena have the potential to entrain large quantities of dust and other particulates. However, their relative rarity, geographically-limited occurrence, and a frequent association with rain and other precipitation hampers the overall dust emission and transport effectiveness of these phenomena.

A second genre of small-scale vertical vortices are those which derive most of their energy from significant wind shear. On Earth, examples include what are colloquially termed "gustnadoes" (along thunderstorm gust fronts; Bluestein 1980) and "mountainadoes" (horizontal vortices tilted into the vertical by significant topographic obstacles; Bergen 1976), with typical near-surface wind magnitudes less than or equal to that of a weak tornado. Like tornadoes, their relative rarity limits the overall significance of their dust entrainment effects.

Such vortices powered by wind shear are theoretically possible on Mars, due to complex topography and widespread slope flows (i.e., winds driven by heating/cooling imbalances along a significant topographic slope). However, no observational detections/identifications have yet been made. This could potentially be due to the limited temporal coverage of relevant high-resolution orbital spacecraft imaging (only in the mid-afternoon), or perhaps due to obscuration from adjacent larger-scale dust lifting phenomena.

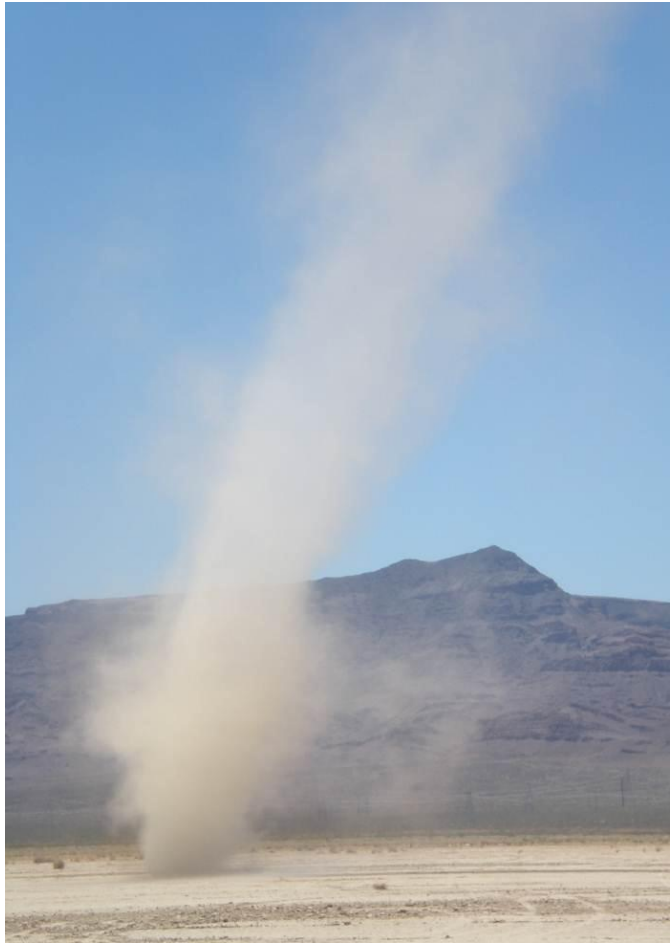

Figure 5.8. The lower portion of a dust devil (~5 m wide at its base) on a playa just southwest of Boulder City, Nevada, USA (June 2010). This particular dust devil persisted for ~15 minutes before dissipating.

### 5.3.2 Dust devils

A third category of small-scale vertical vortices, powered by near-surface thermal buoyancy, are commonly termed *devils* (dust devils being of particular relevance here). Such circulations are more generally referred to as vertical convective vortices, which reflects their causation and control by thermally-forced boundary layer convection (i.e., the structured near-surface turbulence induced by a large temperature imbalance between the ground and the adjacent atmosphere). It is important to note that a given dust devil circulation may not be visible if there is no available dust to entrain, the circulation does not extend fully to the surface, or if the circulation is too weak to entrain dust. An excellent and thorough review of Earth and Mars dust devil literature was published in 2006 (Balme and Greeley 2006), and thus the discussion here will concentrate on comparing the dust devils on the two planets and including recent findings.

Visible dust devils are relatively common during the daytime in the arid/semiarid areas of Earth, and over nearly all of Mars, since the sun-warmed ground provides the energy needed to drive boundary layer convection. The diameter of terrestrial dust devils ranges from less than a meter to tens of meters, and they range in height from a meter to hundreds of meters (e.g., Balme and Greeley 2006, Lorenz 2011; example in Figure 5.8). In contrast, those on Mars (examples in Figure 5.9) can be significantly larger, up to 8 km in height and hundreds of meters in diameter (Fisher *et al.* 2005). These maximum dimensions can largely be explained by a greater solar energy (absorbed by the surface) to atmospheric mass ratio on Mars (tens of times greater than that on Earth), providing martian dust devils with a more effective power source. The duration of these vortices on both worlds is thought to be similar, a few seconds to thousands of seconds (Michaels and Rafkin 2004, Balme and Greeley 2006).

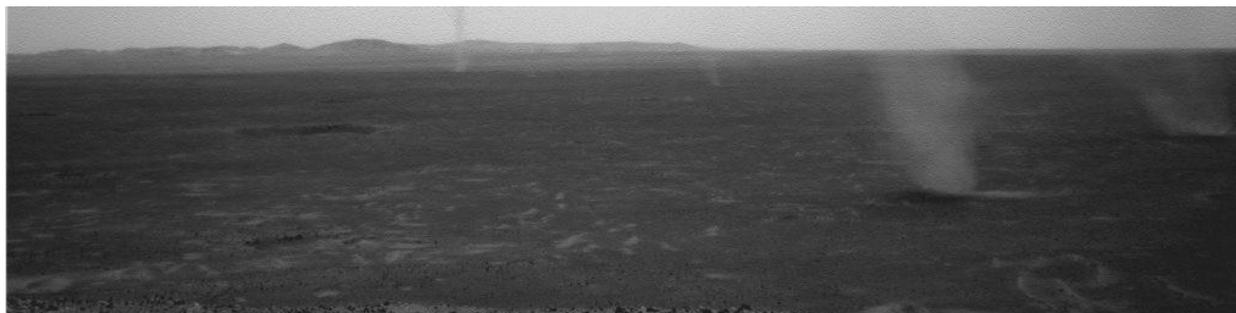
Figure 5.9. Mars Exploration Rover Spirit image from its Navigation Camera, showing multiple simultaneous smaller dust devils of different sizes and intensities. Acquired on Sol 568 within Gusev Crater, Mars. Image credit: NASA/JPL/Cornell/USGS

The structure of a typical dust devil on Earth or Mars consists of a vortex tube in which the walls are composed of an intense dusty helical updraft and the core is relatively dust-free and calm away from the surface. At the top of the vortex tube, dust begins to horizontally disperse while continuing to move upwards, as it becomes entrained in the larger-scale boundary layer convective updraft associated with the dust devil. At the bottom of the vortex, surface material is scoured from the surface largely by large tangential winds within the vortex wall (Balme *et al.* 2003) (see Chapters 2 and 4 for more discussion of surface particle emission). On Earth, the most massive particles (usually sand or small clumps of soil) detrain quickly from the vortex and often create an "ejecta skirt" (Metzger 1999) of material falling back to the surface that surrounds the lowest meters of the vortex. Obvious ejecta skirts have not typically been observed on Mars (Neakrase and Greeley 2010b), although it must be noted that such features are quite difficult to identify in orbital imagery, and *in situ* imaging has been limited to a few sites.

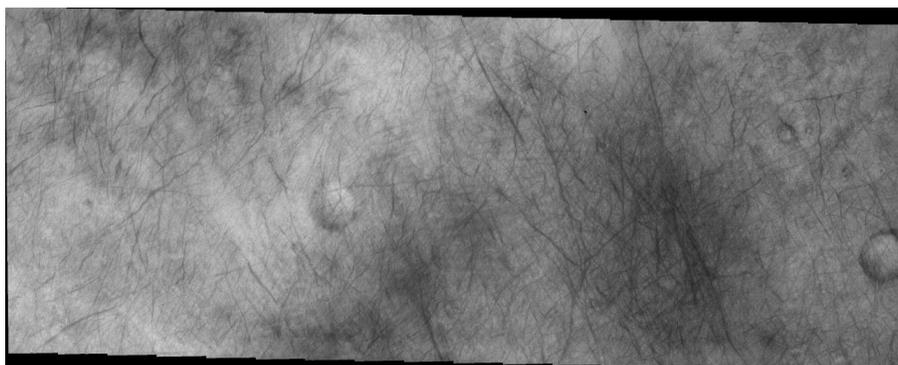
Figure 5.10. Example of orbitally observed dust devil tracks on Mars at ~69.5°S (north is to the left; area of ~66x20 km with a pixel size of ~19 m). Note the concentrations of dark dust devil tracks. Mars Odyssey THEMIS image V33501003 (Christensen *et al.* 2011).

Dust devils can leave an aerially and/or orbitally visible surface trace of their passage, commonly termed a "track". Such features are relatively rare on Earth, likely because the mobile particulates and underlying substrate lack sufficient visible contrast. However, numerous dark (and many fewer bright) dust devil tracks are visible on many areas of the Mars' surface (example in Figure 5.10, often exhibiting complex loops and turns that are indicative of the dynamical convective boundary layer processes that the phenomena were embedded in (e.g., Malin and Edgett 2001, Michaels 2006). Numerical modeling suggests that on Mars, the width of some dust devil tracks may only be representative of the dust devil radius (versus diameter) (Michaels 2006), due largely to the much higher wind speed threshold for particle entrainment on Mars versus Earth (see Section 2.4). Areas where martian dust devil tracks are concentrated exhibit a significantly altered albedo (versus adjacent areas without such tracks), affecting the surface temperature and potentially local- and regional-scale atmospheric circulations (Malin and Edgett 2001, Michaels 2006, Fenton *et al.* 2007).

Due in large part to their common occurrence on Earth and especially Mars, dust devils act as an important conduit for the atmospheric entrainment of dust and other surface particulates within the boundary layer (i.e., to a height from the surface of approximately 1-4 and more km on Earth, and 1-10 and more km on Mars). In regions prone to dust devil activity on Earth, air quality (specifically with regard to particulate content) is reduced due to the dust aerosols generated by these phenomena (Gillette and Sinclair 1990, Mattsson *et al.* 1993), adding significantly to any air quality reduction attributable to non-vortical processes (e.g., dry boundary layer convection) alone. Such aerosols may subsequently have significant effects on the atmospheric energy budget via their modulation of cloud formation processes (acting as cloud condensation and ice nuclei (Twomey 1974, DeMott *et al.* 2003) and their additional scattering and absorption of incoming and outgoing electromagnetic radiation (chiefly in the visible and infrared regions of the spectrum). The significance of such effects attributable solely to dust devils on Earth has not yet been well quantified (Koch and Renno 2005), but is likely far from negligible (e.g., Balkanski et al 2007, Mahowald *et al.* 2010, Kok 2011a). Mars global climate modeling work suggests that these two potential indirect effects of dust devils may have significant roles in the planet's climate (e.g., Newman *et al.* 2002, Basu *et al.* 2004, Wilson *et al.* 2008), since dust devils potentially contribute a large fraction of the ever-present atmospheric dust loading (Basu *et al.* 2004). It has been estimated that martian dust devils may entrain $2.3 \pm 1 \times 10^{11}$ kg of surface material annually, a significant fraction of that likely entrained by local and regional dust storms (Whelley and Greeley 2008).

## 6. Conclusions and remaining questions

As shown in the previous chapters, the past decade has produced considerable progress in understanding the physics of wind-blown sand and dust. Some of the most critical advances include an improved understanding of the characteristics of steady state saltation from advances in theory, numerical models, and wind tunnel experiments (Section 2.3); important insights into the physics of aeolian ripples and dunes and into the origin of complex dune shapes gained through both numerical modeling and cleverly designed experiments (Section 3.3); the discovery of plausible mechanisms to move sand and lift dust on Mars (Sullivan et al. 2008, Kok 2010b, Fenton and Michaels 2010), thereby starting to reconcile the conundrum of widespread aeolian activity (e.g., Cantor *et al.* 2001, Bridges *et al.* 2012a) despite Mars' thin atmosphere; and an improved understanding of the main atmospheric dust-lifting phenomena over arid and semi-arid areas on Earth, including the insight that strong early morning near-surface winds resulting from the breakdown of the low-level jet are likely responsible for a substantial fraction of dust emissions over the Sahara (Knippertz 2008, Schepanski et al. 2009).

Despite these advances, many important questions concerning the physics of wind-blown sand and dust remain. Some of these questions are outlined in the next section.

### *6.1 Important remaining questions regarding the physics of wind-blown sand*

One of the foremost remaining questions in understanding the physics of aeolian saltation is the effect of variability on short time scales (~1 – 100 sec; Baas 2006) on the time-averaged properties of saltation, such as the mass flux. In particular, saltation in the field is often intermittent (Stout and Zobeck 1997), with bursts of saltation flux occurring in *aeolian streamers* (Baas and Sherman 2005). However, theoretical and numerical models of saltation have generally not accounted for this variability. Instead, most models have treated saltation as a uniform and continuous process at a constant wind shear stress (e.g. Anderson and Haff 1988, 1991, Werner 1990, Kok and Renno 2009a). A critical remaining challenge is thus to understand how the turbulence characteristics of the atmospheric boundary layer determine the variability in saltation transport (Schönfeldt 2003, Baas 2006), and how that variability affects the time-averaged properties of saltation. Moreover, in order to help translate insights into variability on short time scales to advances in modeling saltation, dune formation, and dust emission on longer time scales, future research needs to better relate the turbulence characteristics of the near-surface layer in which saltation takes place to the meteorological forcing of the boundary layer (e.g. Hunt and Carlotti 2001).

In addition to the effect of saltation intermittency on time-averaged saltation properties, there are several other important remaining questions in the physics of wind-blown sand. For instance, what is the effect of dust and sand electrification on the characteristics of saltation (Zheng et al. 2003, Kok and Renno 2006, 2008) and possibly on atmospheric chemistry on Mars (Atreya *et al.* 2006, Farrell *et al.* 2006, Kok and Renno 2009b)? What is the effect of saltation on the turbulence intensity (Taniere *et al.* 1997, Nishimura and Hunt 2000, Zhang *et al.* 2008, Li and McKenna Neuman 2012) and the Lagrangian (correlation) time scale of the turbulence in the saltation layer (Anderson 1987b, Kok and Renno 2009a), and how does that affect the properties of saltation? What is the cause of the substantial underestimation of the aerodynamic roughness length measured in wind tunnel studies compared to field studies (Sherman and Farrell 2008) and what limit does this discrepancy impose on the extent to which wind tunnel studies can be used to understand natural saltation? And does the von Kármán constant ($\kappa$) depend on the saltation load (Li *et al.* 2010, Sherman and Li 2012), and, if so, how does that affect theoretical predictions of the mass flux and other saltation properties?

Answering the above questions will improve our ability to quantitatively model saltation transport, which is critical for improving our understanding of the formation and dynamics of ripples and dunes. For instance, one of the key remaining issues for successfully modeling planetary dunes is an understanding of what controls the flux saturation length ($L_s$) that dictates the length scale of dunes (Section 3.1). Although there is strong experimental evidence supporting a scaling of $L_s$ with the grain diameter multiplied by the grain-to-fluid density ratio under different physical conditions (Section 3.1), a significant dependence of $L_s$ on the shear velocity of the driving fluid is observed for subaqueous saltation (Durán et al. 2011a). Because of the large atmospheric density on Venus and Titan compared with that on Earth or Mars, sand transport on Venus and Titan is expected to be much more akin to subaqueous saltation than to its aeolian counterpart (Section 2.4). To correctly model dune and ripple dynamics under different environmental conditions, it is thus critical to achieve an improved physical description of the sand flux and flux saturation length that accounts for the differences in transport mechanisms due to these differences. Furthermore, the dependence of the saltation fluid threshold on the local slope could also play a role for the selection of the minimal dune size (Andreotti and Claudin 2007, Parteli et al. 2007b), and the relevance of this bed slope effect for the dune shape should thus be investigated in future modeling. Also, as discussed in Section 3.2, important weaknesses of current dune models must be addressed in the future, such as the modeling of the separation bubble at the lee and the inclusion of secondary flow effects, which can play a fundamental role for the shape of seif and star dunes (Lancaster 1989a, b, Tsoar 2001). In regards to ripples, the theoretical prediction that the steady-state height and spacing of ripples should be independent of the wind friction velocity (Manukyan and Prigozhin 2009) should be experimentally verified (Andreotti et al. 2006). Also, Martian TARs (Section 3.3) still pose an intriguing puzzle for planetary geomorphologists, as these bedforms could not be unambiguously classified as ripples, megaripples or dunes. Modeling could potentially shed light into the mechanisms at play in the formation of these unique Martian bedforms.

Although numerical modeling has yielded many important insights into the mobility and dynamics of Martian dunes, further critical insights are currently arising from images taken by the HiRISE camera on board the Mars Reconnaissance Orbiter, which has an unprecedented resolution of about ~30 cm. These images show that planet-wide displacements of ripples and dunes from several centimeters up to a few meters are currently taking place within a time-scale of a few months or years (e.g., Hansen et al. 2011, Bridges et al. 2012a), a scenario that contrasts with earlier expectations that saltation events should be rather rare on Mars (Arvidson et al. 1983, Moore 1985, Sullivan et al. 2005). As discussed in Section 2.4, recent models of saltation yield a surprisingly low martian impact threshold velocity (Almeida et al. 2008, Kok 2010a, b) which remains to be verified experimentally using wind tunnel simulations in a Martian atmosphere (White 1979). Could the observed high rates of bedform migration on Mars be explained by a low impact threshold velocity? While it is not yet possible to draw conclusions about dune stability and dynamics from the newest satellite observations, important questions regarding the mechanisms responsible for the surficial modifications detected on the surface of some Martian dunes could be investigated with the help of modeling.

*6.2    Important remaining questions regarding the physics of wind-blown dust*
One of the critical remaining questions in the physics of wind-blown dust is the dependence of the vertical dust flux on wind and soil conditions (Section 4.1.2). Although more physically-based models of the vertical dust flux have been developed over the past decade (Marticorena and Bergametti 1995, Shao *et al.* 1996, Alfaro and Gomes 2001, Shao 2001, 2004, Kok 2011a), substantial uncertainties in their predictions remain (e.g., Shao *et al.* 2011b and Figure 4.2). In particular, we have an inadequate understanding of the dependence of the dust flux on soil moisture, non-erodible elements, biological and physical crusts, and other soil parameters (e.g., Shao 2001, Okin 2008). Moreover, some dust emission schemes require soil parameters that are not usually available on regional or global scales (Alfaro and Gomes 2001, Shao 2001, 2004). As a consequence, most atmospheric circulation model parameterizations of the dust flux are based on semi-empirical relations fitted to measurements of the vertical dust flux as a function of wind speed for specific soil conditions (e.g., Ginoux *et al.* 2001, Zender *et al.* 2003a, Balkanski *et al.* 2007). But since dust emission occurs for a wide range of soils (Gillette 1979, Nickling and Gillies 1993), such parameterizations probably fail to capture a substantial part of the complex dependence of the vertical dust flux on wind and soil conditions (Gillette, 1974, McKenna Neuman *et al.* 2009). In order to facilitate the development of improved physically-based dust flux parameterizations for implementation into atmospheric circulation models, a range of questions relating the movement of soil particles in saltation to the resulting vertical dust flux need to be answered. For instance, how does the strength and variability of interparticle forces acting on dust in soil aggregates depend on soil moisture, the soil size distribution, and soil mineralogy? What is the balance between dust emission due to dust aggregate fragmentation (Kok 2011a) and due to other processes, such as the removal of clay coatings on sand particles (Bullard *et al.* 2004)? How do vegetation characteristics such as porosity and spatial distribution affect the dust flux (Okin 2008, Okin *et al.* 2011)? And how do the turbulent properties of the near-surface flow affect dust emissions? A better understanding of these and other pertinent questions should facilitate more physically realistic parameterizations of the vertical dust flux as a function of wind speed and soil conditions.

In addition to an improved understanding of the small-scale physics of dust emission, improving the fidelity of simulations of the global dust cycle and its role in the Earth system also requires an improved understanding of the meteorological phenomena forcing dust emission. For instance, although numerous atmospheric phenomena for dust lifting have been identified during the last decade (see Section 5.1.2), their relative importance as well as their seasonal and annual frequency remain to be quantified by future studies. Another important remaining concern in modeling the dust cycle in regional and global models is accounting for the occurrence of small-scale atmospheric dust-lifting phenomena (Cakmur et al. 2004, Koch and Renno 2005, Williams 2008). For instance, dust emissions induced by small-scale turbulent winds, such as the strong near-surface winds associated with cold fronts and density currents, need to be better parameterized in models (Cakmur et al. 2004). Furthermore, dust emission associated with convective activity, such as dust fronts induced by mesoscale convective systems, need to be better understood (e.g. Reinfried et al. 2009, Bou Karam et al. 2011).

A substantial number of questions also remain regarding aspects of wind-blown dust and sand particular to Mars. Similar to Earth, there is an inadequate understanding of the influence of ground crusts and other cohesive phenomena on dust and sand transport. Furthermore, due to the existence of only a handful of relevant in situ observations and measurements on the entire planet (e.g., Soderblom *et al.* 2004, Goetz *et al.* 2010), the spatial distribution of such cohesive phenomena, surface sediment size distribution, aerodynamic surface roughness, and other similar properties is not well constrained. What are the sources and sinks for mobile surface particles (e.g., Pleskot and Miner 1981, Christensen 1988)? How well do they stand up to the rigors of participating in aeolian processes over martian geologic timescales (e.g., 10 thousand, 1 million, or even 1 billion years)? Clearly, measurable aeolian sand transport is occurring on contemporary Mars in several locales (e.g., Bridges et al. 2012a), but how much is due to the time-integrated effect of lower-mass-flux day-to-day occurrences versus much rarer high-mass-flux stochastic events (e.g., Chojnacki *et al.* 2011)? What are the relative importance of dust devils, stochastic dust storms, and topographically-induced circulations in maintaining the nearly ever-present atmospheric dust loading on Mars (Newman *et al.* 2002, Basu *et al.* 2004, Kahre *et al.* 2006, Michaels *et al.* 2006, Rafkin 2011)? How are martian dust storms triggered, particularly the rarer planet-encircling variety? How do such dust storms intensify, and what leads to their decay? A combination of further orbital observations and in situ measurements, numerical modeling, and laboratory experiments are needed to address these questions.

**Acknowledgements**

The authors would like to thank Orencio Durán, Thomas Pähtz, Oscar Matthews, Yaping Shao, Thomas Gill, Earle Williams, Christopher Hugenholtz, Rob Sullivan, and Ralph Lorenz for helpful discussions, Mark Gordon for recalculating the splash data in Gordon and McKenna Neuman (2011) for use in Figure 2.6, Haim Tsoar for providing the sand roses used in Figure 3.4, Sebastian Engelstaedter for providing us with Figure 5.1a, and Thomas Pähtz, Jack Gillies, and Rob Sullivan for helpful comments on parts of the manuscript. The constructive reviews of Orencio Durán and an anonymous referee also helped improve the manuscript. Some of this material is based upon work supported by the National Science Foundation under Awards AGS 0932946 and 1137716, by FUNCAP, CAPES and CNPq (Brazilian agencies) and by the German Research Foundation (DFG) through the Cluster of Excellence "Engineering of Advanced Materials".